\documentclass[letterpaper]{article}
\usepackage[margin=1in]{geometry}
\usepackage{achemso}
\setkeys{acs}{articletitle = true}
\usepackage{array}
\usepackage{microtype}
\usepackage[numbers]{natbib}
\usepackage{natmove}
\usepackage{caption}
\usepackage{comment}

\usepackage{graphicx}
\usepackage{amssymb}
\usepackage{multirow}
\usepackage{color}
\usepackage{amsmath}
\usepackage{threeparttable}
\usepackage{subcaption}
\usepackage[nolist,nohyperlinks]{acronym}
\usepackage{hyperref}

\graphicspath{./figures/}

\newcommand{\tg}{$T_\text{g}$}

\newcommand{\cis}{\emph{cis}-1,4}

\newcommand{\ie}{\textit{i.\,e.}}
\newcommand{\eg}{\textit{e.\,g.}}

\begin{document}   

\vspace{4cm}
\begin{center}{\Large \bf  Dynamics and Rheology of Polymer Melts via Hierarchical Atomistic, Coarse-grained, and Slip-spring Simulations}\\
	\vspace{2cm}

	{\fontsize{15}{14} Alireza F. Behbahani$^{\dagger, @}$, Ludwig Schneider$^{\ddagger, @}$, Anastassia Rissanou$^{\dagger}$, Anthony Chazirakis$^{\dagger}$, Petra Ba\v{c}ov\'{a}$^{\dagger}$, Pritam Kumar Jana$^{\ddagger}$, Wei Li$^{\P}$, Manolis Doxastakis$^{\P}$, Patrycja Poli\'{n}ska$^{\S}$, Craig Burkhart$^{\parallel}$, Marcus M\"{u}ller$^{*,\ddagger}$, and Vagelis A. Harmandaris$^{*,\perp,\dagger,\#}$\\}
	\
	\\
	\textit{ $\dagger$ Institute of Applied and Computational Mathematics, Foundation for Research and Technology - Hellas, Heraklion GR-71110, Greece\\
	$\ddagger$ Institute for Theoretical Physics, Georg-August University G\"{o}ttingen, Germany\\
	$\P$ Department of Chemical and Biomolecular Engineering, University of Tennessee, Knoxville, Tennessee 37996, USA\\	
	$\S$ Goodyear S.A., Avenue Gordon Smith, Colmar-Berg L-7750, Luxembourg\\
	$\parallel$ The Goodyear Tire and Rubber Company, 142 Goodyear 
	Blvd., Akron, Ohio   44305, USA\\
	$\perp$ Department of Mathematics and Applied Mathematics, University of Crete, Heraklion GR-71110, Greece\\
	$\#$ Computation-based Science and Technology Research Center, The Cyprus Institute, Nicosia 2121, Cyprus\\}
	\
	\\
	$@$ A.~F.~Behbahani and L.~Schneider contributed equally to this work\\
	$*$ E-mail: mmueller@theorie.physik.uni-goettingen.de, harman@uoc.gr  
\end{center}

\begin{abstract}
     A hierarchical (triple scale) simulation methodology is presented for the prediction of the dynamical and rheological properties of high molecular weight entangled polymer melts. 
     The methodology consists of atomistic, moderately coarse-grained (mCG), and highly coarse-grained slip-spring (SLSP) simulations.
     At the mCG level, a few chemically bonded atoms are lumped into one coarse-grained bead. At this level, the chemical identity of the underlying atomistic system, and the interchain topological constraints (entanglements) are preserved. The mCG interaction potentials are derived by matching local structural distributions of the mCG model to those of the atomistic model through iterative Boltzmann inversion.   
     For matching mCG and atomistic dynamics, the mCG time is scaled by a time scaling factor, which compensates for the lower monomeric friction coefficient of the mCG model than that of the atomistic one. 
     At the SLSP level, multiple Kuhn segments of a polymer chain are represented by one coarse-grained bead. The very soft nonbonded interactions between beads do not prevent chain crossing and, hence, can not capture entanglements. The topological constraints are represented by slip-springs, restricting the lateral motion of polymer chains. A compensating pair potential is used in the SLSP model, to keep the static macromolecular properties unaltered upon the introduction of slip-springs.       
     The static and kinetic parameters of the SLSP model are determined based on the lower level simulation models. Particularly, 
     matching the orientational autocorrelation of the end-to-end vector, we determine the number of slip-springs and calibrate the timescale of the SLSP model. 
     As the test case,  the hierarchical methodology is applied to \cis-polybutadiene (cPB) at $413$ K. Dynamical single-chain and linear viscoelastic properties of cPB melts are calculated for a broad range of molecular weights, ranging from unentangled to well-entangled chains. The calculations are compared, and found in good agreement, with experimental data from the literature.
     
\end{abstract}
 
\section{Introduction}   

Nowadays, with the progress of available computational power, molecular simulations have become a powerful and predictive tool for the investigation of the chain dynamics and the rheology of polymer materials. 
However, molecular simulations of polymer materials of industrial relevance are not straightforward because of the extremely broad range of time and length scales involved in macromolecular chains of high molecular weights.
As an example, relevant timescales of a polymer melt extend from the periods of covalent bond vibrations ($\approx 10^{-14}$ s) to times of entangled chain relaxation that might be of the order of seconds, even at temperatures well above the glass transition temperature, \tg.
Moreover, relevant length scales of a polymer material cover atomic sizes ($10^{-10}$ m) up to the size of a polymer chain ($\approx 10^{-8}$--$10^{-7}$ m) and possibly larger length scales that characterize the structure of hybrid multiphase systems. 
Therefore, a single modeling approach cannot simultaneously address all relevant length and time scales, and it is necessary to combine simulation methods of different spatiotemporal scales.\cite{tschop1998simulation, theodorou2007hierarchical,padding2011systematic, harmandaris2006hierarchical} 

The natural route for increasing the length and time scales accessible to simulations is to use a \ac{CG} representation of polymer chain.\cite{tschop1998simulation,theodorou2007hierarchical,padding2011systematic,sukumaran2009modeling}
A \ac{CG} model is commonly obtained by lumping a group of chemically bonded atoms into one superatom (\ac{CG} bead or particle) and tuning the \ac{CG} model to reproduce some target structural, dynamical, and/or thermodynamic properties of the fine-grained, more detailed, model. A systematic parametrization of CG models includes deriving interaction potentials between \ac{CG} beads and/or calibration of the \ac{CG} model parameters using data from the more detailed atomistic simulations; such a procedure is typically called "bottom-up coarse-graining".  

An alternative to the systematic bottom-up derivation of coarse-grained representations are top-down \ac{CG} models \cite{muller2011studying}. In these approaches, a universal representation (\eg, a bead-spring model) is adopted that only incorporates the relevant interactions, such as the connectivity along the macromolecular backbone and the segmental repulsion, limiting fluctuations of the melt density -- in a computationally efficient way. The strengths of these relevant interactions are related to experimentally measurable quantities, such as the spatial extent of the macromolecules in a melt and the isothermal compressibility. Comparing to experimental data or lower-scale simulations, the strength of the interactions or time and length scales are identified for a specific material. Universal models are also widely used for studying (in a qualitative manner) the general behavior of polymer materials.\cite{kremer1990dynamics}

Because of the averaging over microscopic details, \ac{CG} potentials are softer than the corresponding fine-grained potentials.
However, at low or moderate degrees of coarse-graining (lumping a few atoms in a \ac{CG} bead) the interactions remain strong enough to prevent chain crossing and to preserve entanglement constraints between polymer chains.
So far, various moderately coarse-grained models, which stay close to the microscopic structure, have been proposed in the literature for different polymers through different coarse-graining procedures.
Typical examples include the works by Tsch\"op et al.~\cite{tschop1998simulation,tschop1998backmapping} who introduced a systematic coarse-graining procedure and applied it to derive CG potentials for polycarbonates. They generated long atomistic chains, with a proper distribution of torsional angles,
through \ac{MC} simulations and mapped them onto a \ac{CG} representation by replacing groups of atoms with CG beads. The effective bonded potentials (bond length, angle, torsion) of the CG chain were calculated from the Boltzmann inverse of the corresponding distributions. Finally, the nonbonded interactions between \ac{CG} beads were adjusted to reproduce the correct density of the polymer. Reith et al.~\cite{reith2003deriving} introduced an iterative procedure, called iterative Boltzmann inversion, for deriving effective \ac{CG} potentials.
They also determined nonbonded \ac{CG} interactions by matching \ac{RDF} of the \ac{CG} melt to that of the atomistic melt.  
Harmandaris and collaborators\cite{harmandaris2006hierarchical,harmandaris2009dynamics,fritz2009coarse} devised \ac{CG} models for \ac{PS} chains and studied the structural and dynamical properties of \ac{PS} melts. 
Ohkuma and Kremer\cite{ohkuma2017comparison} studied the properties of two different CG models (with and without pressure correction) for \cis-polyisoprene.
Kempfer et al.~\cite{kempfer2019realistic} 
used a trajectory-matching approach for coarse-graining of \ac{cPB}. In their approach, the parameters of the \ac{CG} model were obtained by optimizing the matching between the trajectory of the \ac{CG} model, produced through DPD simulation, and the reference trajectory of the more detailed model.   
Recently, Shahidi et al.\cite{shahidi2020coarse} used an inverse Monte Carlo method for constructing local-density dependent CG potentials for \cis-polyisoprene; 
in addition to the conventional bonded and pair distributions, the derived CG potentials reproduce the distribution of the nearest-neighbors of the underlying atomistic model.
Moderately CG models have also been used to study interfacial polymer systems, like polyamide/graphene\cite{eslami2011coarse}, \acs{PS}/gold\cite{johnston2013hierarchical}, and polyisoprene/graphite\cite{pandey2014multiscale} systems. 

At high degrees of coarse-graining (lumping many atoms in one \ac{CG} bead), which are necessary for the simulation of high molecular weight polymers, pair interactions are very soft and they can not prevent chain crossing, and hence, the model fails to capture the topological constraints (entanglements) between polymer chains.\cite{padding2011systematic,sukumaran2009modeling,muller2011studying,chappa2012translationally} 
Therefore, to preserve entanglement effects, which are key features of polymer dynamics, additional constraints should be added to the model.\cite{padding2011systematic,padding2001uncrossability,chappa2012translationally,sukumaran2009modeling}
To reintroduce the reptation dynamics into highly \ac{CG} models, single-chain and multi-chain slip-link\cite{hua1998segment1,hua1998segment2,hua1999segment3,masubuchi2001brownian,likhtman2005single,Muller2008Oct,DelBiondo2013May,schieber2014entangled} and \ac{SLSP}\cite{chappa2012translationally,uneyama2012multi,ramirez2018detailed,ramirez2017multi,masubuchi2014simulating,masubuchi2018comparison,masubuchi2018multichain,langeloth2013recovering,masubuchi2016multichain}  models have been developed.

Single chain models restrict the motion of monomers by springs. One end of a spring can slide \cite{likhtman2005single} along the polymer contour or hop from one segment to a neighboring one along the chain contour \cite{Muller2008Oct}. The other end of the spring is anchored to a background. These constraints mimic entanglement effects, including constraint release and contour-length fluctuations. The anchoring to a background, however, requires additional assumptions to describe deformation or flow.\cite{DelBiondo2013May} 

In multi-chain \ac{SLSP} models, the topological constraints are represented by additional springs between beads. Reptation is achieved similarly by sliding the springs on both ends along the backbone of the chains. Since the constraints are not anchored to the background the models are translationally invariant. 
Multi-chain \ac{SLSP} models are particularly useful since they can describe complex systems with spatial heterogeneities.\cite{chappa2012translationally}  

So far, some attempts have been made to parameterize slip-link and SLSP models for specific polymer systems.
Sukumaran and Likhtman\cite{sukumaran2009modeling} compared a single-chain \ac{SLSP} model to the dynamics of the Kremer-Grest bead-spring polymer model.
Theodorou et al.~\cite{vogiatzis2017equation,sgouros2017slip,megariotis2018slip,sgouros2019multiscale} developed an equation-of-state-based multi-chain \ac{SLSP} model, in which the nonbonded interactions are derived from an equation of state (\eg, Sanchez–Lacombe equation of state). They used either experimental data or atomistic simulations to 
parameterize the \ac{SLSP} model for \cis-polyisoprene as well as bulk and interfacial polyethylene systems.
Becerra et al.\cite{becerra2020polymer} used atomistic simulations to parameterize a single-chain slip-link model for poly(ethylene oxide).       
Recently Wu et al.\cite{wu2021atomistic} applied hybrid particle-field atomistic molecular dynamics simulations to a united atom model of PE; in these simulations, the field treatment of the nonbonded interactions make them effectively soft. Entangled dynamics was recovered by the introduction of \ac{SLSP}s.

It is clear from the above discussion that the dynamics of polymer melts have been extensively studied via different simulation methods. However, there are still many challenges related, in particular, to the quantitative prediction of the single-chain dynamics and rheology of well-entangled polymers of given chemistry, and their direct study over a very broad range of spatiotemporal scales. To achieve this, a systematic linking of simulation methodologies across different scales is required. 

To address the above challenges here we propose a hierarchical (triple scale), mainly bottom-up, simulation methodology that involves atomistic, moderately coarse-grained (mCG), and highly coarse-grained \ac{SLSP} simulations to predict the dynamical and rheological properties of polymers, over timescales ranging from a few fs up to ms and molecular weights up to the well-entangled regime. 
As a reference system, we choose \ac{cPB}, which is a well-known elastomer with important industrial applications.
At the atomistic level, \ac{cPB} is described through a united atom model.
In the mCG model, each monomer of \ac{cPB} (containing four united atoms) is mapped into one bead. This model is close to the chemical structure of PB and still allows a significant simulation speedup, relative to the atomistic model. 
In this work, we parameterize a \ac{SLSP} model, which represents a 400-mer chain of \ac{cPB} by just 32 interaction centers, using our \ac{mCG} model.  The parameter passing provides a consistent description of static properties, the single-chain dynamics, and the collective kinetics on time and length scales, where the highly coarse-grained model applies. 
For this specific system, we demonstrate that the \ac{SLSP} model can quantitatively predict rheological properties for high molecular weight systems.

\begin{figure}[!htb]
    \centerline{\includegraphics[width=0.45\textwidth]{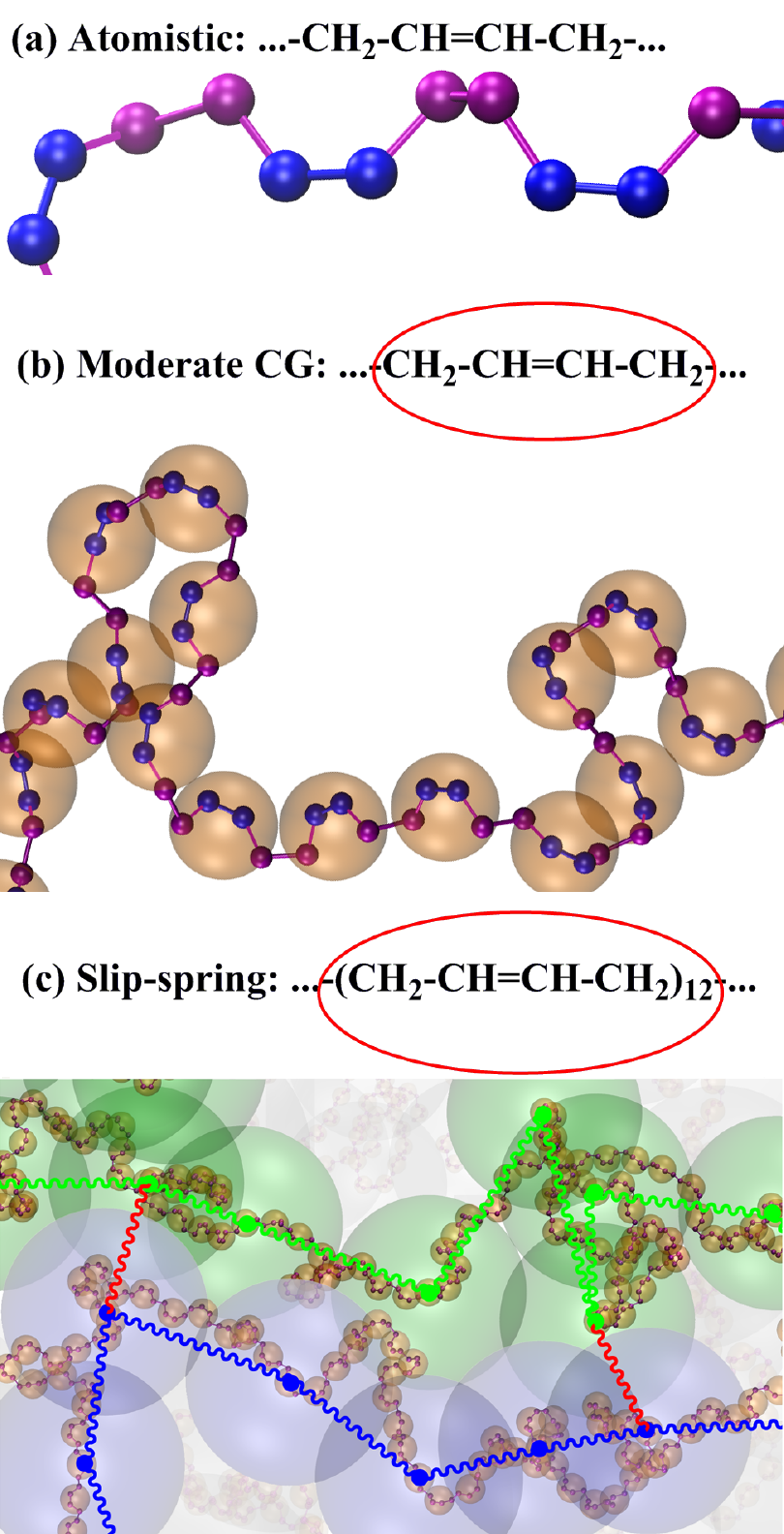}}
    \caption{Different levels of description of cPB. (a) At the finest level, PB is described through a united atom model. (b) At the moderately CG level, each monomer of PB is mapped onto one bead; the CG bead is located on the center-of-mass of the monomer. (c) At the highly CG slip-spring level, $12$ successive monomers of PB are lumped into one CG bead; here, the CG bead is located on the position of the $6^\text{th}$ monomer. 
    At this level of coarse-graining, the polymer chain is modeled as an ideal Gaussian chain, and entanglements are incorporated via slip-springs (red).}
    \label{Fig:schema}
\end{figure}

The computational requirements of each coarse-grained level are orders of magnitudes smaller than the next finer description level.
All the different levels of description of \ac{cPB} are shown in \autoref{Fig:schema}.

\section{Atomistic model}

At the finest level, cPB is described using a united atom model. Each PB monomer has four united atom groups: two CH$_2$ groups (for end monomers: one CH$_2$ and one CH$_3$) and two CH groups that form a double bond (see \autoref{Fig:schema}a).  
A validated force field\cite{smith1998united,smith1999molecular, behbahani2020conformations} proposed by Smith and Paul was chosen for the description of the atomistic interactions. However, to preserve stereochemistry during long runs, the torsional potential of carbon double bonds in this force field was modified.\cite{behbahani2020conformations} 
The MD runs were performed in the NPT ensemble at $T = 413$ K and $P = 1$ atm. 
Temperature and pressure were controlled using Nos\'{e}-Hoover thermostat\cite{nose1984molecular,hoover1985canonical} and Parrinello-Rahman barostat.\cite{parrinello1981polymorphic}
Nonbonded interactions were cut off at $1$ nm and van der Waals tail corrections were applied to energy and pressure. The Leap-frog scheme with $1$ fs time step was used for the integration of equations of motion.            
We used the open-source package GROMACS\cite{pronk2013gromacs} to perform the atomistic and mCG simulations.

\section{Moderately coarse-grained (mCG) model}

\subsection{Parametrization of the mCG model}

At the \ac{mCG} level, each monomer of cPB is mapped onto one coarse-grained bead.
For mapping, one \ac{mCG} bead is placed on the center-of-mass of each monomer of the atomistic chains (see \autoref{Fig:schema}b).  
This model has the advantage of retaining structural features and therefore conserving the chemical identity of the atomistic model. To develop the mCG force field, it is assumed that the effective potential energy (in fact, free energy) of a \ac{mCG}  bead can be separated into bonded and pairwise nonbonded parts.
Furthermore, the bonded interactions involve contributions from bonds, angles, and dihedral angles: 
\begin{equation}
U^\text{CG} = U_\text{nonbonded}^\text{CG}(r) + U_\text{bond}^\text{CG}(l) +  
U_\text{angle}^\text{CG}(\theta) + U_\text{dihedral}^\text{CG}(\phi)
\end{equation}
Here, $r$, $l$, $\theta$, and $\phi$ are the distance between two \ac{mCG} beads, \ac{mCG} bond length, \ac{mCG} bending angle, and \ac{mCG} dihedral angle, respectively. 
Consistent with the above assumption, it is assumed that the probability distribution function of 
 $l$, $\theta$, and $\phi$ can be factorized into independent distribution functions:
\begin{equation}
P(l, \theta, \phi) = P(l) P(\theta)  P(\phi)
\end{equation}

The probability distribution functions are calculated from the atomistic simulations and then converted to \ac{mCG} potentials through the iterative Boltzmann inversion method.\cite{reith2003deriving}
This method iteratively matches the probability distribution functions of the \ac{mCG} model to the respective atomistic distributions, called target distributions.  
Similarly, the nonbonded mCG potentials are obtained by matching the (intra- and inter-molecular) radial pair distribution function, $g(r)$, of the mCG model to that of the atomistic model.
The initial guess for the \ac{mCG} potentials are the Boltzmann inverse of the target distributions:
\begin{eqnarray}
U^\text{CG}_\text{nonbonded, 0}(r) &=& -k_\text{B}T\ln g_\text{target}(r)\\
U^\text{CG}_\text{bond, 0}(l) &=& -k_\text{B}T\ln (P_\text{target}(l)/l^2)\\
U^\text{CG}_\text{angle, 0}(\theta) &=& -k_\text{B}T\ln (P_\text{target}(\theta)/\sin \theta)\\
U^\text{CG}_\text{dihedral, 0}(\phi) &=& -k_\text{B}T\ln P_\text{target}(\phi)
\end{eqnarray}
Here, distributions of bond length and bending angle are normalized by taking into account their respective volume elements, $l^2$ and $\sin \theta$ (ignoring numerical prefactors) .\cite{tschop1998simulation} The \ac{mCG} potentials are then modified in an iterative procedure through:\cite{reith2003deriving}
\begin{equation}
U^\text{CG}_{i+1}(\xi) = U_{i}^\text{CG}(\xi) + k_\text{B}T\ln  \frac{P_i(\xi)}{P_\text{target}(\xi)}
\label{Eq: iter}
\end{equation}
where $\xi$ stands for $r$, $l$, $\theta$, and $\phi$. 
The iteration is performed until the convergence of all \ac{mCG} distributions to their corresponding target distributions is achieved. Convergence of a distribution is confirmed through the minimization of a merit function, defined as: 
\begin{equation}
f_\text{merit} = \frac{\int_{\xi_\text{min}}^{\xi_\text{max}} (P_\text{target}(\xi) - P(\xi))^2 \mathrm{d}\xi}{\int_{\xi_\text{min}}^{\xi_\text{max}} P^2_\text{target}(\xi) \mathrm{d}\xi }
\end{equation}

Furthermore, to correct the pressure of the \ac{mCG} model, a pressure correction step is added to the above iterative procedure (\autoref{Eq: iter}).\cite{reith2003deriving,wang2009comparative}  
Without applying pressure correction, with similar densities, the pressure of the \ac{mCG} model is usually higher than the pressure of the atomistic model. 
For the pressure correction, a linear term is added to the nonbonded potential function:\cite{reith2003deriving,wang2009comparative}
\begin{equation}
\Delta U_\text{nonbonded}^\text{CG}(r) = A (1 - \frac{r}{r_\text{cut}})
\end{equation}   
where $A$ is an empirical constant and $r_\text{cut}$ is the cut-off distance for the calculation of the nonbonded interactions. 

\begin{figure}[!htb]
    \centering
    \begin{subfigure}{0.45\textwidth} % width of left subfigure
        \includegraphics[width=\textwidth]{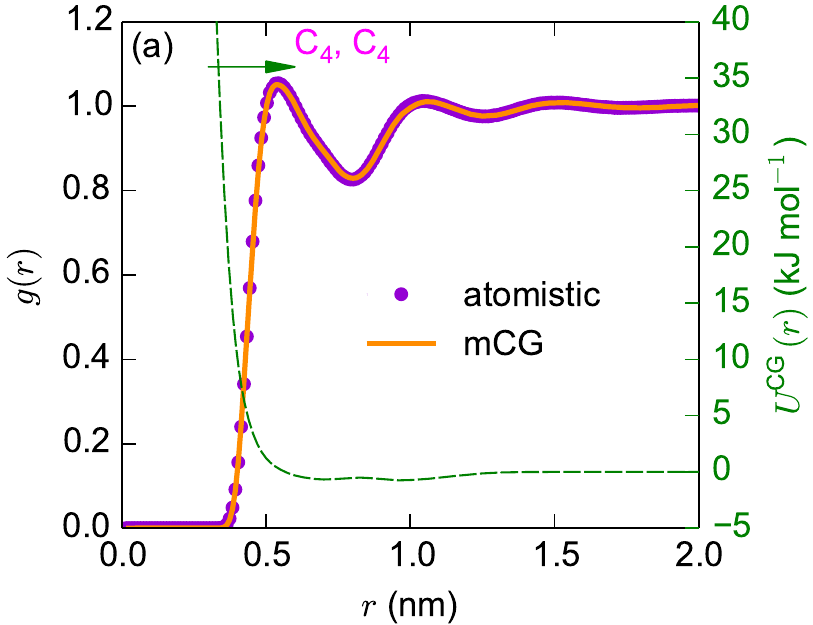}
    \end{subfigure}
    %\vspace{1ev} % here you can insert horizontal or vertical space
    \begin{subfigure}{0.45\textwidth} % width of right subfigure
        \includegraphics[width=\textwidth]{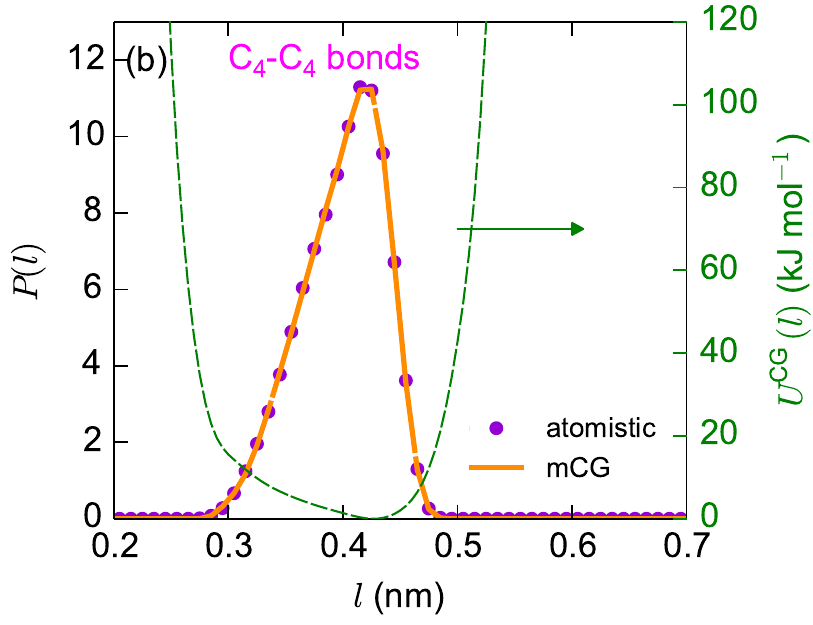}
    \end{subfigure}
    
    \begin{subfigure}{0.45\textwidth} % width of right subfigure
        \includegraphics[width=\textwidth]{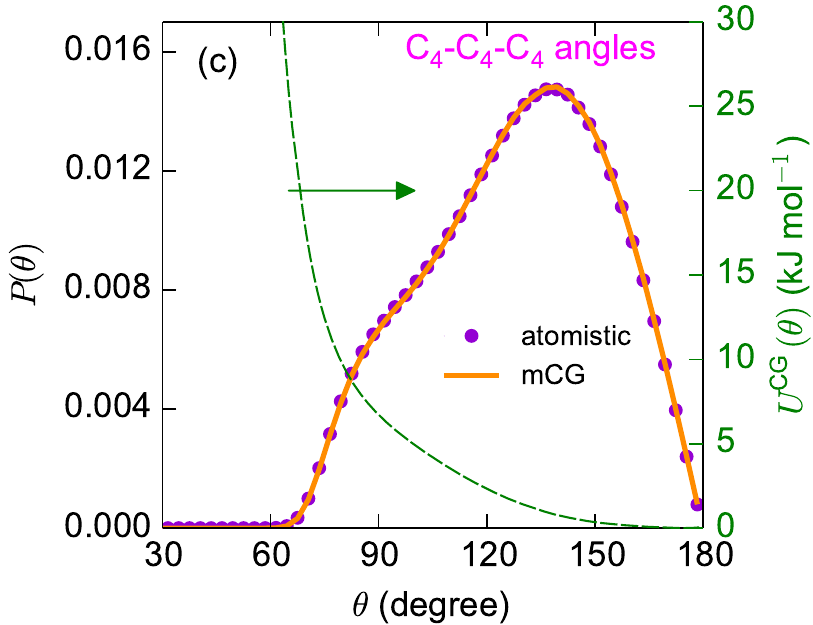}
    \end{subfigure}
    \begin{subfigure}{0.45\textwidth} % width of right subfigure
        \includegraphics[width=\textwidth]{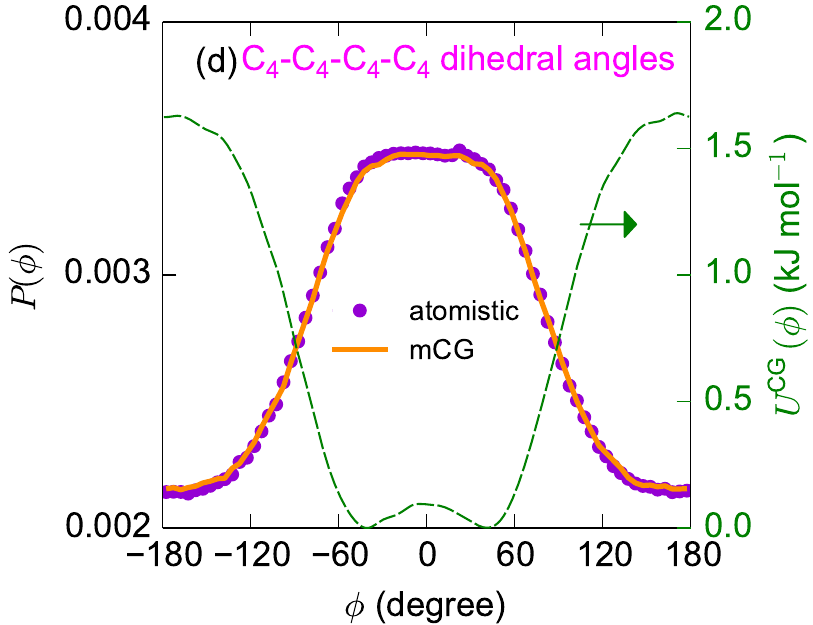}
    \end{subfigure}
    \caption{The developed nonbonded and bonded mCG potentials together with the target atomistic local structural distributions and the distributions obtained from \ac{mCG} simulations. (a) Nonbonded \ac{mCG} pair potential and \ac{RDF} (excluding the first three bonded neighbours) of \ac{mCG} beads (labeled as C$_4$) from atomistic and mCG simulations. (b) Bond stretching mCG potential and the distribution of C$_4$-C$_4$ bond lengths. (c) Bending potential and the distribution of C$_4$-C$_4$-C$_4$ angles. (d) Torsional potential of the mCG model and the distribution of C$_4$-C$_4$-C$_4$-C$_4$ dihedral angles ($\phi = 0$ corresponds to the cis torsional state).}
    \label{Fig:distributions} % caption for whole figure
\end{figure}

For the development of the \ac{mCG} force field for cPB at $413$ K, the reference atomistic simulations were performed for 30-mer chains. The developed bonded and nonbonded potentials together with the
plots of $g(r)$, $P(l)$, $P(\theta)$, and $P(\phi)$ as calculated from the \ac{mCG} and reference atomistic simulations are provided in ~\autoref{Fig:distributions}. 
For all distributions, a good agreement between the results of the \ac{mCG} simulation and the atomistic simulation is observed, indicating the success of the iterative Boltzmann inversion procedure. 
All \ac{mCG} runs were performed in the NPT ensemble. Stochastic rescaling thermostat\cite{bussi2007canonical} and Parrinello-Rahman barostat\cite{parrinello1981polymorphic} were used to control temperature and pressure.
The maximum range of \ac{mCG} nonbonded interactions was $1.5$ nm and nonbonded interactions were excluded for first, second, and third chemically bonded \ac{mCG} beads.  
The elimination of the fast degrees of freedom (\eg, covalent bond vibrations) and the softness of nonbonded potential allow for the increase of the time step in \ac{mCG} simulations relative to atomistic simulations. For mCG simulations, 
a leap-frog integration scheme with a $5$ fs time step was used.

Although the nonbonded interactions of the \ac{mCG} model are softer than the atomistic one, due to the chosen mapping scheme (one mCG bead corresponds to 4 united atoms or one monomer) they are still strong enough to prevent chain crossing. In addition, the bonds between \ac{mCG} beads are finitely extensible (see \autoref{Fig:distributions}b). Hence, the \ac{mCG} model respects the topological constraints, something that is required to describe polymer dynamics, especially of high molecular weight, entangled, systems.

\subsection{Chain dimensions}

\begin{figure}[!htb]
    \centering
    \begin{subfigure}{0.45\textwidth} % width of left subfigure
        \includegraphics[width=\textwidth]{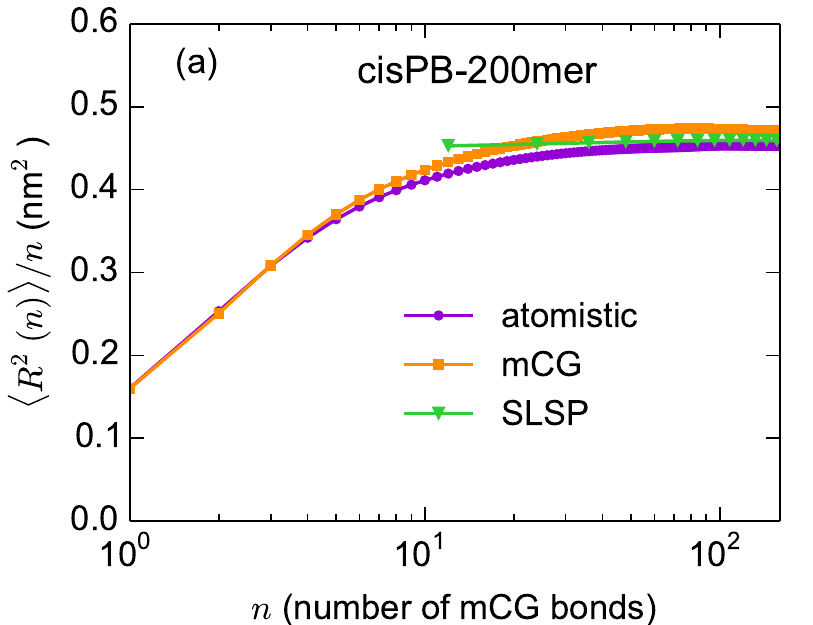}
    \end{subfigure}
    
    \begin{subfigure}{0.45\textwidth} % width of right subfigure
        \includegraphics[width=\textwidth]{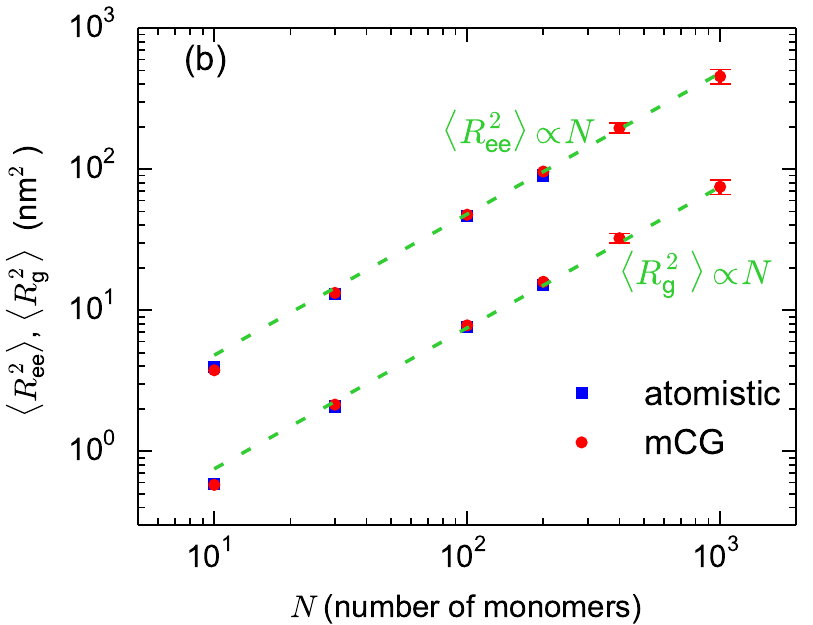}
    \end{subfigure}
    \caption{(a) Mean-squared internal distances, analyzed at the level of monomers, for atomistic, mCG, SLSP models of 200-mer cPB. (b)  Mean-squared end-to-end distance, $\langle R_\text{ee}^2 \rangle$, and mean-squared radius of gyration, $\langle R_\text{g}^2 \rangle$, for the atomistic and \ac{mCG} model chains of different lengths.
        }
    \label{Fig:intdist} % caption for whole figure
\end{figure}

As discussed above, the parameterization of the \ac{mCG} interactions is performed based on the matching of the local structure of the \ac{mCG} model to that of the atomistic model. 
Hence, the \ac{mCG} model reproduces the local structure well.
It is also important to verify the larger scale dimensions of the atomistic and \ac{mCG} chains. 
In Figure~\ref{Fig:intdist}a the mean-squared internal distances, analyzed at the level of monomers, for the atomistic and \ac{mCG} models of 200-mer \ac{cPB} chains, are shown. In Figure~\ref{Fig:intdist}a, $\langle R^2(n)\rangle$ is the  end-to-end distance of a sub-chain containing $n$ \ac{mCG} bonds ($n + 1$ monomers). 
A reasonable agreement between the internal distances of the \ac{mCG} and atomistic models are observed, however, at large scales, the  \ac{mCG} model has a slightly (around $5\%$) longer internal distances than the atomistic model.  
In \autoref{Fig:intdist}b the mean-squared end-to-end distances, $\langle R_\text{ee}^2 \rangle$, and the mean-squared gyration radii, $\langle R_\text{g}^2 \rangle$, of the atomistic and \ac{mCG} chains of various lengths are presented.
Similar to the trend that is observed for 200-mer chain, the end-to-end distances of the \ac{mCG} models are close to those of the atomistic chains.   
Also, with increasing chain length, the chain-length dependence of $\langle R_\text{ee}^2 \rangle$ and the chain-length dependence of $\langle R_\text{g}^2 \rangle$  approach the expected Gaussian behavior ($\langle R_\text{ee}^2 \rangle$, $\langle R_\text{g}^2 \rangle \propto N$).
This effect is important to verify for our highest \ac{CG} model, the \ac{SLSP} model, which models the polymers as flexible chains.
For short atomistic and mCG chains (shorter than 30-mer) the $\frac{\langle  R_\text{ee}^2 \rangle}{\langle R_\text{g}^2\rangle}$ values are larger than $6.2$; however, with increasing chain length $\frac{\langle  R_\text{ee}^2\rangle}{\langle  R_\text{g}^2\rangle}$ tends to $6.0$  which is the expected value for the Gaussian chains (for chains longer than 30-mer, most of the $\frac{\langle R_\text{ee}^2\rangle}{\langle R_\text{g}^2\rangle}$ values are less than $6.05$).

\subsection{Time scaling}
The prediction of the dynamical properties of the underlying atomistic model through \ac{mCG} simulations is of paramount importance. However,
because the friction associated with the fast degrees of freedom that are averaged out is ignored in the \ac{mCG} model (\ie, a calibrated friction is not involved in the CG equations of motion),
the dynamics of the \ac{mCG} system is faster than the dynamics of the atomistic one.\cite{harmandaris2006hierarchical, padding2011systematic}
Therefore, it is necessary to convert the dynamic quantities of the mCG model to those of the atomistic one.  Here, for reproducing the atomistic dynamics, 
the \ac{mCG} time is scaled by a factor $t^*_\text{mCG}$, called, time scaling factor.  A similar procedure has also been used in previous works.\cite{harmandaris2006hierarchical,harmandaris2009dynamics,ohkuma2017comparison}
  
In terms of the Rouse and the tube models, 
the dynamics of unentangled chains are controlled by a monomeric friction coefficient and the dynamics of entangled chains are controlled by a monomeric friction coefficient and a tube diameter.\cite{doi1988the} 
Because the \ac{mCG} model preserves the chemical identity of the atomistic model, its tube diameter, which is a geometrical quantity, is expected to be similar to the tube diameter of the atomistic system (in \autoref{sec:g1(t)} we investigate the crossover from the unentangled to the entangled regime and atomistic and \ac{mCG} models exhibit a similar behavior which is consistent with the development of a similar entanglement network by both models). So, the faster dynamics of the \ac{mCG} system is attributed to its lower monomeric friction coefficient. 
Given that the relaxation times of unentangled and entangled chains are proportional to monomeric friction coefficient ($\tau \propto \zeta$),\cite{doi1988the} 
the time scaling factor that is needed for matching \ac{mCG} dynamics and atomistic dynamics can be seen as the ratio of their monomeric friction coefficients, $t^*_\text{mCG} = \frac{\zeta^\text{AT}}{\zeta^\text{mCG}}$.\cite{harmandaris2009dynamics}
Note that, in practice, it should be checked if a single time scaling factor $t^*_\text{mCG}$ would reproduce all dynamical properties of the underlying atomistic model.

For the determination of $t^*_\text{mCG}$, we scaled the \ac{mCG} time to match the overall chain (translational and orientational) dynamics at the atomistic level.
Translational dynamics is investigated by calculating \ac{MSD} of monomers, $g_1(t)$, and chain centers-of-mass, $g_3(t)$.
Orientational dynamics is studied through the orientational autocorrelation function of the end-to-end vector, $P(t)$, defined as:
\begin{align}
    P(t) = \langle \hat{\boldsymbol{R}}_e(t) \cdot \hat{\boldsymbol{R}}_e(0) \rangle .
    \label{eq:ee-correlaton-definition}
\end{align}
where $ \hat{\boldsymbol{R}}_e$ is the normalized end-to-end vector and $\langle \rangle$ shows averaging over an ensemble of polymer chains, which is approximated by all chains in the simulation box and all time origins. 
In addition, we check the \ac{mCG} model predictions for the shear stress relaxation modulus, $G(t)$, which can be calculated through the autocorrelation of shear stresses:\cite{likhtman2007linear}
\begin{align}
    G(t) = \frac{V}{k_\text{B}T} \langle \sigma_{ij}(t)\ \sigma_{ij}(0)  \rangle\text{,}\ i\neq j
    \label{eq:G(t)}
\end{align}
where $V$, $k_\text{B}$, $T$, and $\sigma_{ij}$ are volume of the system, Boltzmann constant, temperature, and an off-diagonal component of the stress tensor, respectively.
$\langle \rangle$ shows averaging over the off-diagonal components (for isotropic systems) of the stress tensor and time origins.
To improve the statistics, 
one can average over all possible orientations of the coordinate system.\cite{likhtman2007linear}
Multiple-tau correlator algorithm has been used for the calculation of $G(t)$.\cite{ramirez2010efficient}

Typically, the \ac{mCG} time is scaled to match one dynamical quantity, \eg, $g_1(t)$, and then the predictions of the \ac{mCG} simulations for other dynamical quantities (\eg, $g_3(t)$, $P(t)$, or  $G(t)$) are checked.
For short chains, it would be more accurate to match $P(t)$, and $G(t)$ as well, using a time scaling factor that is slightly different from the time scaling factor extracted from fitting $g_1(t)$ and $g_3(t)$; for 30-mer and shorter chains, the time scaling factor needed for matching $P(t)$ and $G(t)$ is around $10$--$15 \%$ smaller than the factor extracted from fitting $g_1(t)$ and $g_3(t)$.    
Such deviations observed for shorter chains are not unexpected and are related mainly to the importance of the chain ends of such systems. Since the interactions of the \ac{mCG} model involve entropic terms, the end monomers should, in principle, be described via a different interaction. However, since the main goal of the \ac{mCG} model is to describe high molecular weight systems, such differences are typically ignored. 
For 100-mer and longer chains, a single scaling factor has been found to describe accurately the above-mentioned dynamical quantities.

\begin{figure}[!htb]
    \centering
    \begin{subfigure}{0.45\textwidth} % width of left subfigure
        \includegraphics[width=\textwidth]{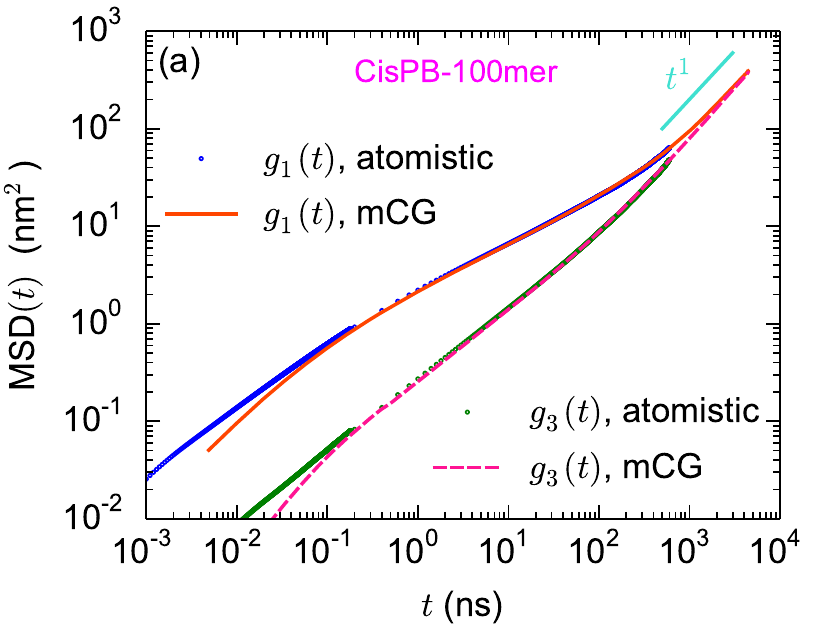}
    \end{subfigure}
    %\vspace{1ev} % here you can insert horizontal or vertical space

    \begin{subfigure}{0.45\textwidth} % width of right subfigure
        \includegraphics[width=\textwidth]{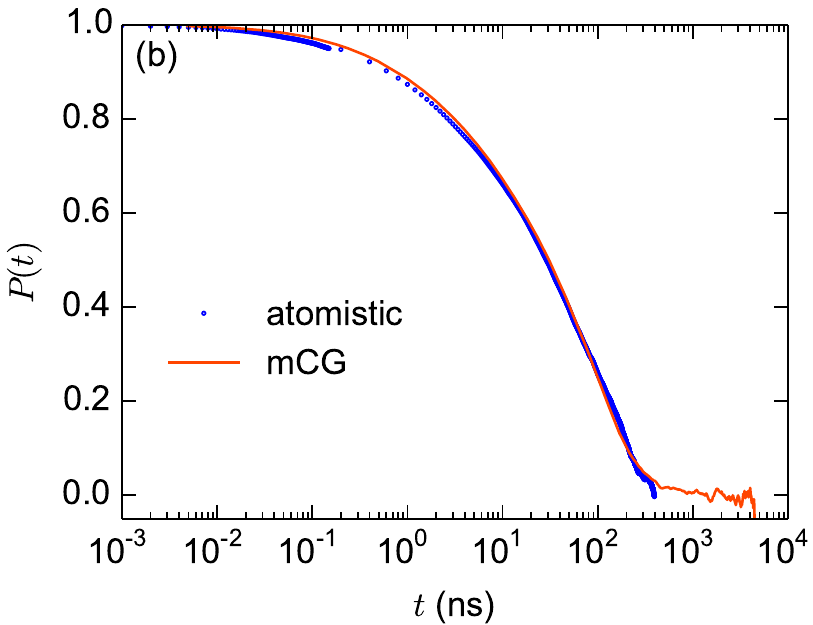}
    \end{subfigure}
    
    \begin{subfigure}{0.45\textwidth} % width of left subfigure
    \includegraphics[width=\textwidth]{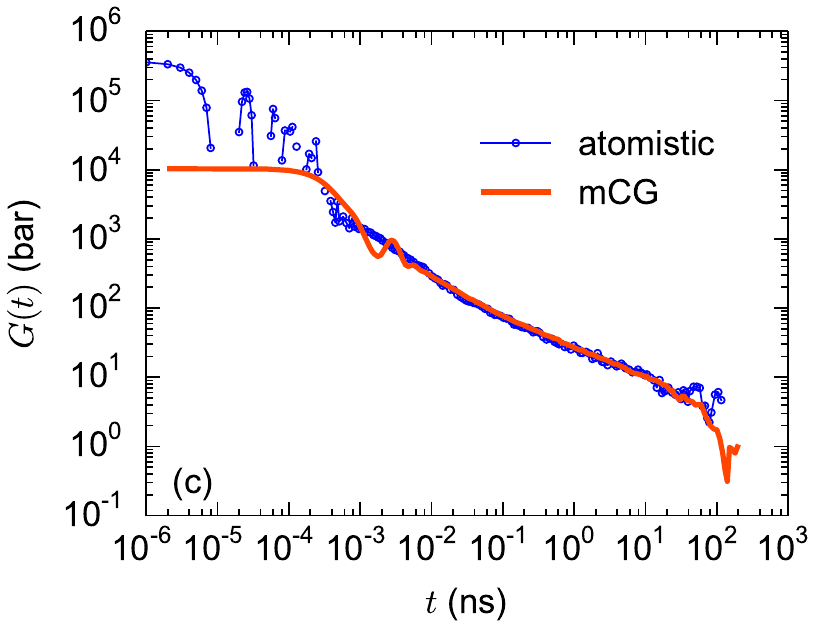}
    \end{subfigure}
    %\vspace{1ev} % here you can insert horizontal or vertical space

    \caption{(a) \ac{MSD} for chain center-of-mass ($g_3(t)$) and monomer center-of-mass ($g_1(t)$) for 100-mer chains of cPB at $413$ K through atomistic and mCG simulations. (b) Orientational autocorrelation function, $P(t)$, of end-to-end vector and (c) stress relaxation modulus, $G(t)$, for 100-mer cPB through atomistic and mCG simulations. In all cases, the mCG times are scaled with a single time scaling factor ($t^*_\text{mCG} = 4.9$), computed by scaling the \ac{mCG} time to match the atomistic $g_1(t)$ in the long time (diffusive) regime.}
    \label{fig:p(t)-msd-100} 
\end{figure}

\autoref{fig:p(t)-msd-100} shows the \ac{MSD} ($g_1(t)$ and $g_3(t)$), $P(t)$, and $G(t)$ for 100-mer cPB, from atomistic and mCG simulations ($t^*_\text{mCG} = 4.9$). 
As shown in \autoref{fig:p(t)-msd-100}a, apart from very short length and timescales (distances less than $5$ \AA\  $\approx$  the size of the mCG beads), the MSD curves calculated through mCG simulations nicely follow the corresponding atomistic ones. 
Also, as shown in \autoref{fig:p(t)-msd-100}b, the $P(t)$ curves calculated through atomistic and mCG simulations exhibit a good agreement.
The $G(t)$ curves of 100-mer cPB from atomistic and mCG simulations are depicted in \autoref{fig:p(t)-msd-100}c. At short times $G(t)$ of the atomistic model shows large oscillations. The oscillations range from large positive values to large negative values (the negative values are not shown in logarithmic scale) and originate from fast motions, \eg, bond vibrations and angle librations, in the atomistic model. After such vibrational motions, segmental relaxation and then, polymeric relaxations (Rouse-like and reptation-like, see also \autoref{sec:LVE}) take place. Because of the presence of rather soft bonds, the $G(t)$ of the \ac{mCG} model does not show large oscillations at very short times; however, mild oscillations happen after segmental relaxation. 
The time of the oscillations corresponds to the period of the bond vibration for the \ac{mCG} chain, which is around $400$ fs, before scaling.
Similar to the results of MSD, after very short timescales (after the segmental relaxation time), the \ac{mCG} model reproduces the $G(t)$ of the atomistic model well.

\begin{figure}[!htb]
    \centering
    \begin{subfigure}{0.45\textwidth} % width of left subfigure
        \includegraphics[width=\textwidth]{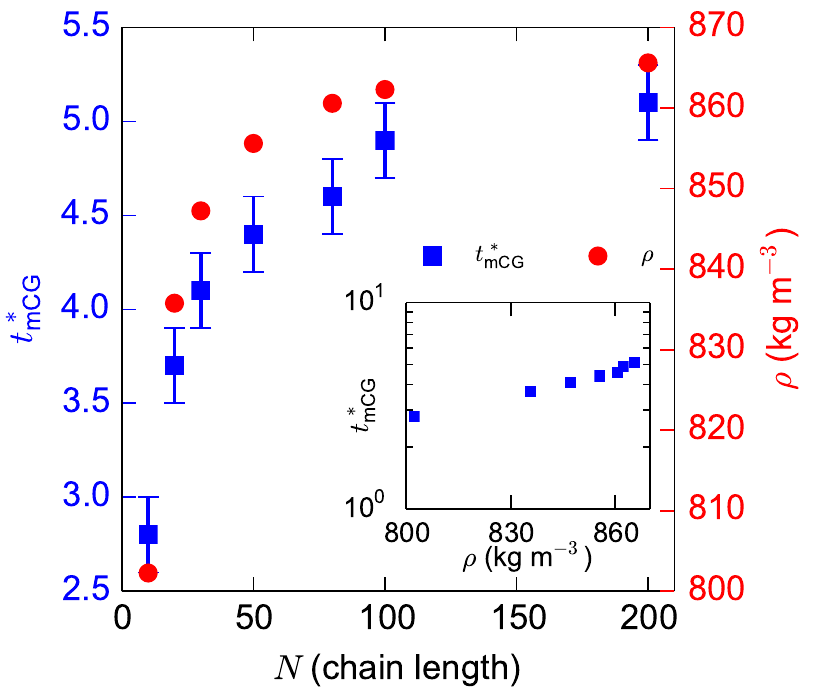}
    \end{subfigure}
    %\vspace{1ev} % here you can insert horizontal or vertical space
    \caption{Chain-length dependence of time scaling factor, $t^*_\text{mCG}$, calculated through matching of atomistic and mCG MSDs, at $413$ K. Additionally, the chain length dependence of density, $\rho$, and (inset) $t^*_\text{mCG}$ vs. $\rho$ are presented. }
    \label{fig:at-cg-time-mapping} % caption for whole figure
\end{figure}

The time scaling factor, $t^*_\text{mCG}$, is chain-length and temperature dependent.
The chain-length dependence of $t^*_\text{mCG}$ is presented in \autoref{fig:at-cg-time-mapping}. 
By increasing chain length, first $t^*_\text{mCG}$ increases and then converges to a constant value. A similar trend has been reported in previous works.\cite{harmandaris2009dynamics,ohkuma2017comparison} 
The chain-length dependence of the density is also shown in \autoref{fig:at-cg-time-mapping}. The densities were calculated from atomistic simulations; however, because of the application of pressure correction, the densities of the \ac{mCG} melts are very similar to the atomistic ones (for all chain lengths, the difference is less than $1.5 \%$). The chain-length dependence of $t^*_\text{mCG}$ is in phase with the chain-length dependence of density.
This behavior can be justified in terms of the free volume theory\cite{ferry1980viscoelastic,harmandaris2002detailed}
which attributes the chain length dependence of 
both density and monomeric friction coefficient, $\zeta$, of a polymer melt to the excess free volume accompanying chain ends;
however, for long chains, the contribution of chain ends in the free volume of the polymer melt becomes negligible and the density and $\zeta$ remain constant.

It is assumed that the chain length dependence of the fractional free volume, $f$, of a polymer melt is similar to the chain length dependence of its specific volume. 
At a fixed temperature, the relations between $\zeta$, $f$, and density
are as follows:\cite{ferry1980viscoelastic,harmandaris2002detailed}
\begin{align}
\zeta(M) = \zeta_{\infty}\exp(B(\frac{1}{f(M)} - \frac{1}{f_{\infty}})),\  \  \
    f(M) - f_\infty = 1 -  \frac{\rho(M)}{\rho_\infty}
\label{eq:freeVol}
\end{align}
where $f_{\infty}$, $\rho_{\infty}$, and $\zeta_{\infty}$ are fractional free volume, density, and monomeric friction coefficient of very high molecular weight melts, and $B$ is a constant of order of unity.\cite{ferry1980viscoelastic,harmandaris2002detailed}
Based on the above expression, it is expected that when $\rho \rightarrow \rho_{\infty}$,  $\zeta^\text{AT}$ and $\zeta^\text{mCG}$, and therefore $t^*_\text{mCG} = \frac{\zeta^\text{AT}}{\zeta^\text{mCG}}$, tend to  constant values.\cite{harmandaris2009dynamics,harmandaris2009dynamics-b} 
Also, when $\rho$ is close to $\rho_{\infty}$, \autoref{eq:freeVol} becomes a linear relation between $\ln \zeta$ (and therefore $\ln t^*_\text{mCG}$) and $\rho$. Inset of \autoref{fig:at-cg-time-mapping} shows that values of $t^*_\text{mCG}$ as a function of $\rho$.
With increasing density, the monomeric friction coefficients of both atomistic and \ac{mCG} melts increase; however, the increase of $t^*_\text{mCG}$ with density shows that the increase of $\zeta$  is larger for the atomistic model than for the \ac{mCG} one.
The constant value of $t^*_\text{mCG}$ for long chains can be used to simulate the dynamics of long chains, for which the atomistic simulation is not feasible.
The above-mentioned procedure for the prediction of the dynamics of the atomistic model through \ac{mCG} simulations applies to the study of the dynamics of structurally complex polymers, e.g., long-chain branched polymers; however, the value of the time scaling factor might be slightly different than that of linear chains.

\section{Slip-spring (\acs{SLSP}) model}

In our highly coarse-grained \ac{SLSP} model, a polymer is represented by a string of beads. 
A single bead or \ac{CG} segment represents multiple monomeric repeating units of an atomistic model.
In the present $\ac{cPB}$ \acs{SLSP} model, a coarse-grained bead 
approximately 
represents $12$ monomers (or 48 united-atom groups).
The Kuhn length of $\ac{cPB}$ chains is approximately equal to $1$nm, corresponding to $\approx 9$ backbone bonds (or $10$ CH$_x$ groups),\cite{behbahani2020conformations} \ie, $12$ monomers are comprised of $5.2$ Kuhn segments.
We chose 12 as the degree of coarse-graining, because it is the smallest number of monomeric repeating units that allows a description via the flexible Gaussian chain model, compare with \autoref{Fig:intdist}.
We should note that it might be advantageous to employ \ac{SLSP} models with a larger degree of coarse-graining to efficiently simulate longer chains on larger spatiotemporal scales (thereby potentially losing resolution on the scale of the tube diameter). However, the parameterization of such even coarser CG models would require data from (computationally expensive) mCG-based simulations of longer chains than what is needed for the \ac{SLSP} model with a smaller degree of coarse-graining. In any case, the parameter-passing strategy would be similar to the strategy used here.  
In the following, we use a chain of $N_{\mathrm{SLSP}} = 32$ beads to describe a 400-mer \ac{cPB} molecule (and account for the mismatch by ignoring the 8 monomeric repeating units at the ends of the mCG chain in the analysis).

The parameterization of the SLSP model is performed based on the results of the mCG simulations. To transfer information between mCG and SLSP levels, first, the mCG coordinates should be mapped to the mesoscopic SLSP presentation. 
 To map mCG coordinates to SLSP
coordinates, for each successive 12 monomers, a CG bead is placed on the position of the monomer that has the median index (here, on the 6$^\text{th}$ monomer).
Let $0\leq i_\text{mCG} < 400$ denote the bead index in the mCG model, then the position of the $i^\text{th}$ bead of the SLSP model coincides with the coordinate of the mCG bead with index $i_\text{mCG} = 13+12i$, 
see \autoref{Fig:schema}.
A similar mapping scheme has been used before.\cite{sgouros2017slip}

\begin{figure}[!htb]
    \centering
    \begin{subfigure}{0.45\textwidth} % width of left subfigure
        \includegraphics[width=\textwidth]{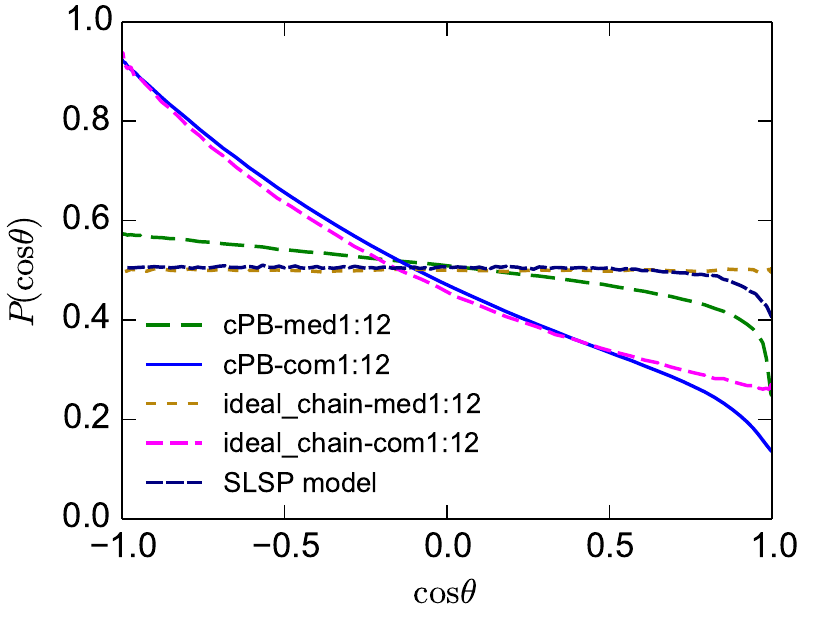}
    \end{subfigure}
    %\vspace{1ev} % here you can insert horizontal or vertical space
    \caption{Distributions of bead-bead-bead angle ($P(\cos\theta)$) through median index (med) and center-of-mass (com) mapping schemes. The results are shown for \ac{cPB} and the ideal chain. For \ac{cPB} one bead represents 12 monomers and for the ideal chain, one bead represents 12 nodes of the chain.}
    \label{fig:angle-dist-map} % caption for whole figure
\end{figure}

One can think of an alternative mapping scheme, in which a CG bead is placed on the center-of-mass of each successive 12 monomers (similar to mapping atomistic coordinates to mCG coordinates). To compare these two mapping schemes, we applied them to \ac{cPB}, as well as to an ensemble of ideal chains (freely jointed chains). 
The distribution of bead-bead-bead angle (after mapping) for \ac{cPB} and ideal chains are provided in \autoref{fig:angle-dist-map}.
It is clear that for both \ac{cPB} and ideal chains, the center-of-mass mapping scheme leads to a non-random distribution of angles between three successive beads. Such behavior has also been previously observed for CG models of polyethylene.\cite{sgouros2017slip}

It is instructive to consider these two mapping schemes for an ideal chain with bead positions ${\bf r}_i$ and uncorrelated bond vectors ${\bf b}_i={\bf r}_{i+1}-{\bf r}_i$. For a degree of coarse-graining of $2$ and the center-of-mass mapping, the scalar product of two successive, coarse-grained bond vectors, ${\bf b}^\text{CG}_0= \frac{1}{2} [{\bf r_3}+{\bf r}_2 - {\bf r_1}-{\bf r}_0]= \frac{1}{2}[{\bf b}_0+2{\bf b}_1+{\bf b}_2]$ and ${\bf b}^\text{CG}_1= \frac{1}{2} [{\bf r_5}+{\bf r}_4 - {\bf r_3}-{\bf r}_2]= \frac{1}{2}[{\bf b}_2+2{\bf b}_3+{\bf b}_4|$ does not vanish
$
\langle {\bf b}^\text{CG}_0 {\bf b}^\text{CG}_1 \rangle = \frac{\langle {\bf b}^2\rangle}{4} \neq 0
$. 
There are no next-nearest or higher bond correlations and the mean-squared length of a coarse-grained bond is $\langle ({\bf b}^\text{CG})^2 \rangle = \frac{3}{2} \langle {\bf b}^2\rangle$.

On the contrary, identifying every second bead position with the position of a coarse-grained bead, we obtain for the coarse-grained bonds ${\bf b}^\text{CG}_0={\bf r}_2-{\bf r}_0 = {\bf b_0}+{\bf b}_1$ and ${\bf b}^\text{CG}_0={\bf r}_4-{\bf r}_2 = {\bf b_2}+{\bf b}_3$, yielding $\langle {\bf b}^\text{CG}_0 {\bf b}^\text{CG}_1 \rangle =0$. There are also no next-nearest or higher bond correlations and the mean-squared length of a coarse-grained bond is $\langle ({\bf b}^\text{CG})^2 \rangle = 2\langle {\bf b}^2\rangle$.

\autoref{fig:angle-dist-map} demonstrates that choosing every $12^\text{th}$ mCG bead position also results in a random angular distribution for the ideal chain. There remains, however, a slight deviation for \ac{cPB} chains. These residual bond-angle correlations arise (i) because a 12-mer only comprises 5.2 Kuhn segments and (ii) due to the incompressibility of the melt, resulting in universal, power-law, bond-bond correlations \cite{Wittmer2004Sep}. The latter universal effect is also present in the highly coarse-grained SLSP model. The former effect could be accounted for in the SLSP model by a bond-angle potential or could be reduced by choosing a larger degree of coarse-graining. In the following, however, we neglect this small deviation for computational convenience. As we base our parameter passing on the end-to-end distance, the large-scale structures of the mCG and SLSP models coincide, but the bond length of the fully flexible SLSP model is slightly longer than an SLSP model with positive correlations between successive bonds.

\begin{comment}
As shown in \autoref{fig:angle-dist-map}, after mapping mCG coordinates to SLSP coordinates (using median index mapping and 1:12 degree of coarse-graining), we still have a slight deviation from the random angular distribution, however, because of the computational efficiency,  we do not consider this distribution in the SLSP model.

Note that if such bead angle distribution deviates strongly from the random one, can be captured via a properly defined bead angle potential.\cite{zhang2019}

The above results can be further discussed in purely geometrical terms. 
A Gaussian bead (or a polymer strand) looks like an elongated ellipsoid rather than a sphere.\cite{theodorou1985shape} 
Therefore, one should put an elongated ellipsoid on the center-of-mass of a polymer strand to preserve its shape.
The use of spherical beads, representing centers-of-mass of a group of monomers, lead to bead angle correlations, as it has been shown even for coarser representations .\cite{zhang2019}
-- MM: This is certainly true but I am not sure if this rationale explains the difference between the center-of-mass mapping and choosing every $\lambda$ segment.
\end{comment}

The interaction potential between \ac{CG} beads is soft, \ie, beads can fully overlap without divergence of the nonbonded energy, in marked contrast to the two previously discussed, finer levels of modeling.
The softness arises from a systematic coarse-graining procedure \cite{Klapp2004Oct}.
The reduction of the degrees of freedom and the softness of the potentials in highly coarse-grained models, allowing for larger time steps, give rise to a significant computational speed-up.

Moreover, softness is necessary to represent the large invariant degree of polymerization, $\bar{\cal N} = (\frac{\rho}{N}R_\text{e0}^3)^2 = (\rho b^3)^2N$ that characterizes experimental systems.\cite{muller2011studying} Here $\rho$, $N$, and $b$ denote the bead density, the number of beads per chain, and the statistical segment length, respectively. To represent a large value of $\bar{\cal N}$, one can either increase $N$ at fixed $\rho$ and $b$ or increase the degree of coarse-graining such that $N$ of the highly coarse-grained chain model remains unaltered but increase $\rho b^3$ instead. The latter is computationally much more convenient.\cite{muller2011studying} Harsh repulsive interactions on a length scale $\sigma$, however, do not allow for an increase of the segment density beyond $\rho \sigma^3 \sim {\cal O}(1)$ because the system either crystallizes or vitrifies. Additionally requiring that the combination of segmental repulsion and finite bond length, $l$, prevents chain crossing, yields the condition $\sigma \approx l \sim b$, where the last relation holds for flexible chains. Thus, it is difficult to increase $\rho b^3$ significant beyond order $1$ for flexible chain models, in which pairwise interactions prevent chain crossing and large invariant degrees of polymerization necessitate a fine discretization of the chain contour.

We perform all simulations of the \ac{SLSP} model with the GPU accelerated simulation software HOOMD-blue\cite{anderson2008general,phillips2011pseudo,glaser2015strong} with custom \ac{SLSP}-plugins.
The combination of a highly coarse-grained model with modern, GPU-accelerated computers expands the study possibilities of the two previous models to higher molecular weights and longer timescales.

The nonbonded interaction of the model employed in our molecular dynamics simulation is an empirical pairwise potential.
The potential has a two-fold purpose:
(i) It suppresses density fluctuations and (ii) controls the repulsion of unlike bead types (in multicomponent systems).
The potential is soft and quadratic, and it takes the form 
\begin{align}
    \label{eq:md-pair}
    V_{\text{nb}}(r) = A_{ij} \frac{k_\text{B}T}{2} \left( 1- \frac{|r|}{\sigma}\right)^2 \text{ for } |r| < \sigma .
\end{align}
$\sigma$ defines the interaction range of this potential.
The parameter $A_{ii}$ correlates to the inverse compressibility and $A_{ij}$ controls the repulsion of unlike bead types.
The subscript $ij$ indicates the type of interacting particles.
In this work we only describe beads for $i=$cPB, hence there is only a single interaction parameter $A_{ii}$. 
To convert the simulations from the microcanonical ensemble (NVE) to the canonical ensemble (NVT) we add random forces and friction according to the \ac{DPD} thermostat \cite{espanol1995statistical}.
We use a friction and fluctuation value $\gamma_\text{DPD} = 1/\tau$ for all reported simulations with the \ac{SLSP} model.
We choose this value to allow an early transition from ballistic motion to overdamped motion.
For a more detailed discussion of this DPD parameter and its effect in the context of the \ac{SLSP} model, we refer to Ref~\cite{heck2017}.

For this work we selected a nonbonded interaction parameter of $A_{ii} = 5$.
In previous works, we have used values $A_{ii}$ smaller by as much as an order of magnitude $A_{ii}$\cite{chappa2012translationally}.
Using the higher value does not guarantee a clear separation in strength between bonded and nonbonded interactions.
However, the advantage of this higher value is that the compressibility characteristics of the melt are closer to the \ac{mCG} and atomistic description.
The empiric potential does not feature an attractive tail so the measured pressure is large compared to the experiment.
Adjusting the pressure would require a longer-ranged attractive interaction that does hardly influence the liquid structure but significantly adds to the computational costs. 
If required the effect of such an additional attraction could be accounted for by thermodynamic perturbation theory.\cite{Allen1989Jun}

The $N_\mathrm{cg}$ beads are bonded along the backbone by a bond-length potential that not only captures the Gaussian equilibrium statistics but, additionally, mimics the reduced extensibility for strong elongations and accounts for corrections to the limit of large degree of coarse-graining.
\begin{align}
    V_\text{b}(r) = \frac{k_2}{2} r^2 + \frac{k_4}{2} r^4 . \label{eq:bond-potential}
\end{align}
with $r$ being the distance between bonded neighbors.
In the limit $ k_4 = 0$ the potential coincides with the discretized Edwards Hamiltonian of a Gaussian chain\cite{matsen2001standard}.

The harmonic potential is a direct consequence of the coarse-graining procedure in the limit of infinitely long chains. If the degree of coarse-graining is much larger than the Kuhn length, a highly coarse-grained bead is comprised of many Kuhn segments. The end-to-end distance of such a subchain that is represented by a coarse-grained bead is the sum of many, independent, identically distributed contributions and its distribution converges towards a Gaussian.  Therefore top-down models utilize harmonic bonds that represent the relevant property -- chain connectivity -- in a universal and computationally efficient way.

Here we go beyond this top-down strategy. The additional fourth-order term in \autoref{eq:bond-potential} is a first, systematic attempt to incorporate deviations from the asymptotic Gaussian behavior that occurs for small degrees of coarse-graining.
The specific shape of using a fourth-order expansion is inspired by the original description of Kuhn et. al.\cite{kuhn1942beziehungen} for an inverse Langevin, which were appropriate if we model a \ac{FJC} model.
Moreover, in nonequilibrium situations that include
the stretching of chains, the Gaussian chain model may no longer be appropriate.
An intuitive example is the contour length. As it is well-known \cite{kuhn1942beziehungen, rubinstein2003polymer}, any real chain has a fixed contour length but this constraint is not enforced in a Gaussian chain. 

We examine possible deviations from the Gaussian behavior by monitoring the bond extension, $z$, along a specified axis (for isotropic bulk polymer melts, all axis are equivalent). 
In the Gaussian chain model with harmonic bonds this distribution, $p(z)$, is Gaussian. However, the distribution of distances between 12-mers -- the length of an \ac{SLSP} bead -- in a 400-mer chain shows a stronger decay for large bond lengths.

\begin{figure}
    \centering
    \includegraphics[width=0.75\textwidth]{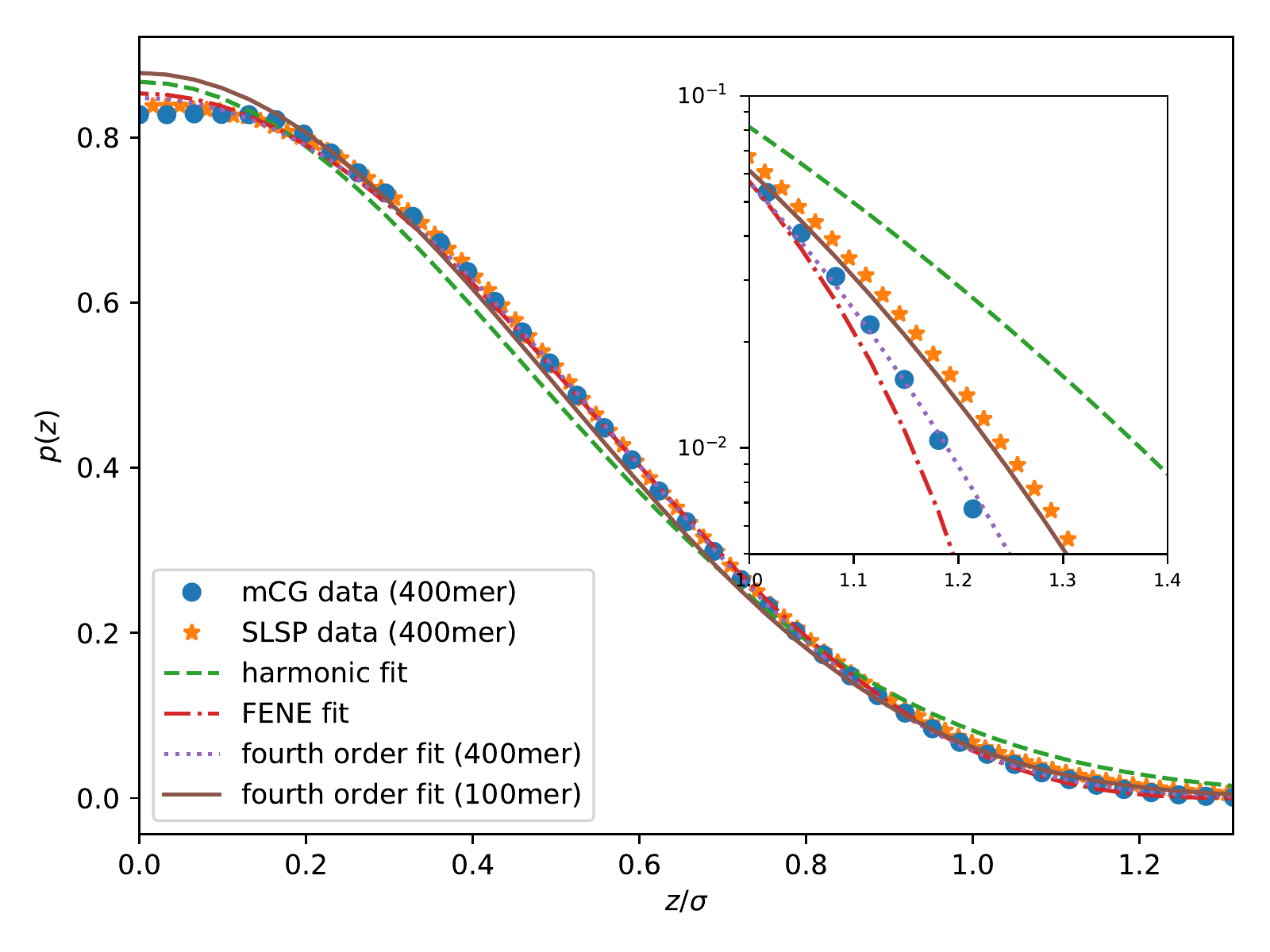}
    \caption{Distribution, $p(z)$, of the extension of 12-mer subchain in a polymer of 100 or 400 monomeric repeating units, obtained by simulations of the mCG model. For the highly coarse-grained \ac{SLSP} model a 12-mer is represented by a single bead.
    The specific shape of the \ac{mCG} distribution depends on the chain length, as demonstrated with fits to 400-mer and 100-mer data set. 
    We assume a Boltzmann distribution $\exp(-V_\text{b}(r)/k_\text{B}T)$ for the bond-length distribution to fit the free parameters of different bond potentials. This assumption is verified by data of the SLSP model shown as stars.}
    \label{fig:slsp-pz-C}
\end{figure}

\autoref{fig:slsp-pz-C} visualizes the results of the mCG model in comparison to the Boltzmann distributions of single bonds in the \ac{SLSP} model.
Fitting the distribution to the data from the mCG simulations allows us to determine the optimal parameters for our bond potential.
We illustrate this procedure for the harmonic, expanded fourth-order, and \ac{FENE} potential. All three potentials provide a reasonable fit to the mCG data, yet small but systematic deviations can be appreciated in the inset that highlights the tails of the distribution on a logarithmic scale.
The harmonic potential overestimates the weight of the tails in the distribution.
The \ac{FENE} potential has a finite length as it is required to obtain a finite contour length and, in turn, underestimates the weight of the tails of the distribution.
We obtain the best fit with the fourth-order potential from \autoref{eq:bond-potential}.

The parameters of the fourth-order potential slightly depend on the chain length in the \ac{mCG} model system.
However, we are interested in a universal, chain-length independent bond potential.
As a compromise we utilize the parameters $k_2 \approx 3.30 k_\text{B}T/\sigma^2$ and $k_4 = 2.41 k_\text{B}T/\sigma^4$ for the fourth order potential.
The resulting distribution provides a good agreement in the center as well as in the tails of the distribution and improves upon the harmonic or \ac{FENE} potential. We identify the basic length scale, $\sigma$, of interactions in the SLSP model by the root-mean-squared bond length, $b_0 = 0.75\sigma$

The use of soft, \ac{CG} polymer models to study dynamics poses challenges that are addressed by specific extensions to the standard bead-spring model:
To analyze the dynamical characteristics of long polymers it is vital to include entanglement effects.
Unfortunately, the inevitable softness of the nonbonded interactions in a highly \ac{CG} model does not implicitly prevent backbone-crossing.
To compensate this effect, we use the \ac{SLSP} model \cite{chappa2012translationally}, which represents entanglements via additional bonds -- \acfp{SLSP}.
The dynamics of \acp{SLSP} is designed to accurately mimic the reptation motion of long polymer chains in a dense melt, going beyond the classical tube model \cite{doi1988the} by including constraint release and contour-length fluctuations \cite{likhtman2005single} and allowing the study of spatially inhomogeneous systems.\cite{ramirez2018detailed}

The \ac{SLSP} model \cite{chappa2012translationally} adds additional bonds to the system, which enforce spatial constraints similar to entanglements. As will become apparent later, the bond potential has to be of finite range, so we use the \ac{FENE} potential
\begin{align}
\label{eq:vss}
    V_{ss}(r) = -\frac{k_{ss} r_{ss}^2}{2} \log\left[1 - \left(\frac{r}{r_{ss}}\right)^2\right].
\end{align}
The maximum extension is set for efficiency reasons to $r_{ss} = \sigma$, \ie, the interaction range of the nonbonded potential.
The origin of the \ac{SLSP} force is the same as the backbone potential, however, for technical reasons the potential has to be of finite range so we opt for the \ac{FENE} potential, with force constants that mimics the backbone potential \cite{chappa2012translationally}.
The insertion and movement of the additional slip spring bonds are governed by kinetic Monte-Carlo moves, sampling the grand-canonical ensemble.
The chemical potential $\mu$ of the slip springs dictate the average number of \acp{SLSP}.
More commonly we use the fugacity $z = \exp(\beta \mu) \propto n_{ss}$, which is directly proportional to the number of \acp{SLSP}.
The introduction of additional bonds into the systems changes the system properties, especially the pressure.
To eliminate this effect, an additional, repulsive pair potential is introduced
\begin{align}
    \label{eq:comp}
    V_\text{comp}(r) = z k_BT\exp(-\beta V_{ss}(r) ).
\end{align}
This potential exactly compensates the introduction of the slip springs in the partition function \cite{chappa2012translationally}.
Thus, all static properties remain unchanged, independent of the SLSP fugacity, $z$. Only the dynamical properties change to incorporate the effect of entanglements.

The dynamics of the \acp{SLSP} is governed by kinetic \ac{MC} moves, designed to mimic reptation dynamics. The \ac{MC} move that attempts to slide each \ac{SLSP} along the backbone (reptation) and the \ac{MC} move that transfers \ac{SLSP} from one chain end to another (dis/re-entanglement and constraint release) are applied with a frequency of $\Delta t = 0.5\tau$. This timescale corresponds to the relaxation time of a bead in our highly CG model, \ie, beads, and \ac{SLSP} move on the same timescale. Specifically, the \ac{MSD} of beads within $10\tau$ is about the average length of the backbone bonds. The attempt frequency of the \ac{MC} move is high enough to not significantly change the dynamics if the frequency were further increased. These two \ac{MC} moves -- SLSP sliding and end-transfer -- suffice to study the dynamics but the number of SLSP remains constant.

Additionally, we apply a \ac{MC} move that attempts to create or delete new \ac{SLSP} at the chain ends. This \ac{MC} move realizes the grand canonical SLSP ensemble. The \ac{MC} move is computationally more expensive and thus is attempted only every $\Delta t= 10\tau$.
The \ac{MC} that control the creation and deletion of \acp{SLSP} at the end of the polymer molecules controls the tube renewal in the entangled systems.
The chosen update frequency is high enough that the tube renewal in accordance with the diffusion of the individual particle beads.
Increasing the frequency further does not change the result, but a lower frequency eventually leads to delayed tube-renewal dynamics.
As a result, the transition into free diffusion of the tube would happen later than justified by the number of topological constraints present in the system.

\subsection{Parametrization of the SLSP models}

Utilizing the \ac{SLSP} model to predict properties of real polymeric systems requires a mapping of the model parameters to physical units.
In our systematic approach, this parameter matching builds upon the results of two finer-grained models -- the atomistic and \acf{mCG} model.
In this section, we detail how we use the results of the \ac{mCG} model to identify the length and timescale of the \ac{SLSP} model for a specific polymer, \ac{cPB}.
We use the results of the mCG model for a 400-mer melt because this longest chain length, accessible to the lower-scale models, exhibits the most pronounced entanglement effects.

One particularly important characteristics of our soft, \ac{CG} modeling strategy is the independent mapping of static equilibrium properties and of dynamical properties of polymer melts. First, we identify the length scale to match the static single-chain properties of a 400-mer \ac{cPB} system. The configurations of long flexible polymer chains in a dense melt are characterized by a single relevant length scale -- the mean-squared end-to-end distance, $\langle R_\text{ee}^2 \rangle$. In the absence of nonbonded interactions we obtain $R_{e0}^2 = N b_0^2 = N (0.75\sigma)^2$. By matching this chain extension to the result of the \ac{mCG} model, $\sqrt{\langle R_\text{ee}^2\rangle} = 13.446$ nm for a 400mer \ac{cPB}, we identify the internal unit of length, $\sigma$, in our simulation model, $\sigma = 3.07 \pm 0.06$ nm. 

The invariant degree of polymerization $\sqrt{\mathcal{\bar{N}}} = \frac{\rho}{N} R_{e0}^3$ quantifies the chain density and is independent from the degree of coarse-graining. The density and chain extension of the \ac{mCG} model corresponds to $\sqrt{\mathcal{\bar{N}}} \approx 59.3$. This density requires $n=926$ chain in the \ac{SLSP} model for a cubic simulation box with length $L=2.5 R_{e0}$. In a melt, the ratio between the bare $R_{e0}^2$ and the actual measured end-to-end distance slightly depends on chain length and density, due to the chain end effect. For the specific system the ratio adopts the value, $\langle R_\text{ee}^2 \rangle / R_{e0}^2 \approx 1.0552$. This minor, state-dependent effect is neglected in the following.

Second, we adjust the parameters of our model to match the dynamics of the \ac{mCG} model. We note that the parameters that dictate the dynamics are independent of those that determine the static properties. The unentangled dynamics require the identification of segmental friction and the timescale mapping, $t^*_\text{SS}$, between our soft, \ac{CG} model, and the results of the \ac{mCG} model. We have reduced the monomeric friction but the crossover from ballistic to diffusive segment motion still occurs on short time scales that are irrelevant to the analysis. For the longer, entangled chains, we additionally adjust the \ac{SLSP} parameters, \ie, the \ac{SLSP} fugacity, $z$.

In the spirit of top-down modeling, we could adjust $t^*_\text{SS}$ and $z$ to match experimental data. Here, instead, we follow a different approach using the available data of the mCG model to identify the timescale factor and \ac{SLSP} fugacity. This parameter passing is based on the expectation that the \ac{SLSP} model does not only capture the dynamics of well-entangled polymer melts but also accurately describes the broad crossover from unentangled to entangled polymer dynamics. 

To identify the timescale and \ac{SLSP} fugacity, we decided to focus on the decorrelation of the normalized end-to-end vector $P(t)$ (cf.~\autoref{eq:ee-correlaton-definition}).
This quantity is particularly suited to determine the timescale matching between the \ac{mCG} model and \ac{SLSP} model because it can be determined with high accuracy in both models and allows an accurate timescale matching due to its characteristic decay.

For the two extreme descriptions of entanglement effects -- the Rouse model \cite{rouse1953a} and the tube model \cite{doi1988the} -- there exist explicit predictions for $P(t)$.
In the Rouse model, appropriate for unentangled, short polymer, $P(t)$ adopts the form \cite{doi1988the}:
\begin{align}
    P_\mathrm{Rouse}(t|\tau_R) = \sum_{p : \mathrm{odd}} \frac{8}{p^2 \pi^2} \exp(-p^2t/\tau_R)
\end{align}
with $\tau_R$ being the Rouse time. The tube model \cite{doi1988the} predicts the identical shape, $P_\mathrm{tube}(t) = P_\mathrm{Rouse}(t|\tau_d)$, but the characteristic timescale is the disentanglement time, $\tau_d$.

For the timescale mapping, we scale the intrinsic simulation time of the \ac{SLSP} model by a factor $t^*_\text{SS}$ to match the results of the \ac{mCG} model. Since we allow this time scaling, the Rouse model and the tube model predict the identical shape of $P(t)$, making it a robust quantity to match the timescale over the entire regime of molecular weights.

\begin{figure}
    \centering
    \begin{subfigure}{0.46\textwidth}
        \centering
        \includegraphics[width=\textwidth]{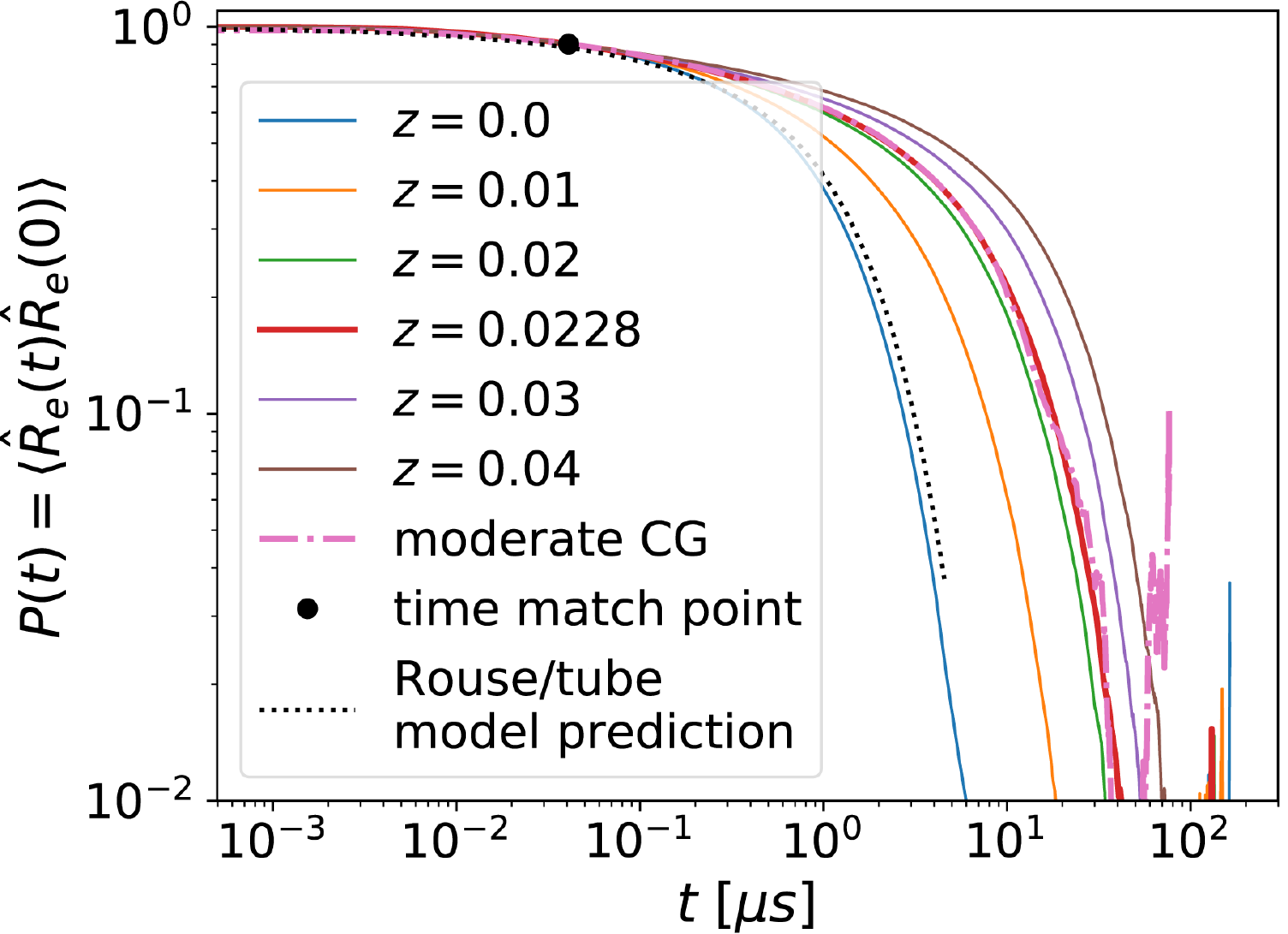}
        \caption{Matching the timescale $t^*_\text{SS}$ for short timescales.}
        \label{fig:slsp-rec-short}
    \end{subfigure}
    \hfill
    \begin{subfigure}{0.46\textwidth}
    \centering
        \includegraphics[width=\textwidth]{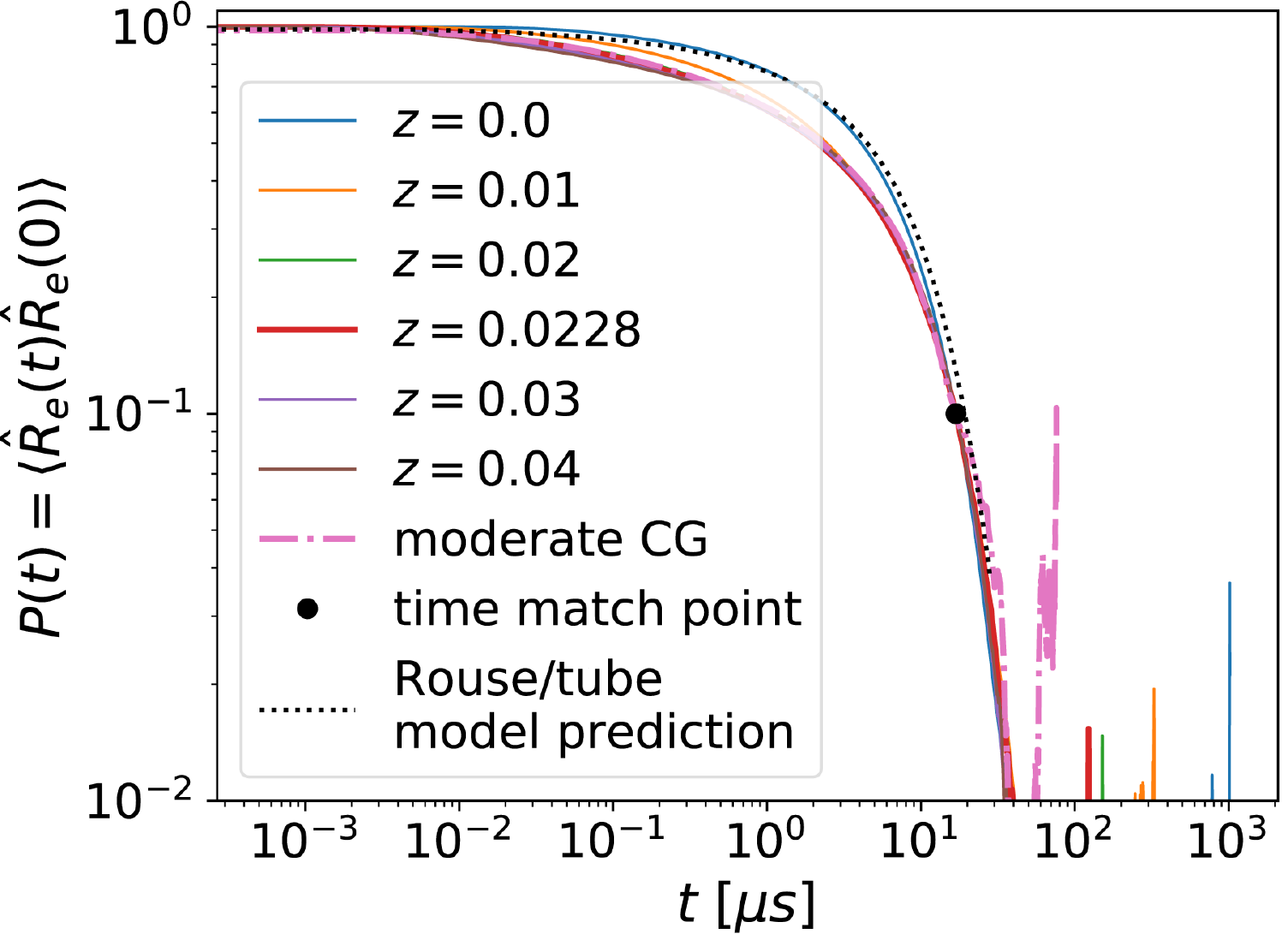}
        \caption{Matching the timescale $t^*_\text{SS}$ for long timescales.}
        \label{fig:slsp-rec-long}
    \end{subfigure}
    \caption{
    Timescale matching \textit{via} the end-to-end vector correlation $P(t)$. The matching is obtained by horizontally shifting results of the \ac{SLSP} model (on a logarithmic scale) until they overlay the \ac{mCG} results at the "match point". In general, agreement at different "match points", $t_m$, is observed for a different combination of timescale, $t^*_\text{SS}$, and \ac{SLSP} fugacity, $z$, because the shape of the curve depends on the entanglement density. Only for the specific \ac{SLSP} fugacity $z\approx 0.0228$, is the resulting timescale mapping, $t^*_\text{SS}$, independent of the "match point", $t_m$. For reference, we also present $P(t)$ of the Rouse and tube model. It matches the results of $z=0$, \ie, no \acp{SLSP}.
    }
    \label{fig:slsp-rec}
\end{figure}

\autoref{fig:slsp-rec} illustrates this timescale matching \textit{via} the end-to-end vector correlation $P(t)$. By choosing multiple (here two) different "match points", $t_m$, to overlay the results, we visually demonstrate that the shape of $P(t)$ depends on the entanglement density. This deviation is present for the results of the \ac{mCG} model and all \ac{SLSP} results, with $z > 0$.

A deviation is expected because the 400-mer polymer chains are long enough to deviate from the Rouse model but yet not sufficiently long to comply with the idealization of the tube model, which only applies to strongly entangled melts since the tube model neglects constraint release and contour-length fluctuations. The deviation of the mCG model indicates that these effects are relevant, particular for the rather short chains. The fact that these deviations are also observed in the SLSP model demonstrates that the SLSP description goes beyond the Rouse and tube models respectively.

We can use the shape changes, as depicted in \autoref{fig:slsp-rec}, to determine not only a timescale mapping but also the optimal number of \ac{SLSP} per chain, controlled by the fugacity, $z$, in our \ac{SLSP} model.
The premise is: For the optimal number of \ac{SLSP} per chain is the mapping factor $t^*_\text{SS}$ independent of the "match point", where the result is overlaid.
Hence, we use the two chosen "match points", for short and for long times, and plot the matching factor $t^*_\text{SS}$ as a function of the \ac{SLSP} fugacity, $z$.

\begin{figure}
    \centering
    \begin{subfigure}{0.46\textwidth}
        \centering
        \includegraphics[width=\textwidth]{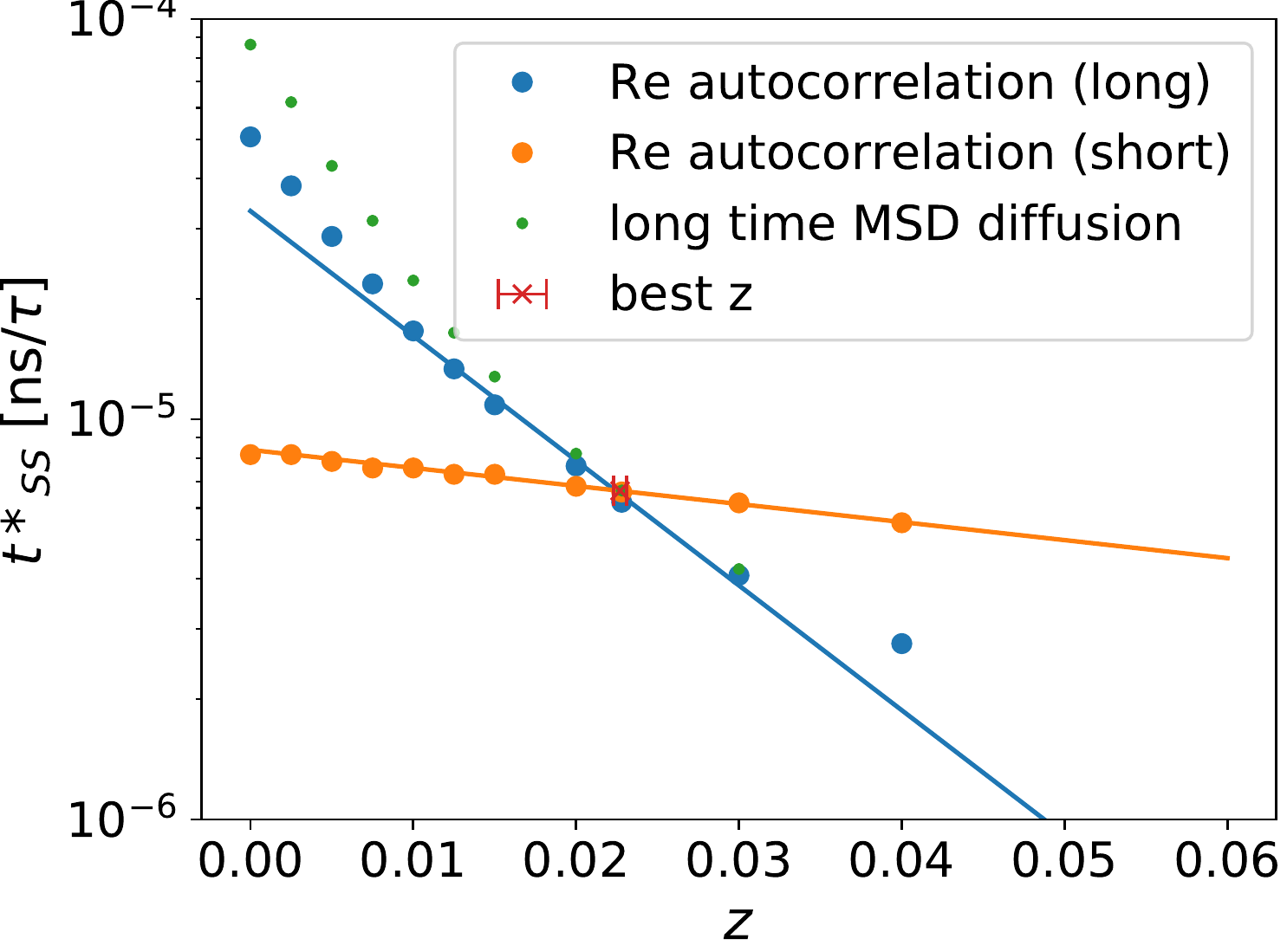}
        \caption{Time mapping factor $t^*_\text{SS}$ obtained via different procedures as a function of the fugacity $z$. The $z$ where all methods agree, determines the optimum.}
        \label{fig:slsp-zfactor}
    \end{subfigure}
    \hfill
    \begin{subfigure}{0.46\textwidth}
        \centering
        \includegraphics[width=\textwidth]{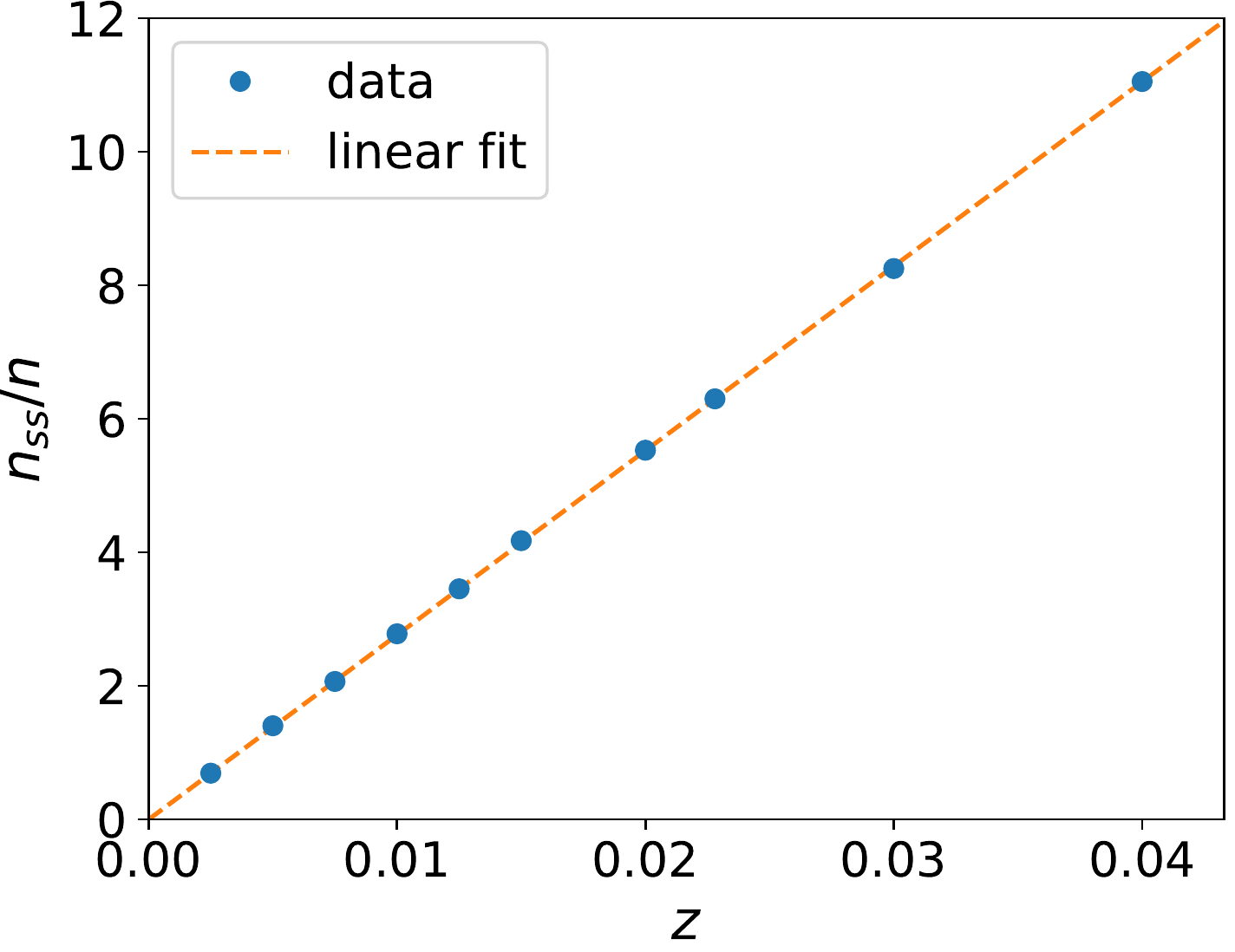}
        \caption{Number of \acp{SLSP} as a function the \ac{SLSP} fugacity $z$. For the optimum $z=0.0228$, we have $n_{ss}/n\approx6.3$. The linear prediction of Ref.~\cite{chappa2012translationally} is replicated.}
        \label{fig:slsp-nss}
    \end{subfigure}
    \caption{
    Timescale mapping $t^*_\text{SS}$ for the \ac{SLSP} model obtained \textit{via} matching different properties.
    The end-to-end vector correlation, $P(t)$, offers two different times "match points" (see \autoref{fig:slsp-rec}) and the long-time diffusion allows for a timescale mapping as well.
}
    \label{fig:slsp-z-mapping}
\end{figure}

\autoref{fig:slsp-zfactor} demonstrates how the timescale mapping varies for different \ac{SLSP} fugacities $z$, depending on the property, for which the mapping is constructed. For the already discussed mapping of $P$ at different match points, $t_m$, this gives rise to different shapes of $P(t)$ (because $z$ is a function of $t_m$). If the shape were optimally reproduced by the \ac{SLSP} model the time mapping factor $t^*_\text{SS}$ would be independent of $t_m$. This strategy identifies the optimal combination of $z$ and $t^*_\text{SS}$, \ie, $z\approx 0.0228$ and $t^*_{SS}= 0.0132\ \mu \mathrm{s} /\tau$.
There exists an alternate approach to identify the timescale \textit{via} the long-time diffusion, quantified by the \ac{MSD}. This strategy has been discussed earlier for the mapping between the atomistic model and the \ac{mCG} model and also provides a relation $t^*_{SS}(z)$ that is presented in \autoref{fig:slsp-zfactor}. Gratifyingly, this curve passes through the optimal combination identified by matching $P(t)$, indicating that we have obtained a consistent timescale mapping for multiple, single-chain properties.

This optimal fugacity, $z\approx 0.0228$, corresponds to roughly $n_{ss}/n \approx 6.3$ \ac{SLSP} per chain (see \autoref{fig:slsp-nss} that depicts a linear dependence of $n_{ss}$ on $z$, verifying our \ac{SLSP} simulation code). Each anchor of an \ac{SLSP} poses a topological constraint for the polymer motion. Although, there is no one-to-one correspondence between the \ac{SLSP} constraints and entanglements\cite{chappa2012translationally,ramirez2018detailed,ramirez2017multi}, the number of \ac{SLSP} constraints $\approx 12.6$ is in the same order of magnitude as the number of entanglements expected for a 400-mer \ac{cPB} polymer. For cPB at $T = 298$ K, $M_\text{e} \approx 2900$ g mol$^{-1}$ has been reported for the molecular weight of a chain portion between entanglements.\cite{ferry1980viscoelastic}. Based on this value of $M_\text{e}$, 400-mer chains ($M = 21.6$ kg mol$^{-1}$) have around $7.5$ entanglements per chain. Due to the limited number of experimental data for pure cPB, we also compare to the experimental value $M_\text{e} \approx 1800$ g mol$^{-1}$ for 1,4-PB (mainly containing cis and trans bonds) at $413$ K \cite{fetters1994connection}; based on this estimate, a 400-mer chain has around $12$ entanglements.

\begin{figure}
    \centering
    \begin{subfigure}{0.46\textwidth}
        \centering
        \includegraphics[width=\textwidth]{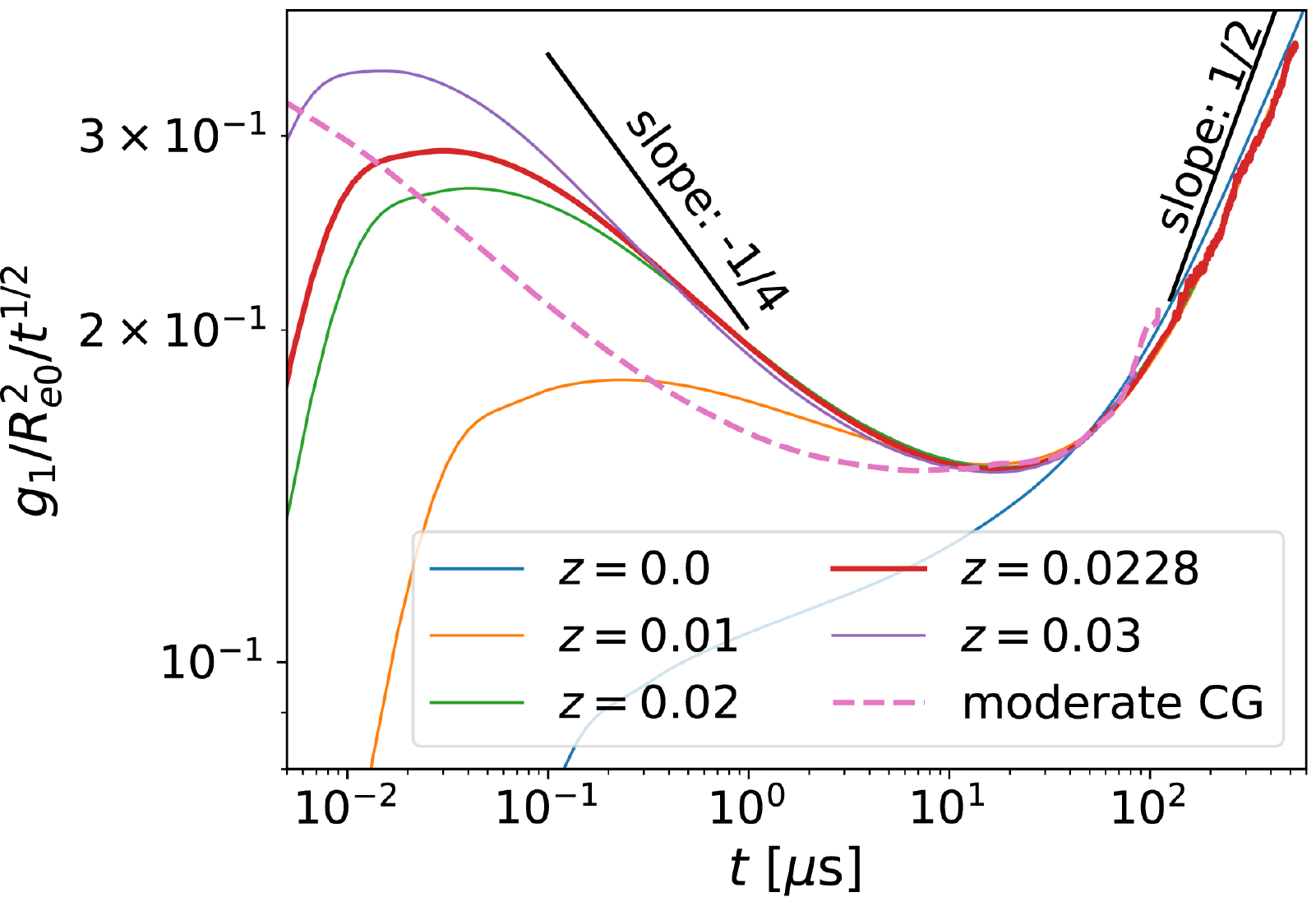}
        \caption{Monomer \ac{MSD} $g_1$ shows power-law behavior characteristic of entangled dynamics for different \ac{SLSP} fugacities $z$ and identifies the shortest timescale, when the \ac{SLSP} model coincides with the \ac{mCG} model.}
        \label{fig:slsp-msd-z}
    \end{subfigure}
    \hfill
    \begin{subfigure}{0.46\textwidth}
        \centering
        \includegraphics[width=\textwidth]{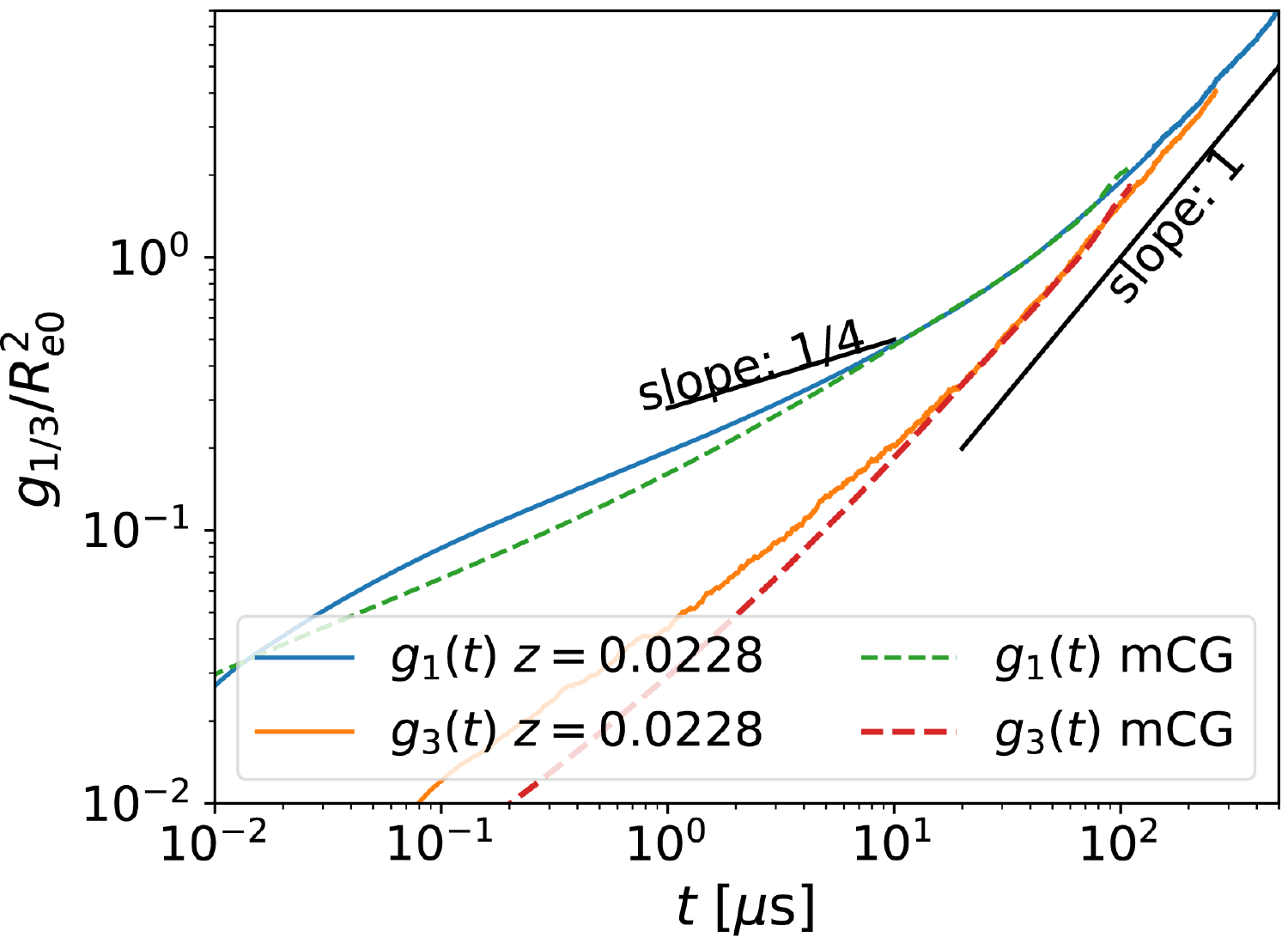}
        \caption{Monomer $g_1$ and center of mass $g_3$ \ac{MSD} with the optimal \ac{SLSP} fugacity $z=0.0228$ compared to the data of the \ac{mCG} model.}
        \label{fig:slsp-msd-3}
    \end{subfigure}
    \caption{Comparison of \ac{MSD} of the \ac{mCG} and \ac{SLSP} models. }
    \label{fig:slsp-msd}
\end{figure}

\autoref{fig:slsp-msd} compares the \ac{MSD} of the \ac{SLSP} model for different fugacities $z$ to the results of the \ac{mCG} model, using the aforementioned long-time \ac{MSD} to match the timescales. 
We observe the power laws that are the hallmark of entangled dynamics for systems with large $z$, already for the small contour discretization, $N_{\mathrm{cg}}=32$. Plotting $g_1(t)/\sqrt{t}$, we highlight exponents smaller than $1/2$.  

The comparison between the \ac{mCG} results and the data of the \ac{SLSP} model, however, reveals deviations on short timescales.
Specifically, the \ac{SLSP} model exhibits a larger \ac{MSD} at times shorter than $t\lesssim {\cal O}(\mu$s$)$, corresponding to the length scale of a bead of the \ac{SLSP} model or the tube diameter, \ie, $g_1(1\mu$s$)\approx 3.56\sigma^2$.
\footnote{The crossing of the data sets for very short times, $t\approx 10^{-2}\mu$s, stems from (i) the small number of Rouse modes in the \ac{SLSP} model and (ii) the deferred crossover between the ballistic and overdamped dynamics in the \ac{SLSP} model where the segmental friction does not only depend on the interactions but additionally on the strength of the \ac{DPD} thermostat.}

We observe non-negligible deviations between the \ac{MSD} of mCG and the \ac{SLSP} model for time scales below $t>10\mu$s whereas the the relaxation modulus $G(t)$ and the end-to-end relaxation $P(t)$ already agrees for $t>1\mu$s. 
The difference between the observable is that the monomer \ac{MSD} ($g_1$) is sensitive to the motion of individual beads or segments, while for the $G(t)$ and $P(t)$ characteristics of the orientation of the chain as a whole are more relevant.
It is natural to expect that the more coarse-grained models can represent the large molecule scales more accurately.
This is a reason why we selected the end-to-end distance correlation $P(t)$ to match time scales and entanglement density.
This relation of accuracy discrepancy is not unique to the model transition from mCG to SLSP, we observe similar characteristics for the transition between the atomistic and mCG model.
The mCG model provides a faithful description of the MSD for $t>0.2$ns, whereas $G(t)$ agrees with the atomistic model already for $t>0.01$ns (\autoref{fig:p(t)-msd-100}).

We note that smaller fugacities, $z\leq 0.1$, appear to agree somewhat better on short timescales. Such small values of $z$, however, are not consistent with the dynamics of the end-to-end vector. We hypothesize that (i) the fluid-like packing structure and the more complex bonded interactions of monomeric repeating units in the \ac{mCG} model affect this short-scale dynamics and (ii) the constraining tube formed by the \ac{SLSP} may be somewhat softer compared to the effect of true non-crossability. This comparison identifies the smallest timescale, above which our \ac{SLSP} model for cPB can provide accurate predictions. The dynamical properties on shorter scales can be covered by more detailed models.

\subsection{Rheological properties}
Having identified the parameters and scales of the \ac{SLSP} model by comparing the single-chain statics and dynamics to the mCG model, we can now investigate the collective, rheological properties to validate our model.

The autocorrelation of the non-diagonal elements of the stress tensor (shear stress relaxation modulus, $G(t)$) is of particular interest to many applications related to the rheology of polymer systems. For the \ac{SLSP} the bonded and non-bonded stress is calculated for each chain. 
To assign the stresses to individual chains, first, all virials of conservative forces are attributed to individual particles. In our calculation, we ignore the force contribution from the DPD thermostat. The virial of an interacting pair is equally assigned to the involved particles. In a second step, the particle virials are summed up in each polymer chain.
For more details on this chain average technique and its implication for the stress autocorrelation $G(t)$, refer to appendix \ref{sec:slsp-appendix}.

For a comparison between the models, we adjust them to match the shear modulus, $G_0$, because of the softness of the interactions in the \ac{SLSP} model. To this end, we select a match point and vertically shift the curves on a logarithmic scale.

\begin{figure}
    \centering
    \includegraphics[width=\textwidth]{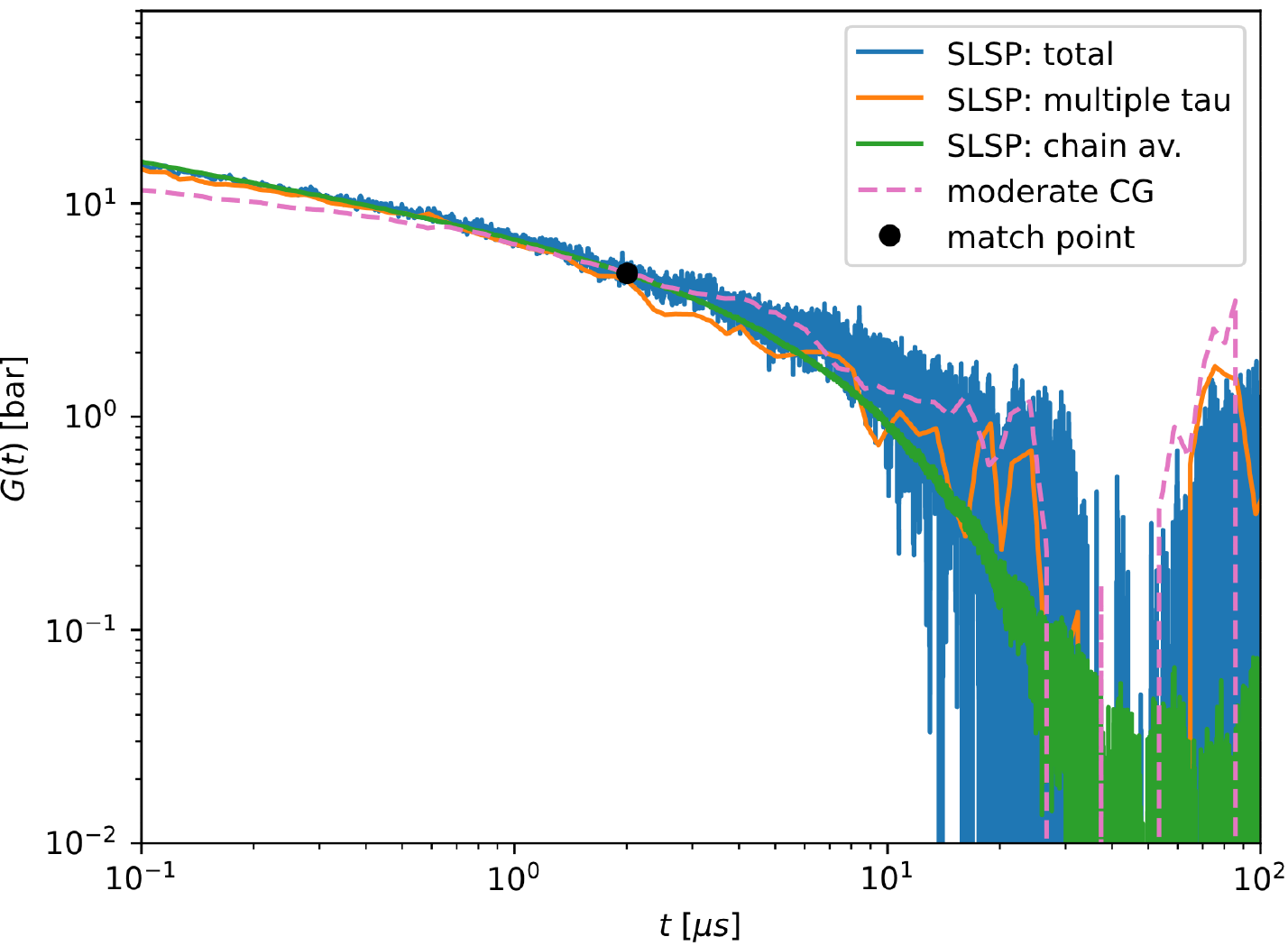}
    \caption{
    Stress-autocorrelation of the non-diagonal stress tensor elements $G(t)$ obtained by the mCG model and the \ac{SLSP} model for the optimal \ac{SLSP} fugacity. A vertical shift (on the logarithmic scale) is performed to match the the shear modulus $G_0$ of the \ac{mCG} and \ac{SLSP} models. For comparison, we also provide the total stress result with less statistics. We also include a stress autocorrelation calculated from the total stress, but post processed with the multiple-tau-correlator \cite{likhtman2007linear} as the mCG data is generated with this algorithm as well.
    }
    \label{fig:slsp-gt}
\end{figure}

Data are shown in \autoref{fig:slsp-gt}. We observe reasonable agreement between the mCG and the \ac{SLSP} model for the previously determined entanglement density $z=0.0228$. In particular, this agreement extends to the terminal decay time of $G(t)$, validating our timescale matching. Note, however, that even with advanced averaging methods such as the multiple-tau-correlator algorithm\cite{likhtman2007linear} $G(t)$ is plagued by rather high uncertainty. Only at shorter times, the curves slightly deviate. This is, however, expected because the mCG model includes local dynamical modes that are integrated out in the \ac{SLSP} model. The shortest timescale, on which agreement is expected, is on the order of $\mu$s and the observed deviations are on shorter timescales.

\subsection{Transferability of slip-spring model parameters}
The idea of the hierarchical (consecutive) systematic coarse-graining is to be able to predict quantitatively material properties, of high molecular weight, entangled, systems, with the highest \ac{CG} model because of its computational efficiency.
For this goal, it is important to achieve the transferability of the obtained matching parameters to other molecular weights.
In the previous section, we demonstrated how we can obtain the matching parameters between the  \ac{mCG} model and the \ac{SLSP}.
Before moving on to higher molecular weights we demonstrate this transferability by simulations of 200-mer melt because it is still tractable by both, the \ac{mCG} model and the \ac{SLSP} model.

 \begin{figure}
    \centering
    \begin{subfigure}{0.3\textwidth}
        \centering
        \includegraphics[width=\textwidth]{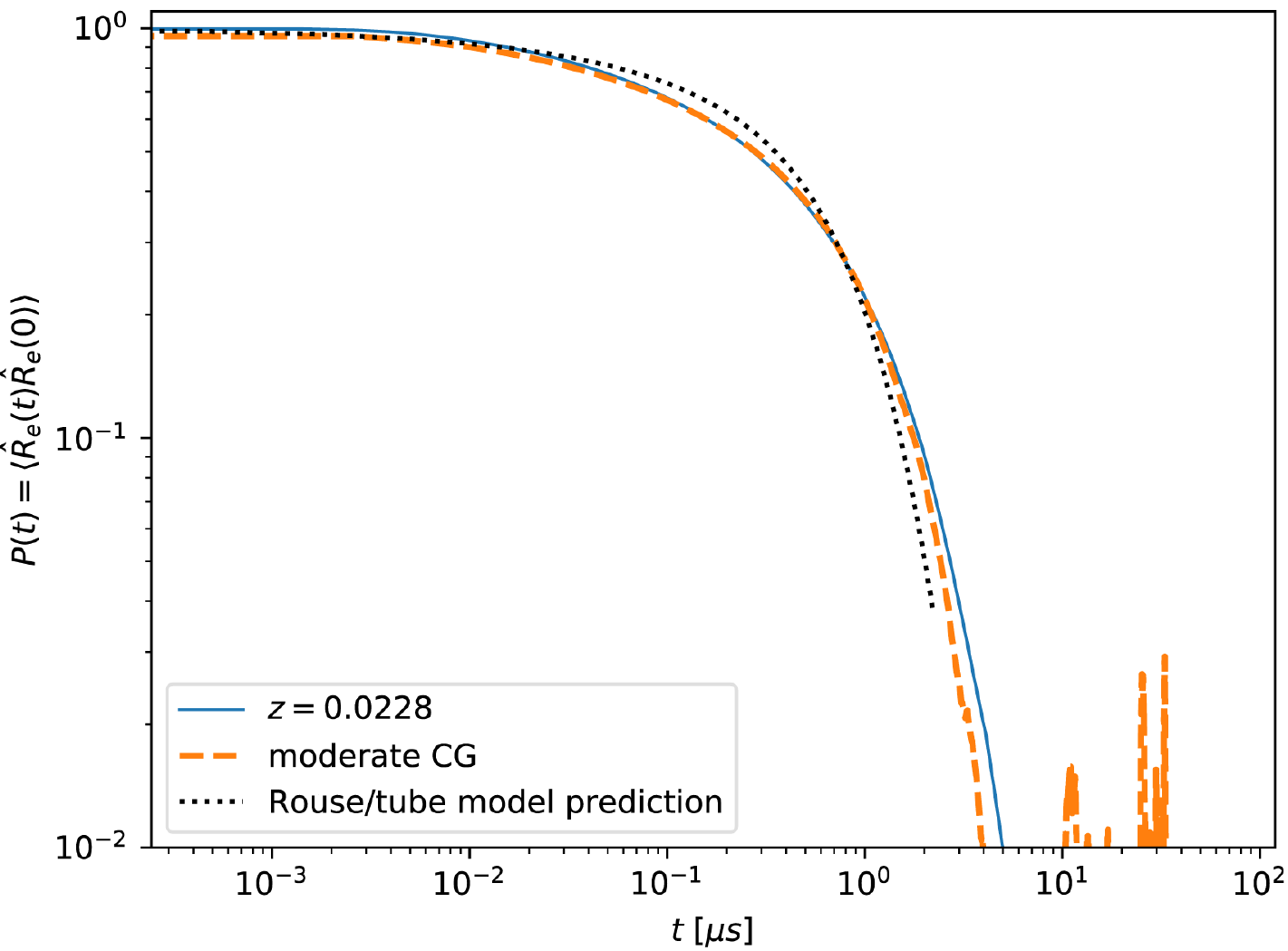}
        \caption{End-to-end vector correlation, c.f. \autoref{fig:slsp-rec}}
        \label{fig:slsp-200-rec}
    \end{subfigure}
    \begin{subfigure}{0.3\textwidth}
        \centering
        \includegraphics[width=\textwidth]{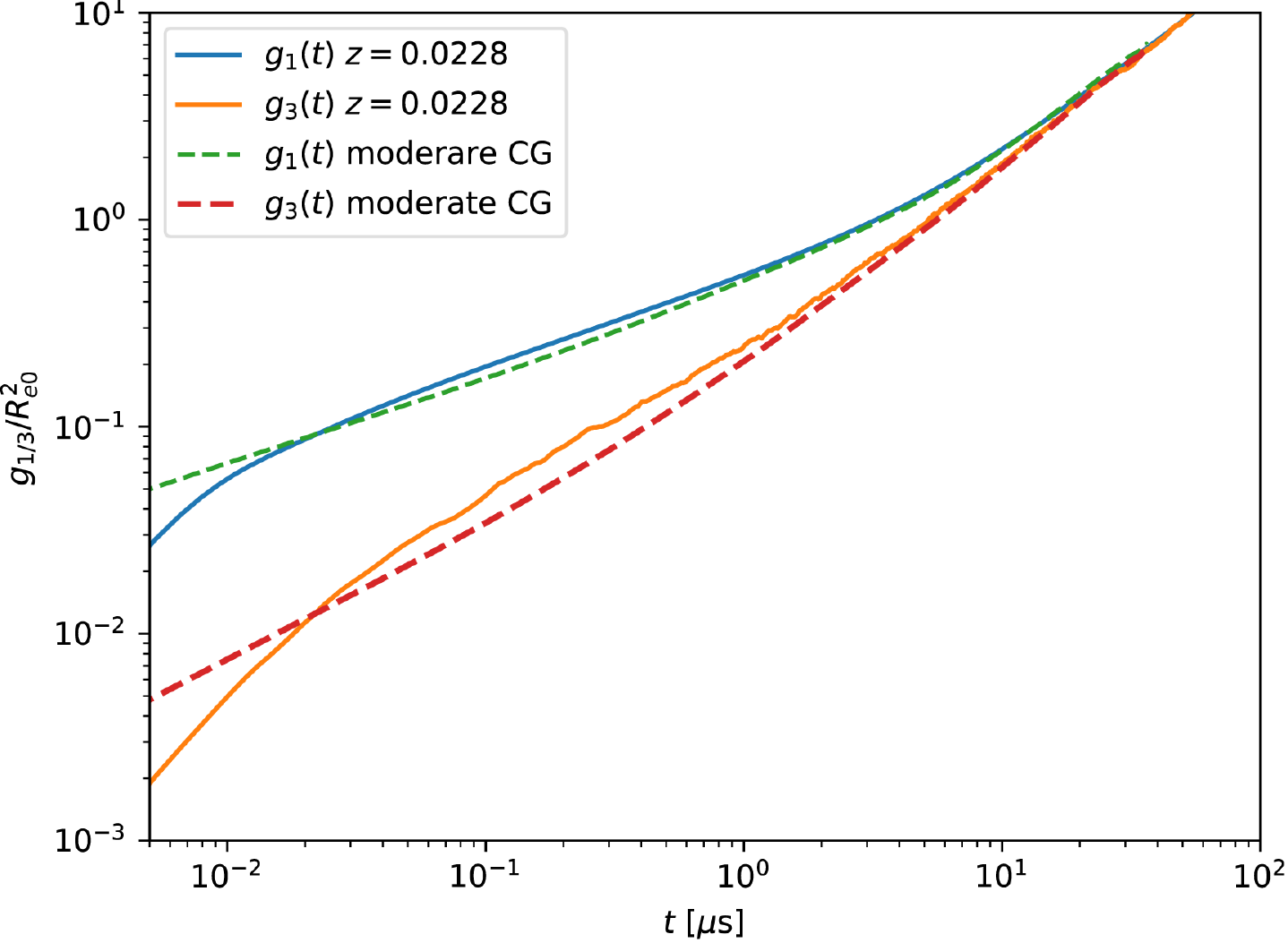}
        \caption{Monomer $g_1$ and center of mass $g_3$ \ac{MSD}, c.f. \autoref{fig:slsp-msd}.}
        \label{fig:slsp-200-msd}
    \end{subfigure}
    \begin{subfigure}{0.3\textwidth}
        \centering
        \includegraphics[width=\textwidth]{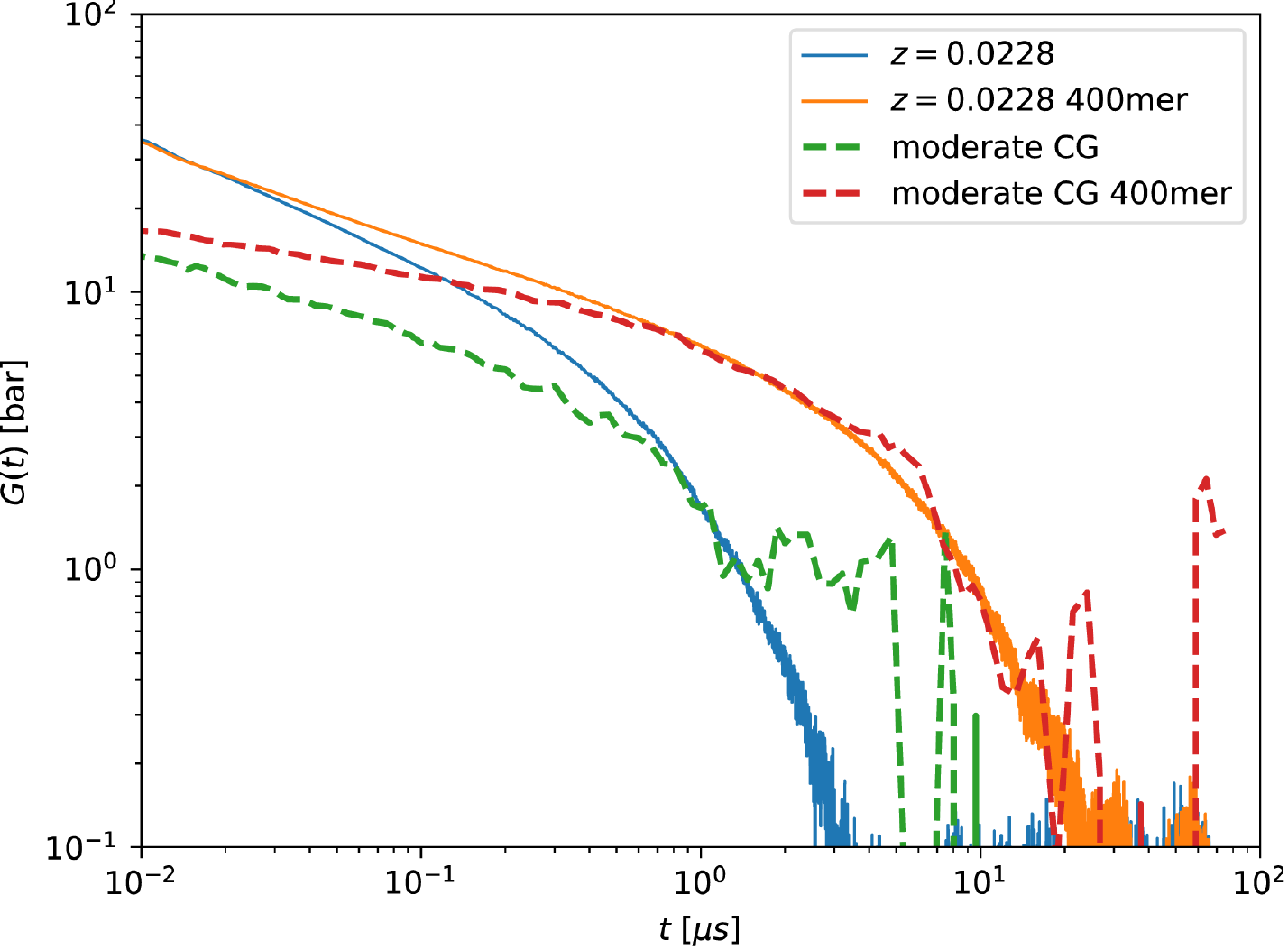}
        \caption{Stress auto-correlation $G(t)$, c.f. \autoref{fig:slsp-gt}.}
        \label{fig:slsp-200-gt}
    \end{subfigure}
    \caption{Transferability between the mCG and \ac{SLSP} model.
    The matching parameters are obtained with 400-mer \ac{cPB} and transferred to simulations of a 200-mer melt without additional adjustments.
    This test demonstrates good agreement between the models with molecular weights other than the reference 400mer. It even shows that the time and length scales at which we can expect accurate predictions are independent of the molecular weight.}
    \label{fig:slsp-200-transfer}
\end{figure}
 
\autoref{fig:slsp-200-transfer} presents the three most characteristic properties that we have used to quantify the dynamics of polymer materials: the end-to-end vector correlation $P(t)$, the \ac{MSD}, and the stress autocorrelation function $G(t)$ for the 200-mer systems.
Note, that as mentioned above all model parameters have been obtained from the 400-mer melt. Thus, the calculation of the data for the 200-mer melt in \autoref{fig:slsp-200-transfer} serves as a direct check of the transferability of the SLSP model parametrization, with no additional adjustment.
Overall, the agreement is very good, in particular for the $P(t)$ and the \ac{MSD}. For the stress autocorrelation function, $G(t)$, the overlap of the two curves does not cover a large time interval. The simulations of the 400-mer melt indicate that the \ac{SLSP} model only applies for $t>1\mu$s, independent from the molecular weight. On this timescale, however, $G(t)$ cannot be obtained with high accuracy in the simulation of the mCG model. Within rather large uncertainties, the data suggest that the time of the terminal decay is correctly predicted by the \ac{SLSP} model. 

Taken together, the comparison of the mCG and \ac{SLSP} models for melts of 200-mer and 400mer cPB shows the \ac{SLSP} model can accurately predict a long and significant portion of $G(t)$ for higher molecular weights because the relevant timescale increases with molecular weight. 

\section{Dynamics of PB melts}

In this section, the dynamical and rheological properties of cPB melts are discussed, using the simulation results from the atomistic, mCG, and SLSP models. Dynamical properties are presented for different chain lengths from oligomers up to the molecular weights of industrial relevance (100 kDa), predicted by the different models.

\subsection{Translational dynamics}
\subsubsection{Self-diffusion coefficient}

\begin{figure}[!htb]
    \centering
    \begin{subfigure}{0.45\textwidth} % width of left subfigure
        \includegraphics[width=\textwidth]{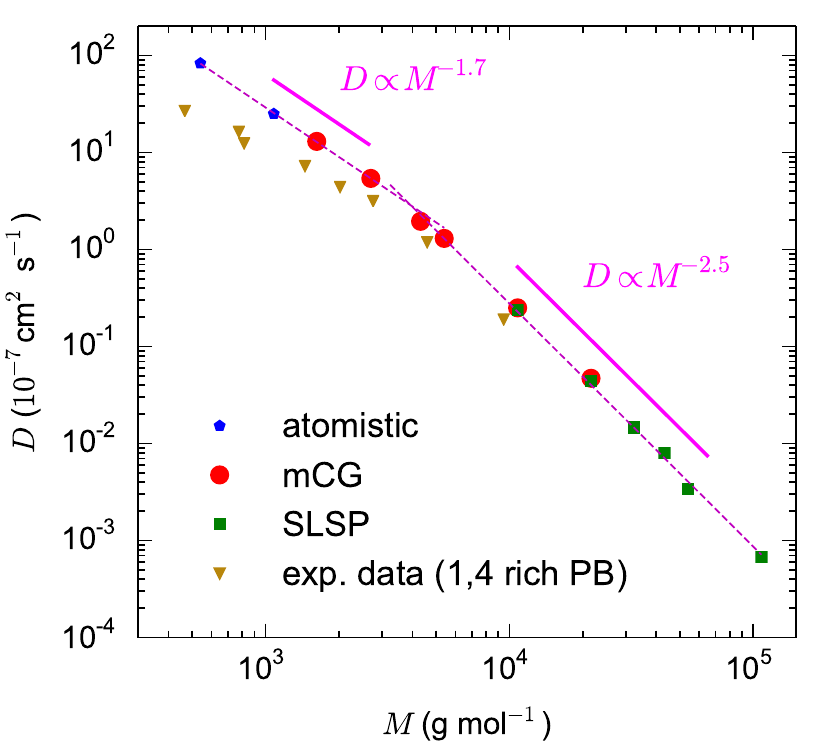}
    \end{subfigure}
    %\vspace{1ev} % here you can insert horizontal or vertical space
    \caption{Molecular-weight dependence of the self-diffusion coefficient for bulk melts of cPB at $413$ K. Depending on molecular weight, the results of the atomistic, mCG, or SLSP models are reported (for  $M = 10.8$ and $21.6$ kg mol$^{-1}$ the results of both mCG and SLSP models are shown). The experimental self-diffusion coefficients for samples of 1,4 rich PB with narrow molecular-weight distributions are also shown.\cite{meier2013long}    }
    \label{Fig:D} % caption for whole figure
\end{figure}

The self-diffusion coefficient, $D$, can be calculated from the linear part of the MSD of chain center-of-mass, $g_3(t)$, through the Einstein relation, $D = \lim\limits_{t\rightarrow \infty} \frac{g_3(t)}{6 t}$.
The molecular-weight dependence of $D$ for cPB at 413 K is presented in \autoref{Fig:D}. 
Depending on the molecular weight, $M$, results from the atomistic, mCG, or SLSP models are presented. For the 200-mer and 400-mer chains ($M = 10.8$ and $21.6$ kg mol$^{-1}$), the results of both mCG and SLSP simulations are shown.
The experimental values of $D$,\cite{meier2013long} calculated through ﬁeld-cycling $^1$H NMR relaxometry, for a series of nearly monodisperse 1,4 rich PB at $413$ K are also shown in \autoref{Fig:D}.  
Considering the presence of trans segments, which are slower than cis ones,\cite{behbahani2020conformations} in the experimental samples, a fair agreement between simulation and experimental results is observed.

Two different scaling regimes are observed for $D$, corresponding to the unentangled and the entangled regimes. The gradual crossover between these two regimes takes place around chain lengths of 
$50$ to $80$ monomeric repeating units ($M_\text{c} \approx 2.7$ to $4.3$ kg mol$^{-1}$).  
The Rouse and the tube models predict $D \propto M^{-1}$ and $D \propto M^{-2}$ scaling relations for unentangled and entangled chains, respectively.\cite{doi1988the,rubinstein2003polymer} 
Here the scaling exponents of the unentangled and entangled regimes are not exactly equal to the predictions of the Rouse and the tube models. 
These deviations are the results of the chain-end effect (which lead to a change of density with chain length)\cite{harmandaris1998Rouse,harmandaris2003PE}
and contour length fluctuations and constraint release for (mildly) entangled chains.\cite{lodge1999reconciliation,harmandaris2009dynamics}
Similar deviations have been observed for other polymer melts as well.\cite{lodge1999reconciliation,harmandaris2009dynamics,vogiatzis2017equation, sgouros2017slip}
The calculated scaling exponent in the entangled regime of cPB at $413$ K (around $2.5$) is in good agreement with the experimental scaling exponent for hydrogenated PB at $413$ K (around $2.45$).\cite{tao2000diffusivity}

The key result from comparing the chain diffusion properties, derived from the three models is the seamless transition between the models as higher molecular weights are considered.
Without the three coarse-graining steps, it is intractable to perform these simulations and predict the long-time diffusion characteristics.
The fact that the \ac{SLSP} model can seamlessly continue the scaling, highlights its ability to capture the entanglement effects correctly.
The \ac{SLSP} model includes contour-length fluctuation and constraint-release effects and, indeed, shows very similar scaling as the experimental results.
The matching between the models allows us to use the \ac{SLSP} model not only for qualitative but also for quantitative predictions.

\subsubsection{Segmental mean-squared displacement}
\label{sec:g1(t)}
\begin{figure}[!htb]
    \centering
    \begin{subfigure}{0.45\textwidth} % width of left subfigure
        \includegraphics[width=\textwidth]{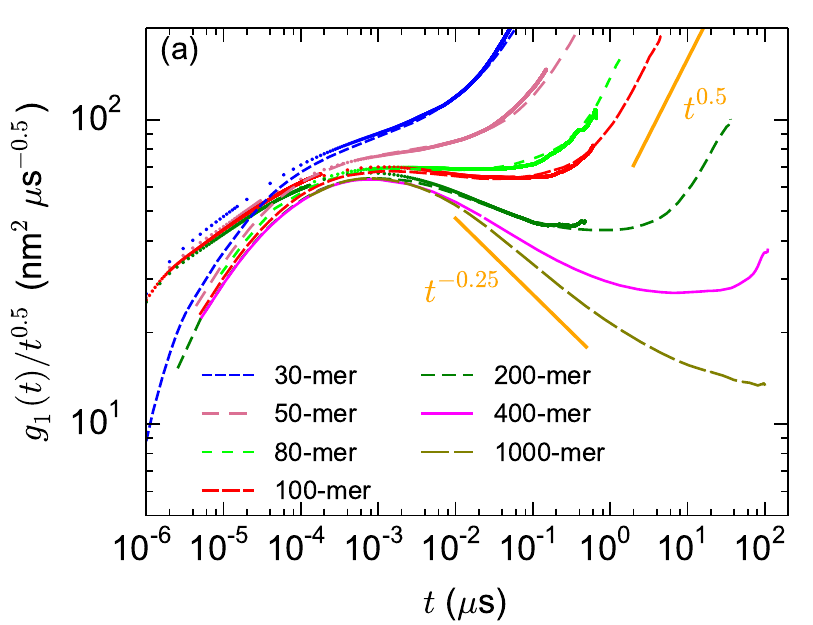}
    \end{subfigure}
    \begin{subfigure}{0.45\textwidth} % width of left subfigure
        \includegraphics[width=\textwidth]{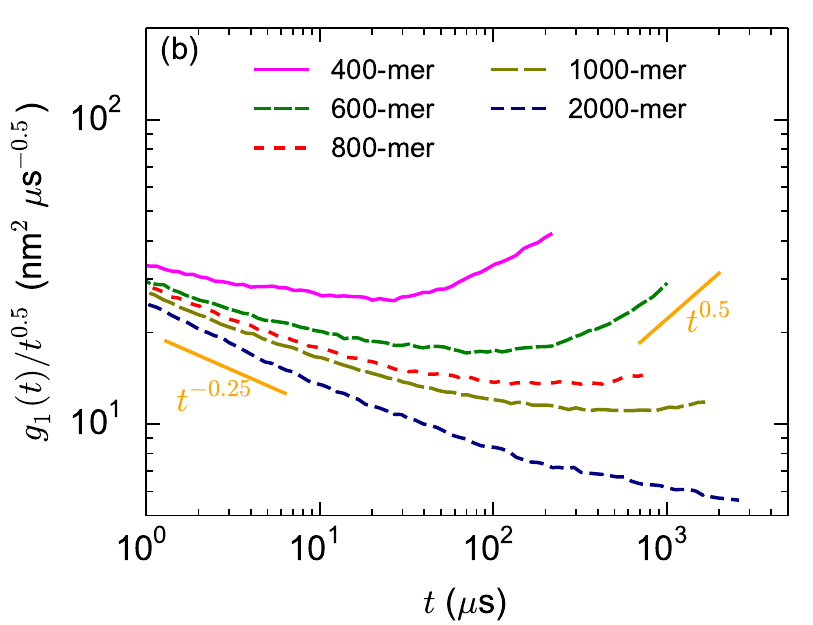}
    \end{subfigure}
    %\vspace{1ev} % here you can insert horizontal or vertical space
    \caption{Segmental mean-squared displacement, $g_1(t)$, normalized by $\sqrt{t}$ for cPB chains of different lengths, obtained by simulations of: (a) the atomistic (dots), and \ac{mCG} (lines) models  and (b) the \ac{SLSP} simulations. 
    For a direct comparison between the \ac{MSD} of the \ac{mCG} and \ac{SLSP} models see \autoref{fig:slsp-msd-3} and \autoref{fig:slsp-200-msd}.
       } 
    \label{fig:mcg-g1} % caption for whole figure
\end{figure}

The gradual crossover from unentangled to entangled polymer dynamics regime can also be observed through the calculation of the segmental MSD, $g_1(t)$, as a function of chain length. The Rouse model predicts the following scaling regimes for unentangled polymer chains (in the limit of very long chains):\cite{doi1996introduction} 
\begin{equation}
g_1(t) \propto 
\begin{cases}
t^{\frac{1}{2}}\ \  \  \  \  \ t<\tau_\text{R}\\
t^{1}\ \  \  \  \  \ t>\tau_\text{R}
\end{cases}
\end{equation}     
In the case of entangled chains, the tube model predicts:\cite{doi1988the}
\begin{equation}
g_1(t) \propto 
\begin{cases}
t^{\frac{1}{2}}\ \  \  \  \  \ t<\tau_\text{e}\\
t^{\frac{1}{4}}\ \  \  \  \  \ \tau_\text{e}<t<\tau_\text{R}\\
t^{\frac{1}{2}}\ \  \  \  \  \ \tau_\text{R}<t<\tau_\text{d}\\
t^{1}\ \  \  \  \  \ t>\tau_\text{d}
\end{cases}
\end{equation}     
where $\tau_\text{e}$, $\tau_\text{R}$, and $\tau_\text{d}$ are the entanglement time, the Rouse time and the reptation (disentanglement) time, respectively. The deviations from the Rouse behavior become more clear if $g_1(t)$ is normalized by $t^{\frac{1}{2}}$, which is the expected asymptotic Rouse slope. 
In this presentation, a negative slope at intermediate times is a sign of entanglements.\cite{likhtman2007linear} 
The plot of $\frac{g_1(t)}{\sqrt{t}}$ for cPB chains of various molecular weights is presented in \autoref{fig:mcg-g1}.

In panel (a) the results of atomistic, \ac{mCG},and \ac{SLSP} models are displayed. 
$\frac{g_1(t)}{\sqrt{t}}$ of 50-mer and shorter chains do not have a negative slope. However, before the normal diffusion regime, $g_1(t)$ evolves faster than $t^{\frac{1}{2}}$ (asymptotic Rouse behaviour).
The deviation is more pronounced for shorter chains.
The Rouse model for short chains (\ie, a finite number of modes) predicts this observed behavior. In \autoref{Fig:g1-Rouse} (\autoref{sec:g1-rouse}), the $g_1(t)$ curves of unentangled atomistic and mCG chains are compared with the predictions of the Rouse model. For 10-mer cPB, the Rouse model describes the $g_1(t)$ of atomistic and mCG models; particularly, for mCG chains, the agreement is very good and the Rouse model nicely fit the  $g_1(t)$ curve for times larger than the short time ballistic regime.
For 30-mer and 50-mer chains, a deviation between the simulation results and the Rouse model is observed. 
The ignorance of non-crossability of chains in the Roue model is the probable origin of this deviation (for a more detailed discussion see \autoref{sec:g1-rouse}).

For 80-mer and longer chains, an interval of negative slope is observed in the plot of $\frac{g_1(t)}{\sqrt{t}}$. The negative slope shows the effect of entanglement constraints on the motion of monomers. With increasing chain length, the negative slope decreases and tends to the prediction of the tube model, \ie, $-0.25$.
The deviations from the prediction of the tube model are due to the rather short lengths of the studied chains that belong to the region of the gradual crossover from the unentangled to the entangled regime.
It is worth mentioning that the onset of the appearance of the negative slope in $g_1(t)$ coincides with the breaking point in the slope of $D$ vs. $N$, both occur around $50$ to $80$ monomers.   
$g_1(t)$ of both atomistic and mCG models exhibit similar behavior (similar slopes) in the region of transition from the unentangled to the entangled regime (consider that scaling of mCG time does not affect the scaling exponents of $g_1(t)$). This similarity originates from the preservation of the chemical identity of the atomistic model in the mCG one.     

In \autoref{fig:mcg-g1}b results about $\frac{g_1(t)}{\sqrt{t}}$ from the \ac{SLSP} simulations are presented. 
The \ac{SLSP} model significantly extends the range of molecular weights and times that can be investigated. 
Utilizing the \ac{SLSP} model enables us to verify the prediction that higher molecular weights show a more pronounced $1/4$ scaling of $g_1(t)$ at early-intermediate times.
We can also observe the transition to the scaling behavior, $g_1(t) \propto t^{1/2}$, at later intermediate times and molecular weights up to a 1000-mer.
The last transition into free diffusion, $g_1(t) \propto t$, is still a computational challenge for high molecular weights. The present data only reach this regime for less than 600 monomeric repeating units per chain. 

\subsection{Linear viscoelastic properties}
\label{sec:LVE}
\begin{figure}[!ht]
    \centering
    \begin{subfigure}{0.45\textwidth}
        \includegraphics[width=\textwidth]{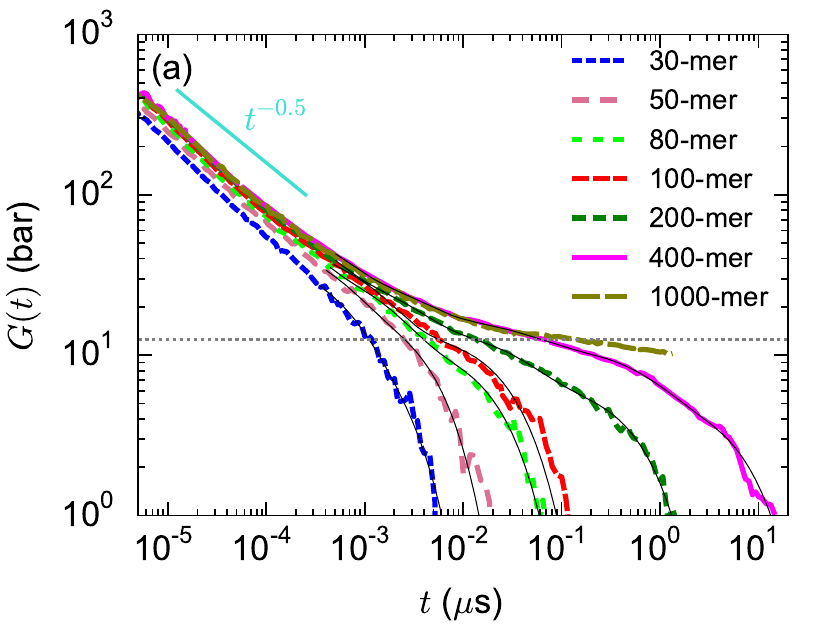}
        \end{subfigure}
    \begin{subfigure}{0.45\textwidth} % width of right subfigure
        \includegraphics[width=\textwidth]{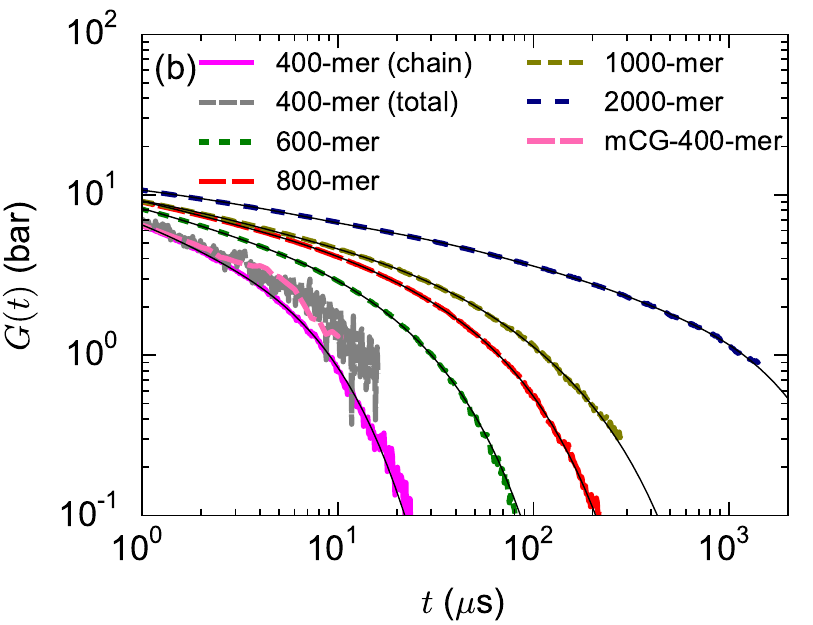}
    \end{subfigure}
    \caption{Chain-length dependence of shear-stress relaxation modulus, $G(t)$, for \ac{cPB} at $413$ K from (a) \ac{mCG} and (b) \ac{SLSP} models.
    In panel (a), the experimental\cite{fetters1994connection} plateau modulus of 1,4-PB at 413 K is shown by a dashed line.
    The $G(t)$ curves of the \ac{SLSP} model are calculated through chain averaging procedure; however, for 400-mer chains, $G(t)$ from the calculation of the autocorrelation of total stress is also presented.    The black solid lines are the results of fitting Maxwell modes. To improve visibility, these lines are only shown at long times.
    The \ac{SLSP} model is not able to resolve the short time characteristics, however, the behavior on longer timescales ($>1\mu$s) has been verified in \autoref{fig:slsp-gt}.}
    \label{fig:gt} % caption for whole figure
\end{figure}

The typical measure for the study of the linear viscoelastic properties of polymers is the stress relaxation modulus, $G(t)$. The $G(t)$ curves, calculated from the mCG and \ac{SLSP} models of cPB chains with various lengths are provided in \autoref{fig:gt}.
\autoref{fig:gt}a and \autoref{fig:gt}b present the results of the \ac{mCG} and \ac{SLSP} model, respectively.  
Both the Rouse and the tube models predict a decay of $G(t)$ with the slope of $t^{-0.5}$ at short times ($t < \tau_\text{R}$ and $t < \tau_\text{e}$ for unentangled and entangled chains, respectively) and exponential decay of $G(t)$ at long times ($t > \tau_\text{R}$ and $t > \tau_\text{d}$ for unentangled and entangled chains).\cite{doi1988the,rubinstein2003polymer} However, the tube model predicts a plateau for $G(t)$ at intermediate times which is not predicted by the Rouse model.\cite{doi1988the,rubinstein2003polymer} 
It is clear from \autoref{fig:gt}a that all $G(t)$ curves decay with the slope of $t^{-0.5}$ at short times; at intermediate times, with increasing chain length, convergence to the expected behavior (plateau of $G(t)$) is observed.

Due to the high degree of coarse-graining, the \ac{SLSP} model cannot resolve the behavior at short time scales, $t<1\ \mu$s. Therefore we present in \autoref{fig:gt}b the stress relaxation modulus only for longer times; the short-time decay $G(t) \propto t^{-1/2}$ is omitted.  The \ac{SLSP} model, however, enables us to unlock the long timescales that are necessary to predict rheological properties of higher molecular weights. We can access chain lengths up to 2000-mer ($M = 108$ kg mol$^{-1}$).
Even for the highest molecular weight, the 2000-mer cPB, we do not observe a plateau in $G(t)$, as it is expected in the limit of infinite molecular weight according to the tube model. Instead, we observe a slight, but continuous decay of $G(t)$ that highlights the liquid properties of the material on all timescales. This effect can be partially rationalized by contour-length fluctuations of the tube.

\begin{figure}[!htb]
    \centering
    \begin{subfigure}{0.45\textwidth} % width of left subfigure
        \includegraphics[width=\textwidth]{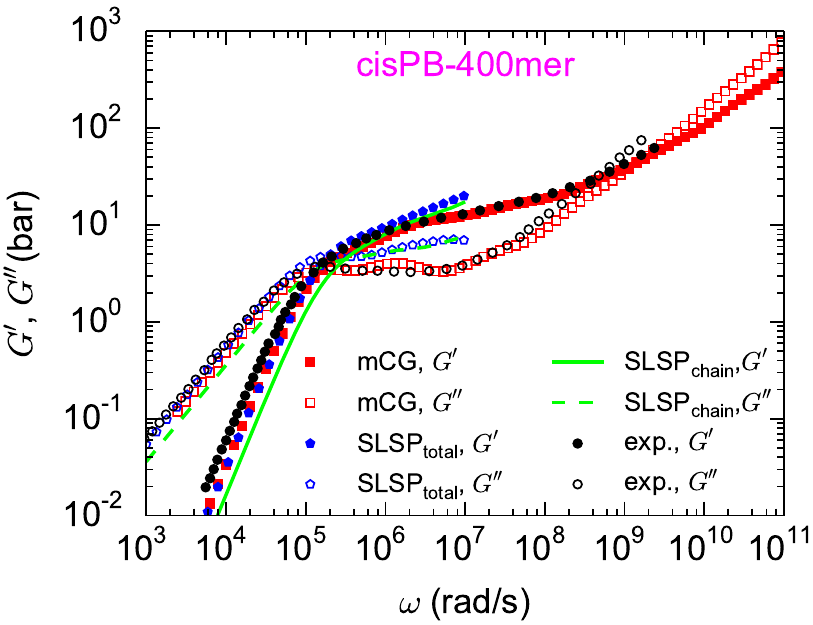}
    \end{subfigure}
    %\vspace{1ev} % here you can insert horizontal or vertical space
    \caption{Calculated (througth mCG and SLSP models) $G'(\omega)$ and $G''(\omega)$ for 400-mer cPB ($M = 21.6$ kg mol$^{-1}$) together with the experimental data\cite{liu2006probe} for a nearly monodisperse 1,4 rich PB ($M_\text{n} = 21.7$ kg mol$^{-1}$). The experimental data have been horizontally and vertically shifted by the experimentally determined shift factors.\cite{liu2006probe,colby1987melt}}
    \label{fig:G'-exp} % caption for whole figure
\end{figure}

The complex modulus of oscillatory shear, $G^*(\omega)$, can be calculated from $G(t)$ through:
\begin{equation}
G^*(\omega) = G'(\omega) + \mathrm{i} G''(\omega) = \mathrm{i}\omega \int_0^\infty \mathrm{e}^{-\mathrm{i}\omega t} G(t) \mathrm{d}t
\label{Eq:G*}
\end{equation}
where $\omega$ is the angular frequency, and $G'(\omega)$ and $G''(\omega)$ denote the storage and loss moduli. 
The above Laplace transform can be calculated by fitting $G(t)$ to a series of Maxwell modes and analytic calculation of the integral:
\begin{equation}
G(t) = \sum_{i=0}^n g_i \mathrm{e^{-t/\lambda_i}}
\label{Eq:maxwell}
\end{equation}
\begin{equation}
G'(\omega) = \sum_{i=0}^n \frac{g_i \omega^2 \lambda_i^2}{1 + \omega^2 \lambda_i^2},\ 
G''(\omega) = \sum_{i=0}^n \frac{g_i \omega \lambda_i}{1 + \omega^2 \lambda_i^2}
\label{Eq:G'}
\end{equation}
where $g_i$ and $\lambda_i$ are the modulus and relaxation time of the $i^\text{th}$ Maxwell mode, respectively. 
Note that, we found agreement between the  $G'(\omega)$ and $G''(\omega)$ calculated through fitting Maxwell modes on the $G(t)$ curves of mCG simulations with those obtained from the direct analytical transformation of $G(t)$ using the i-Rheo \emph{GT} tool.\cite{tassieri2018rheo} 
In \autoref{fig:G'-exp}, the calculated (through mCG and SLSP simulation) $G'(\omega)$ and $G''(\omega)$ curves for 400-mer cPB ($M = 21.6$ kg mol$^{-1}$) are presented together with the experimental data\cite{liu2006probe} for a nearly monodisperse 1,4 rich PB with an almost similar molecular weight ($M_\text{n} = 21.7$ kg mol$^{-1}$).
The experimental spectra were reported at $T = 298$ K; thus, here they have been horizontally and vertically shifted by the experimentally determined\cite{liu2006probe,colby1987melt} shift factors to generate the spectra at $413$ K ($a_T = 31$ and $b_T = 1.25$ were used for horizontal and vertical shifts, respectively).
Taking into account the different microstructures of the simulated and experimental samples and also uncertainties of the shift factors, a good agreement between simulation and experiment is observed.

\begin{figure}[!ht]
    \centering
    \begin{subfigure}{0.45\textwidth}
        \includegraphics[width=\textwidth]{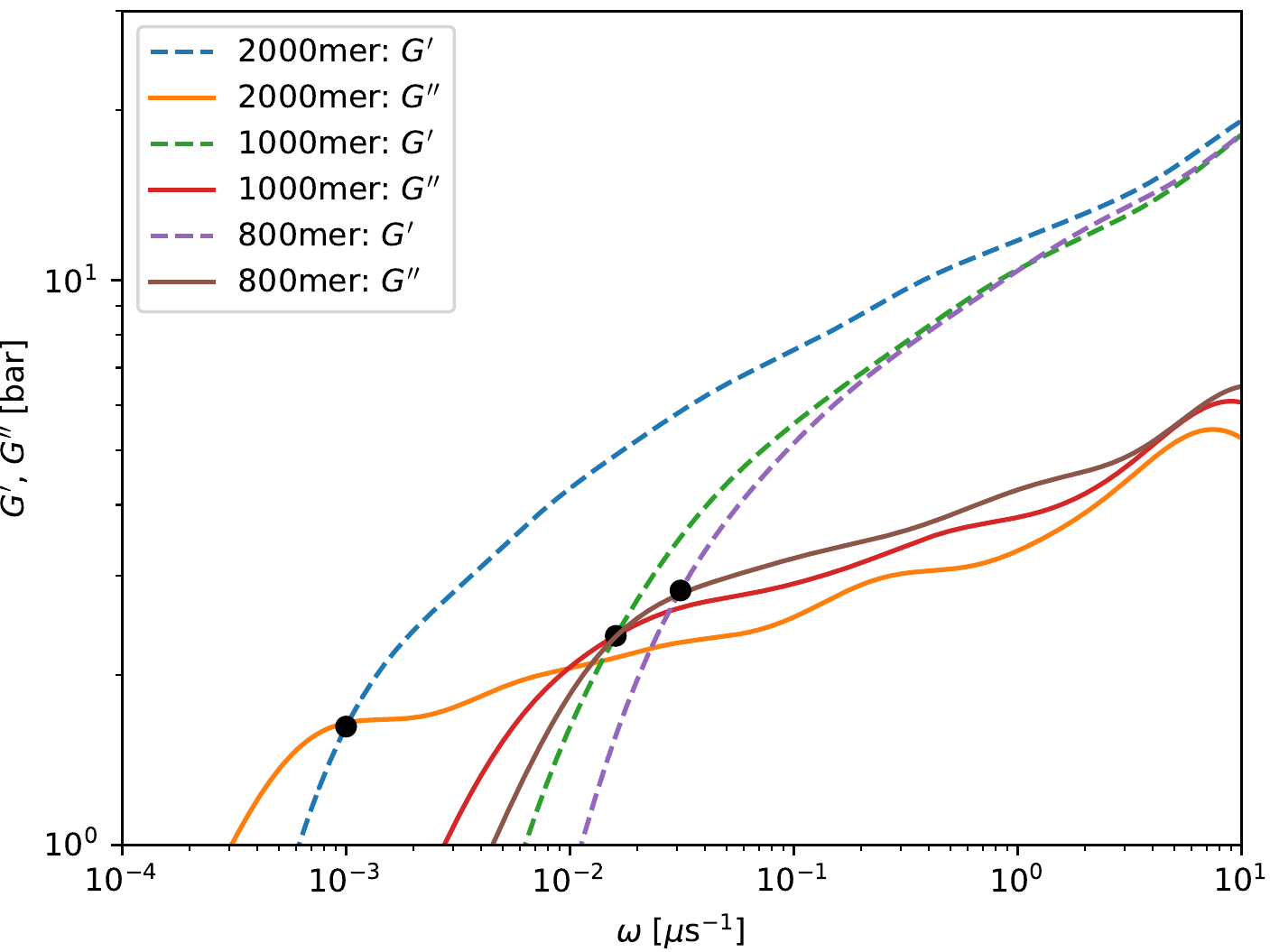}
        \end{subfigure}
    \begin{subfigure}{0.45\textwidth} % width of right subfigure
        \includegraphics[width=\textwidth]{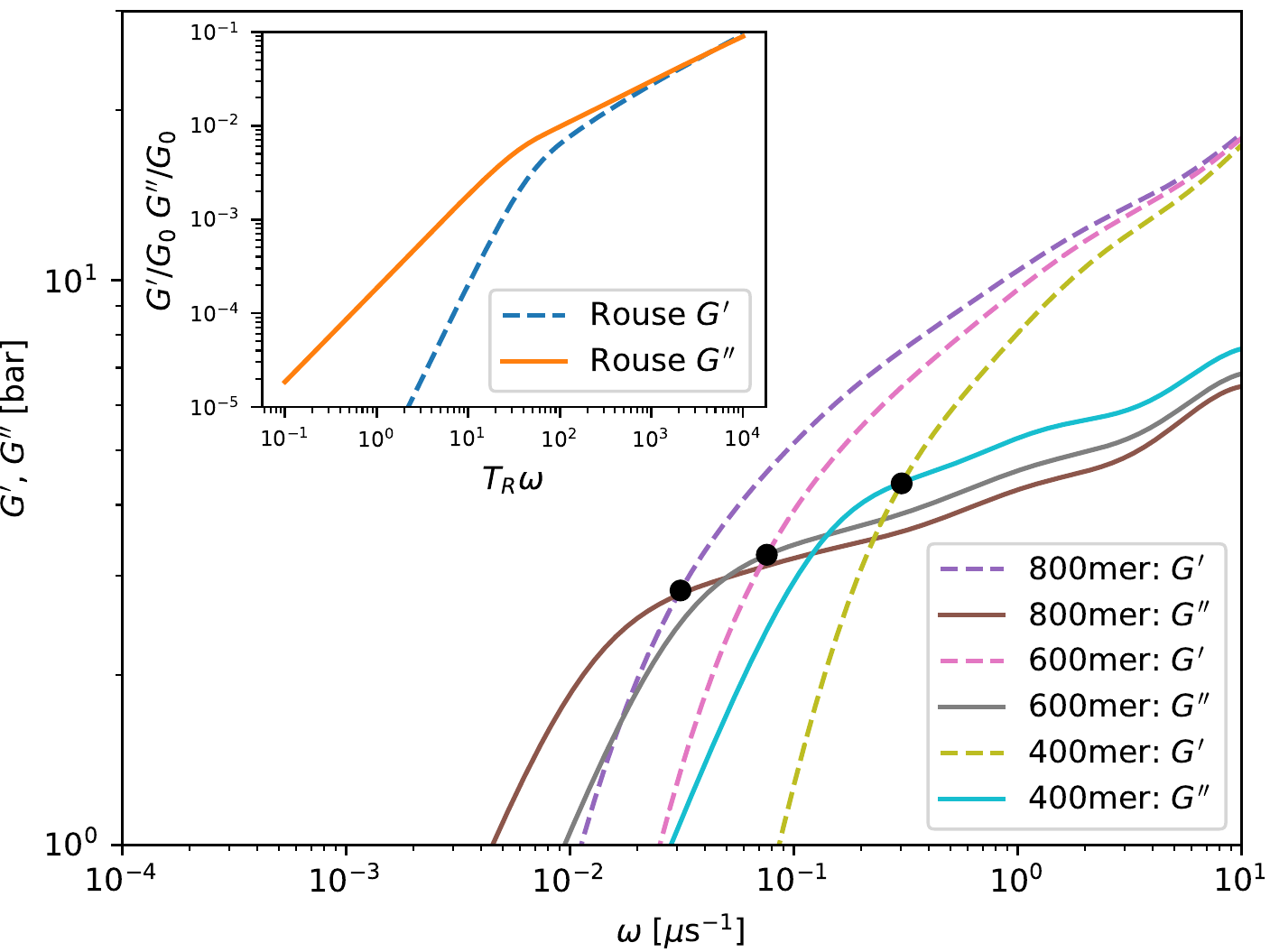}
        \end{subfigure}
    \caption{Molecular-weight dependence of storage modulus, $G'(\omega)$, and loss modulus, $G''(\omega)$ calculated through the \ac{SLSP} model for \ac{cPB} at $413$ K. For clarity we split the molecular weights to two panels. The longest relaxation time can be identified by the crossing point of $G'$ and $G''$ at each molecular weight.
    The validity range of the \ac{SLSP} model, $t>1\mu$s, limits our predictions to low frequencies. The quality of the Maxwell model fitting affects the quality of the predicted moduli. The inset display the Rouse characteristics for reference.}
    \label{fig:slsp-gcomplex} 
\end{figure}

Fitting the Maxwell modes to the \ac{SLSP} model allows us to calculate the complex moduli for higher molecular weights. The range validity of the \ac{SLSP} model limits the range of frequencies that we can examine to below $10^7$rad/s. For higher frequencies, we can use the \ac{mCG} model as described earlier.
\autoref{fig:slsp-gcomplex} plots the obtained moduli for the 400-mer up to the 2000-mer cPB. Whereas the high-frequency behavior is not significantly affected by molecular weight, we observe that the low-frequency moduli strongly increase with chain length and that the curves extend to lower frequencies. This behavior is expected because cPB with higher molecular weight exhibits a more solid-like behavior at large, intermediate timescales, and the terminal relaxation time increases with $M_\text{w}$.

While the exact shape of the dynamic moduli is influenced by the chain averaging (appendix \ref{sec:appendix-slsp-gt}) and the quality of the Maxwell model fit, we can still determine the longest relaxation time by the first crossing point of $G'$ and $G''$ as in indicated in \autoref{fig:slsp-gcomplex}.
Even as the $G''$ does not feature a negative slope, the crossing is a signature of the entanglement dynamics.
The Rouse model prediction from the moduli (inset of \autoref{fig:slsp-gcomplex}) does not feature a crossing point of the moduli.

\subsubsection{Zero-shear viscosity}

\begin{figure}[!htb]
    \centering
    \begin{subfigure}{0.45\textwidth} % width of left subfigure
        \includegraphics[width=\textwidth]{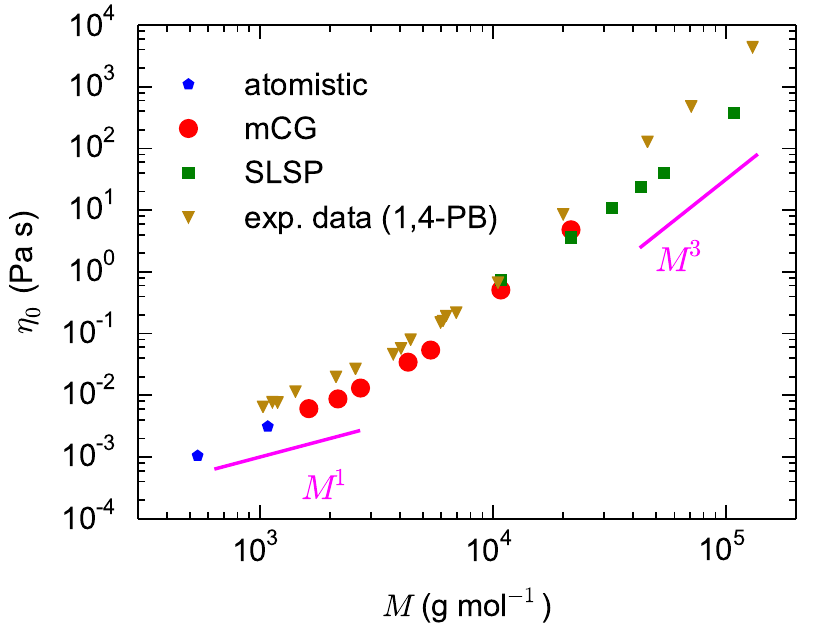}
    \end{subfigure}
    %\vspace{1ev} % here you can insert horizontal or vertical space
    \caption{Molecular-weight dependence of zero-shear viscosity for a bulk melt of cPB at $413$ K. Depending of molecular weight, the results of the atomistic, mCG, and SLSP simulations are reported (for $M = 10.8$ and $21.6$ kg mol$^{-1}$ the results of both mCG and SLSP models are shown). The experimental viscosities (adjusted to $413$ K) for samples of nearly monodisperse 1,4 rich PB  are also shown.\cite{colby1987melt} }
    \label{Fig:eta} % caption for whole figure
\end{figure}

The last part of our analysis concerns zero-shear viscosity, $\eta_0$, which can be calculated from $G(t)$ as:
\begin{equation}
\eta_0 = \int_0^\infty \mathrm{d}t \; G(t) 
\label{Eq:eta}
\end{equation}
The molecular-weight dependence of $\eta_0$, is presented in \autoref{Fig:eta}. Depending on chain length, the results of atomistic, mCG, or SLSP simulations are reported; however, for 200-mer and 400-mer chains, the viscosities from both mCG and SLSP models are presented, that exhibit very good agreement.
The experimental\cite{colby1987melt} values of $\eta_0$ for nearly monodisperse 1,4-PB  samples (less than $10\%$ vinyl content, containing both cis and trans carbon double bonds) are also presented; the experimental values of $\eta_0$ were measured at lower temperatures than $413$ K and have been shifted to $413$ K, using the reported shift factors.\cite{colby1987melt} Considering different microstructure of the simulated and experimental PB chains, a fair agreement between simulation and experimental data is observed. Consistent with the data of diffusion coefficients (see \autoref{Fig:D}), the viscosities of the simulated \ac{cPB} samples are lower than those of the experimental 1,4-PB samples. The latter is understandable if we consider that the experimental data concern samples containing vinyl- and trans- PB components, both of which have a larger glass transition temperature than cPB.  

The Rouse and the tube model predict $\eta_0 \propto M$ and $\eta_0 \propto M^3$ scaling relations for unentangled and entangled chains, respectively.\cite{doi1988the,rubinstein2003polymer} 
As can be seen from \autoref{Fig:eta} for both experimental and simulation data, deviations from the predictions of the Rouse and tube models are observed.
Note that, at the entangled regime, the simulation results still do not show a constant scaling exponent, and the exponent increases with molecular weight. 
As mentioned for the data of $D$ (\autoref{Fig:D}), the chain-end effect in the unentangled regime and contour length fluctuations and constraint release in the entangled regime are well-known mechanisms affecting the scaling exponent.\cite{colby1987melt}

\subsection{Computational efficiency}

In this section, we briefly highlight why it is necessary to have the three tiers of modeling and the computational advantage of the hierarchical multi-scale methodology, by comparing the computational cost of the three used models. 
The parameter passing from the atomistic \textit{via} the \ac{mCG} to the \ac{SLSP} model offers the opportunity to compare the computational efficiencies of the three particle-based models. 

As a reference system, we chose 926 chains of a 400-mer cPB and calculate the estimated physical time it takes to propagate the system for $1\mu s$  at $413$ K. Such a system involves $1\,481\,600$ CH$_x$ groups in the united-atom model, $370\,400$ beads in the mCG representation and $29\,632$ particles in the \ac{SLSP} scale.
This system size can be straightforwardly simulated by the \ac{SLSP} model. However, it is difficult to simulate such system sizes with the \ac{mCG} model and it is out of the scope of atomistic simulations. Therefore we have extrapolated the required times from systems of smaller size for these two models.

The \ac{SLSP} model is implemented for HOOMD-blue, which is optimized for the execution of Nvidia \acp{GPU}. To simulate the described benchmark we require approximately $590$ s of computing time on an Nvidia V100 GPU.
The \ac{mCG} and atomistic simulations are executed on CPUs. Running with 500 particles per core, the \ac{mCG} simulation of the above-mentioned system takes around $10$ hours on 740 cores (Intel Xeon Gold 6148 @ 2.4 GHz) and the atomistic simulation takes around $10$ days on 2960 cores.

 \begin{figure}
    \centering
    %\begin{minipage}{0.49\textwidth}
    \includegraphics[width=0.75\textwidth]{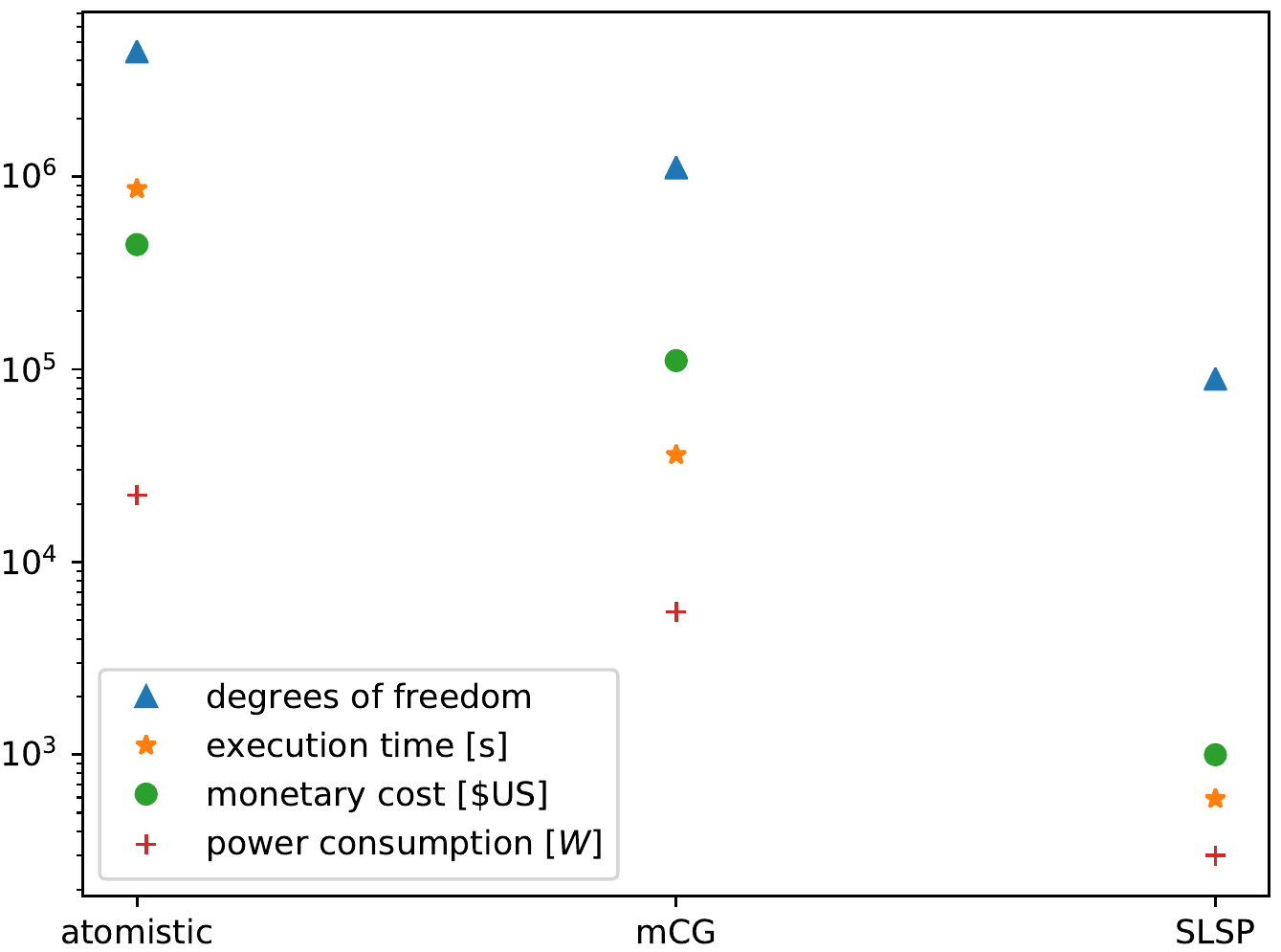}
    %\end{minipage}
    %\hfill
    %\begin{minipage}{0.5\textwidth}
    \begin{tabular}{|l|c|c|c|}
    \hline
    model: & atomistic &      mCG     &    SLSP\\
    \hline
    simulation time & $1\mu$s & $1\mu$s & $1\mu$s \\
    particles       & $1\,481\,600$ & $370\,400$ & $29\,632$\\
    execution time [s] & $\approx 864\,000$ & $\approx 36\,000$ & $590$\\
    CPUs [Intel Xeon Gold 6148] & $148$ & $37$ & - \\
    GPUs [Nvidia V100] & - & - & $1$ \\
    cost of CPUs [$\approx\$$US] & $444\,000$& $111\,000$ & - \\
    cost of GPU [$\approx\$$US] & - & - & $10\,000$\\
    power consumption CPUs [TDP $W$] & $22\,200$ & $5\,550$ & - \\
    power consumption GPU [max $W$] & - & - & $300$\\
    \hline
    \end{tabular}
    %\end{minipage}
    \caption{Overview of the computational costs of the three different models in comparison (at $413$ K).}
    \label{fig:computation}
\end{figure}

\autoref{fig:computation} summarizes important aspects of the three different models. The united-atom simulations involve a factor of $4$ more degrees of freedom than the \ac{mCG} model but require a factor of $\approx 100$ more computational resources. This additional speedup by a factor of $\approx 25$ stems from its lower friction coefficient ($S \approx 5$) and the larger time step permissible ($5$ fs).
Since the atomistic and \ac{mCG} models use CPUs whereas the \ac{SLSP} model employs GPUs a quantitative comparison is not straightforward. Nevertheless, we can appreciate from the data that the speedup by a factor of $61 \times 37$ in seconds (1 Xeon CPUs \textit{vs} V100 GPU) from the \ac{mCG} model to our \ac{SLSP} model is more than the mere reduction of the number of degrees of freedom, a factor $12.5$, and the GPU-acceleration of molecular dynamics codes, roughly a factor $6.5$.
\footnote{From \url{https://developer.nvidia.com/hpc-application-performance} we estimate that 13 Dual Xeon Gold 6240@2.60GHz correspond to 4 V100-GPUs for GROMACS or LAMMPS.} The additional speedup by a factor of $\approx 28$ stems from the soft interactions.
Note that, it is expected that the computational efficiencies of the coarser models are temperature-dependent and exponentially increase with decreasing temperature.

An alternative to set the different architectures in perspective consists of comparing the approximate acquisition costs or power consumption of the key components (only CPU or GPU but not the supporting infrastructure). For instance, the models behave like $65\,000$ : $677$ : $1$ in terms of acquisition costs.

The comparison of the different models emphasizes that atomistic models are required to account for the chemical specificity but also indicates that they cannot access the time and length scales relevant to the rheology of long macromolecules. Using a three-model parameter-passing strategy we can successfully bridge from the atomistic description to our \ac{SLSP} model that can investigate the rheological properties of macromolecular materials. From \autoref{Fig:D} and \autoref{Fig:eta} it is apparent that even a direct mapping from atomistic to the soft, \ac{CG} (SLSP) model is not yet feasible with the available computational resources; the scales are too far apart. Thus, the \ac{mCG} description is required for sufficient overlap. \autoref{Fig:D} and \autoref{Fig:eta} illustrates that the \ac{mCG} model covers an intermediate regime of molecular weights between the atomistic and \ac{SLSP} model. 

\section{Summary} 

We propose a systematic, mainly bottom-up, hierarchical coupling of simulation models from three different scales to consistently model-specific polymer melts over broad spatiotemporal scales and predict their rheological behavior directly from the monomeric structure.
As a reference system, we apply the methodology to \acf{cPB} melts. 
We start with a validated force field for a united-atom model.
With a standard structural-based method (here iterative Boltzmann inversion), we can transfer the static properties of the united \acf{cPB} to a  moderately \ac{CG} (\ac{mCG}) model.
To match \ac{mCG} and atomistic dynamics, the \ac{mCG} time is scaled by a proper time scaling factor, $t^*_\text{mCG}$, which compensates for the lower monomeric friction coefficient of the \ac{mCG} model than that of the atomistic model. $t^*_\text{mCG}$ is chain length and temperature-dependent; its chain length dependence is in phase with the chain-length dependence of melt density and both, $t^*_\text{mCG}$ and density, tend to constant values for sufficiently long chains.

\begin{figure}[!htb]
    \centering
        \includegraphics[scale=0.8]{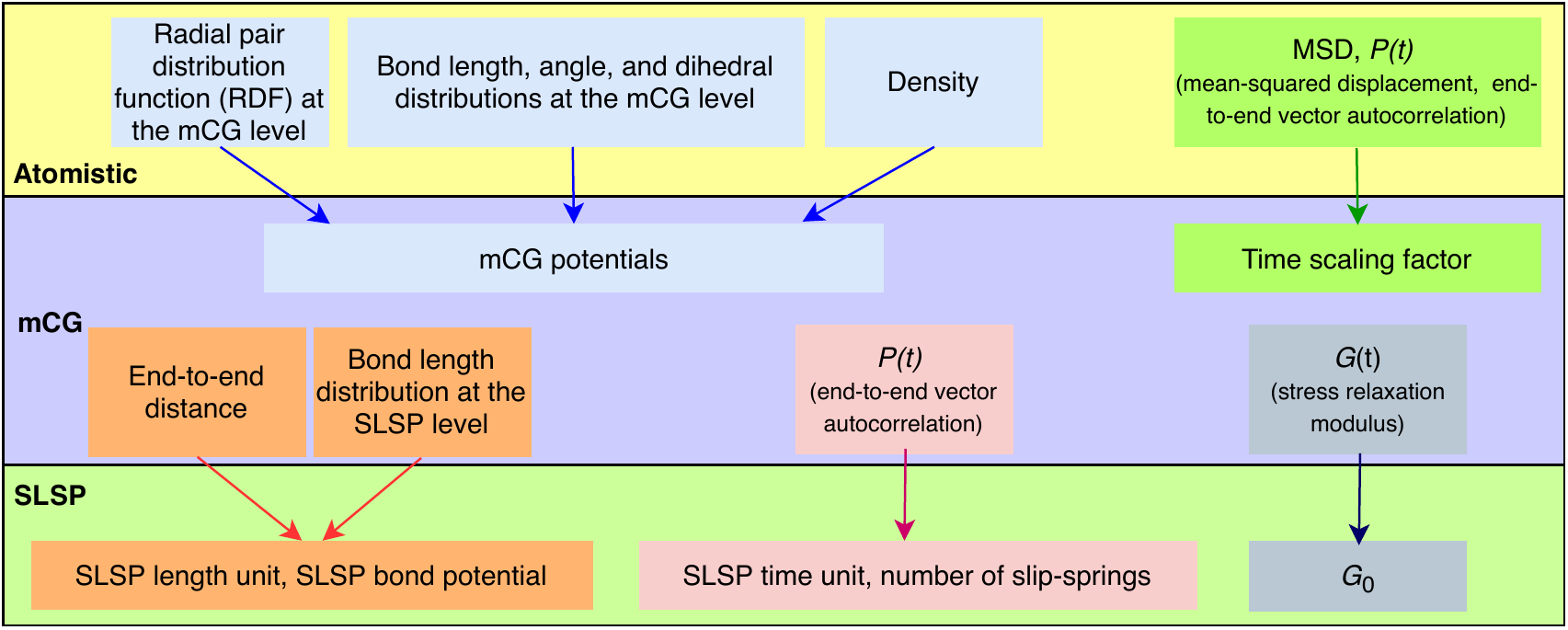}
    \caption{Summary of the parameterization steps for the mCG and SLSP models.
    The mCG potentials are derived by matching bulk density and local structural distribution of the mCG chains to those of the atomistic chains. To match \ac{mCG} and atomistic dynamics, the \ac{mCG} time is scaled by a proper time scaling factor, $t^*_\text{mCG}$, which is calculated by comparing MSD and $P(t)$ (orientational autocorrelation function of the end-to-end vector) of the two models. With increasing chain length, $t^*_\text{mCG}$ tends to a constant value.
    At the SLSP level, the calibration of the length unit and the determination of the shape of the bond length distribution are performed based on the end-to-end distance and bond length distribution of the mCG chain. The number of slip-springs and the time unit is set based on the end-to-end vector correlation, $P(t)$, of the more detailed model. Finally, stress relaxation modulus at $t = 0$, $G_0$, is determined based on the short time behavior of $G(t)$ of the lower-level model.     
    }
    \label{Fig:flowchart} % caption for whole figure
\end{figure}

For the third (mesoscopic) layer of the polymer models, we use the \acf{SLSP} model.
The transferred static properties from \ac{mCG} model to \ac{SLSP} model are at long length scales; 
so, in the \ac{SLSP} model, the chain configurations can be considered as a Gaussian chain, simplifying the handover of static parameters from the moderately \ac{CG} to the \ac{SLSP} model.
However, because the \ac{SLSP} model serves at a high degree of coarse-graining, the nonbonded interactions are soft and do not prevent backbone crossing. 
Therefore, without additional \acp{SLSP}, the model would not be able to capture the correct entangled dynamics. 
Hence, we have two kinetic parameters to pass from the moderately \ac{CG} model to the \ac{SLSP} model.
i) a timescale mapping, and ii) the number of entanglements, which are represented via \acp{SLSP} in the latter model.
We find that the end-to-end vector correlation, $P(t)$, allows a good mapping of both parameters.
We could verify the mapping with a comparison of the single-chain \ac{MSD} and the collective stress-relaxation modulus, $G(t)$.
The mappings are transferable to higher molecular weights, which enables us to not only access longer time and length scales with the coarse-grained models, it also allows us to explore industrially relevant high molecular weights. A summary of the parameterization steps for the mCG and SLSP models
are provided in \autoref{Fig:flowchart}.

Finally, we verify our simulation results by measuring rheological properties such as diffusion and viscosity as a function of molecular weight.
We could verify our results against selected experimental results as well as the predicted power-law scalings of these properties.

Overall, we construct three levels of modeling \acf{cPB}.
By matching static and kinetic parameters between the models we can make quantitative predictions from the level of individual atoms ($1$ \AA) up to high molecular weight macromolecules ($100$ nm),
and explore timescale from picoseconds up to several hundred microseconds.

\section{Acknowledgments}
The work was supported by computational time granted from the Greek Research \& Technology Network (GRNET) in the National HPC facility ARIS under a project named POL-COMP-TIRE.
V.H. acknowledges support by project “SimEA”, funded by the European Union’s Horizon 2020 research and innovation programme under Grant Agreement No. 810660.
This work is supported by the Goodyear Tire and Rubber Company.

\appendix
\section{Appendix}\label{sec:slsp-appendix}
\subsection{Stress autocorrelation in SLSP model}\label{sec:appendix-slsp-gt}
An accurate calculation of $G(t)$ is challenging because of the high noise in the virial tensor. For the SLSP model, we attack this problem by calculating the stress for the polymer chains individually and average the results of these single-chain stresses.

All forces in the model are pairwise forces, hence the virial is calculated overall interacting particle pairs. This includes all bonded and nonbonded interactions. For each pair, the stress is equally distributed to both participating particles. Summing up all contributions in a molecule results in the individual chain stresses.
This allows for better statistics because the result of each chain can be averaged.
We note that the summation of all chain stresses results in the same as the total stress, however, the autocorrelation is not the same, as interchain correlations are sacrificed for noise reduction.

The standard method, employed for the atomistic and \ac{mCG} model simulations, is the calculation of the total stress in the system and its autocorrelation.

\begin{figure}
    \centering
    \begin{subfigure}{0.46\textwidth}
        \centering
        \includegraphics[width=\textwidth]{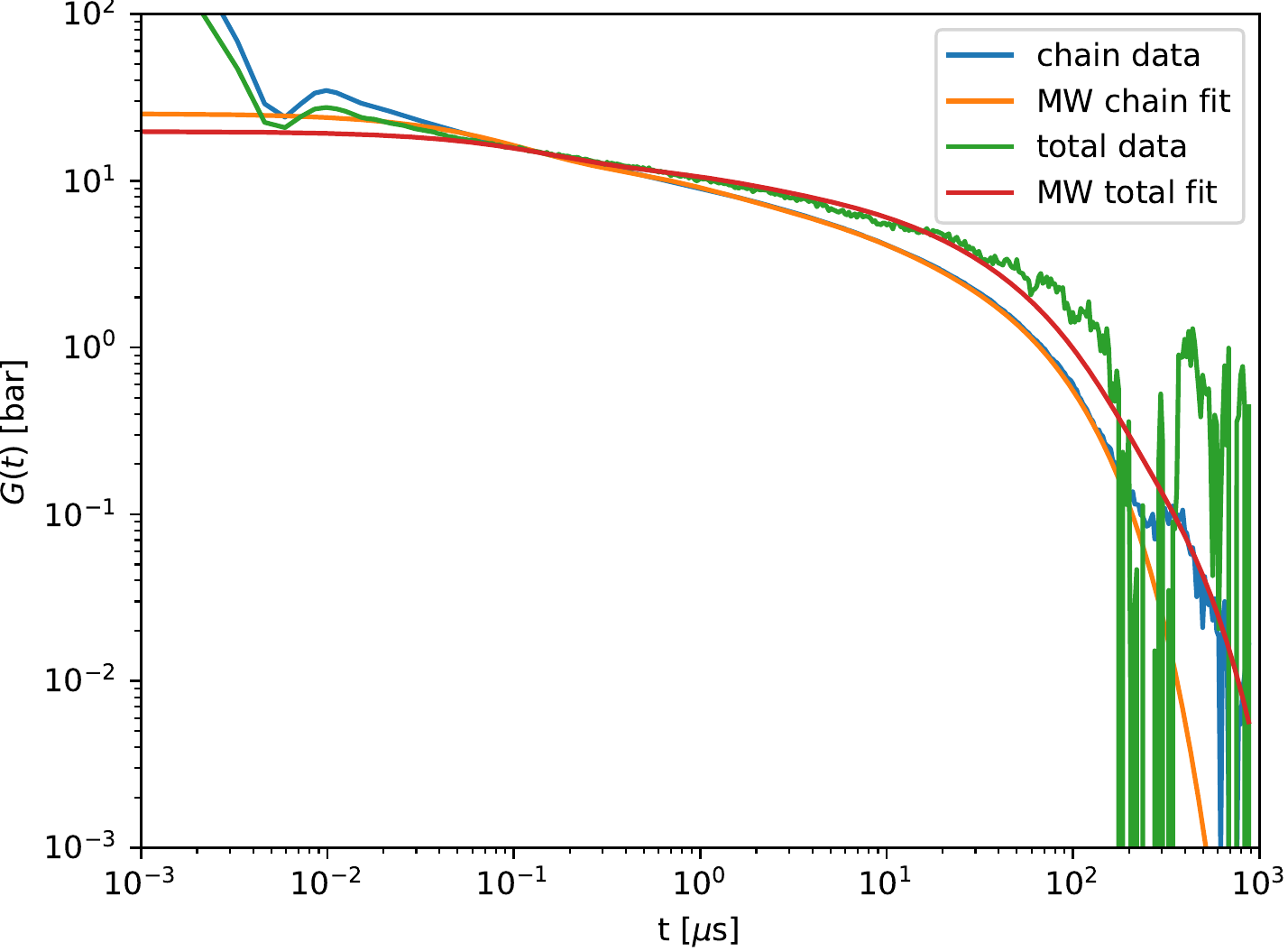}
        \caption{$G(t)$ calculated per chain and in the total system.}
        \label{fig:appendix-slsp-Gt-total}
    \end{subfigure}
    \hfill
    \begin{subfigure}{0.46\textwidth}
        \centering
        \includegraphics[width=\textwidth]{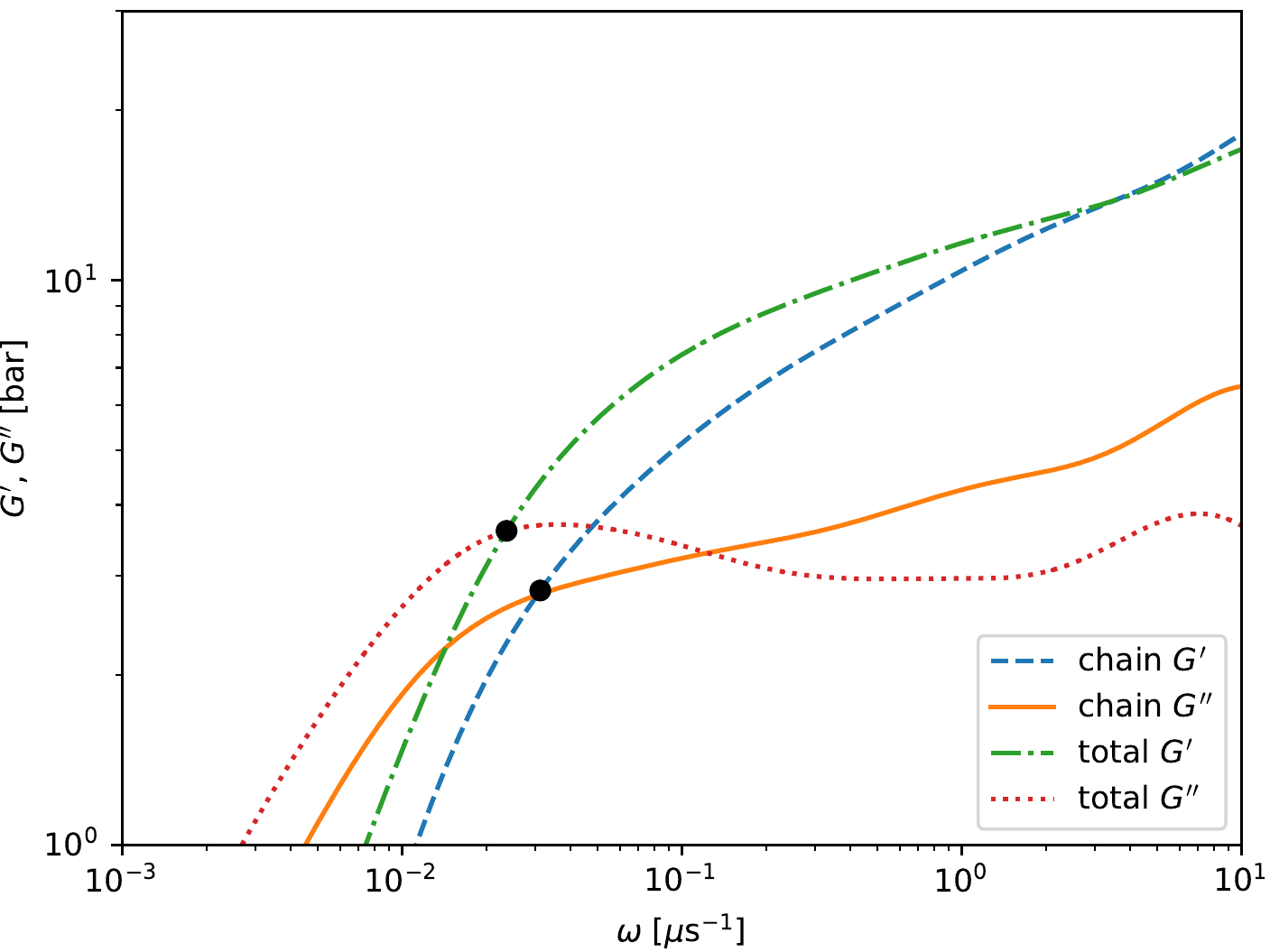}
        \caption{Dynamic moduli of both $G(t)$ option in comparison.}
        \label{fig:slsp-dynamic-total}
    \end{subfigure}
    \caption{Comparison of the $G(t)$ calculated as chain average versus the calculation via the total stress with a 800mer chain. The chain average offers the better statistics, but has a slightly softer decay. The important longest relaxation time is captured correctly with both models.}
    \label{fig:slsp-total}
\end{figure}

\autoref{fig:slsp-total} compares the two methods. The chain-averaging strategy offers significantly better statistics, which results in better fits for the Maxwell model. The decay of $G(t)$, however, is softer than that obtained by the total stress correlation. There is no pronounced plateau of $G(t)$, obtained \textit{via} the chain-averaging method. As a consequence, the loss modulus for the considered chain lengths does not feature the negative slope after the first maximum. Both, a plateau of $G(t)$ and a negative slope of $G''(\omega)$, are considered key signatures of reptation dynamics of entangled chains. We note that this observation is not a shortcoming of the underlying \ac{SLSP} model but rather of the measurement of the stress autocorrelation $G(t)$. The total stress autocorrelation captures these effects in our simulations.

For this manuscript, we opt for the higher precision of the chain-averaging method. We focus on the important longest relaxation time, which is captured in both scenarios but is more easily identified with better statistics. The longest relaxation time of an entangled polymer melt can be identified as the lowest frequency where the dynamic moduli $G'$ and $G''$ cross each other as marked in \autoref{fig:slsp-dynamic-total}.

\subsection{Segmental mean-squared displacement of unentangled chains}
\label{sec:g1-rouse}
In this section, we compare the segmental mean-squared displacements, $g_1(t)$, of unentangled cPB chains with the predictions of the Rouse model for the $g_1(t)$ of short chains (\ie, few Rouse modes):\cite{doi1996introduction}
\begin{equation}
\frac{g_1(t)}{R_\text{ee}^2} = \frac{2}{\pi^2} \left[ 
\frac{t}{\tau_\text{R}} + \sum_{p=1}^{N_\text{R}-1} \frac{1}{p^2} \left(1-e^{-p^2t/\tau_\text{R}} \right)\right]   
\label{Eq:g1-Rouse}
\end{equation}
here $N_\text{R}$ is number of Rouse segments of the chain, $\tau_\text{R}$ is the Rouse time, and $R_\text{ee}^2$ is the end-to-end distance.
\autoref{Eq:g1-Rouse} predicts three scaling regimes for $g_1(t)$: at short times, $g_1(t) \propto t$; at intermediate times $g_1(t) \propto t^x$, where $0.5 \leq x \leq 1$ depending on $N_\text{R}$ (for large values of $N_\text{R}$, $x = 0.5$); finally, at long times $g_1(t) \propto t$ (normal diffusion regime).  
For using \autoref{Eq:g1-Rouse}, we calculate $\tau_\text{R}$ from diffusion coefficient, $D$, through:\cite{doi1996introduction}
\begin{equation}
\tau_\text{R} = \frac{R_\text{ee}^2}{3\pi^2 D}
\label{Eq:D-tau_r}
\end{equation}

\begin{figure}[!htb]
    \centering
    \begin{subfigure}{0.45\textwidth}
        \includegraphics[width=\textwidth]{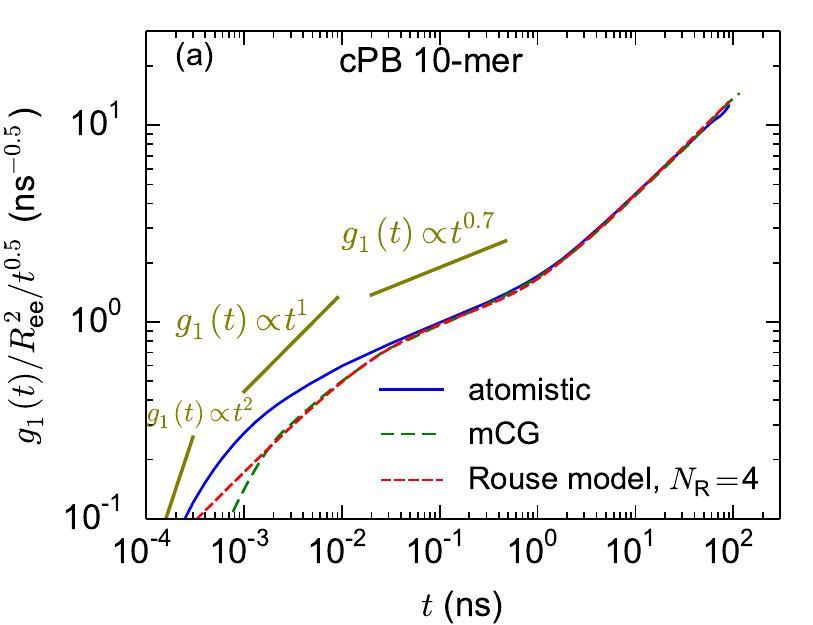}
        \end{subfigure}
    
    \begin{subfigure}{0.45\textwidth} % width of right subfigure
        \includegraphics[width=\textwidth]{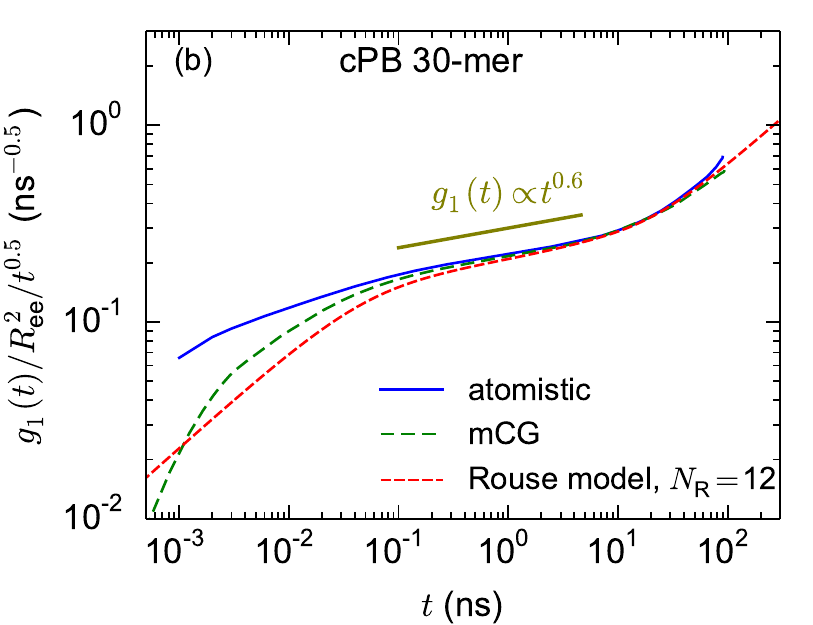}
    \end{subfigure}
    
    \begin{subfigure}{0.45\textwidth} % width of right subfigure
        \includegraphics[width=\textwidth]{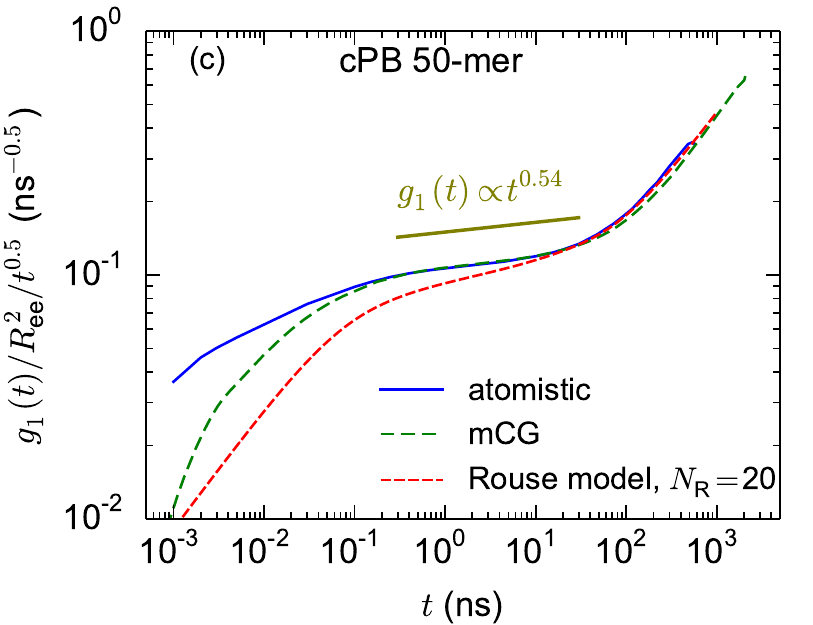}
    \end{subfigure}
    \caption{Mean-squared displacement of monomers, $g_1(t)$, normalized by $t^{0.5}$    for (a) 10-mer, (b) 30-mer, and (c) 50-mer cPB chains from atomistic and mCG simulations. The predicted $g_1(t)$ curves of the Rouse model for chains containing finite number of segments, $N_\text{R}$, are included; in all cases, one Rouse segment corresponds to around $2.5$ monomeric repeating units ($\approx$ one Kuhn segment).}
    \label{Fig:g1-Rouse} % caption for whole figure
\end{figure}

The only free parameter for fitting is $N_\text{R}$, the number of Rouse segments. 
To determine $N_\text{R}$ we fitted 
\autoref{Eq:g1-Rouse} to the $g_1(t)$ of 10-mer chains and determined $N_\text{R}$ for them; for other chain lengths, the $N_\text{R}$ values were calculated from the $N_\text{R}$ of 10-mer chains ($N_\text{R}$ is proportional to chain length).           

\autoref{Fig:g1-Rouse}a shows the $g_1(t)$ of 10-mer 
atomistic and mCG chains together with the prediction of the Rouse model with $N_\text{R} = 4$.
This $N_\text{R}$ value is in agreement with the number of Kuhn segments per a 10-mer chain.
One Kuhn segment of the model cPB chains contains around $9$ backbone bonds;\cite{behbahani2020conformations} 
therefore, a 10-mer cPB chain (containing $39$ backbone bonds) contains around $4$ Kuhn segments.
After the very short time ballistic regime (in which $g_1(t) \propto t^2$), $g_1(t)$ of 10-mer mCG chains exhibit a perfect match with the result of the Rouse model. 
After the ballistic regime, $g_1(t)$ of mCG model shows a distinct $g_1(t) \propto t^1$ regime, which is in agreement with the Rouse model; such a short time linear regime is not clearly seen in the $g_1(t)$ of the atomistic chains. However, the Rouse model well describes the subdiffusive regime of the  $g_1(t)$ of 10-mer atomistic chain.    

\autoref{Fig:g1-Rouse}b-c shows the $g_1(t)$ of  30-mer and 50-mer atomistic and mCG chains together with the predictions of the Rouse model with  $N_\text{R} = 12$ and $20$, respectively.
Here, as in the case of 10-mer chains, one Rouse segment corresponds to approximately one Kuhn segment.
For 30-mer and 50-mer chains  
the agreement between the simulation results and the Rouse model is not as good as for 10-mer cPB.
Particularly, for these cases, the Rouse model overestimates the slope of $g_1(t)$ at the intermediate subdiffusive regime (the effect is more pronounced for 50-mer cPB).  
The probable origin of this behavior is the ignorance of the non-crossability of chains in the Rouse model. The effect of non-crossability is higher for longer chains and finally leads to the appearance of entangled dynamics (as discussed in the main manuscript,
for cPB, the signs of the entangled dynamics emerge around chain lengths of $50$--$80$ monomers).

\bibliographystyle{achemso}
\bibliography{references}

\begin{acronym}[RNEMDS]
  \setlength{\parskip}{0ex}
  \setlength{\itemsep}{0ex}
  \acro{AFM}{atomic force microscopy}
  \acro{ALU}{arithmetic logic unit}
  \acro{API}{application programming interface}
  \acro{BCC}{body-centered cubic}
  \acro{CG}{coarse-grained}
  \acro{CLT}{central limit theorem}
  \acro{CPU}{central processing unit}
  %\acro{CUDA}{compute unified device architecture}
  \acro{DFG}{Deutsche Forschungs Gemeinschaft}
  \acro{DPD}{dissipative particle dynamics}
  \acro{DSA}{directed self-assembly}
  \acro{FENE}{finite extensible nonlinear elastic}
  \acro{FJC}{freely jointed chain}
  \acro{FRC}{freely rotating chain}
  \acro{GCD}{greatest common divisor}
  \acro{GPGPU}{general purpose computing on graphics processing units}
  \acro{GPU}{graphics processing unit}
  \acro{HDF5}{hierarchical data format version 5}
  \acro{HOOMD}{Highly Optimized Object-oriented Many-particle Dynamics}
  \acro{HPC}{high performance computing}
  \acro{IO}{Input/Output}
  \acro{JSC}{Jülich Supercomputing Centre}
  \acro{KIT}{Karlsruhe Institute of Technology}
  \acro{LAOS}{large amplitude oscillatory shear}
  \acro{LEBC}{Lees-Edwards boundary condition}
  \acro{LGPL}{GNU lesser general public license}
  \acro{MC}{Monte-Carlo}
  \acro{mCG}{moderately coarse-grained}
  \acro{MCS}{Monte-Carlo steps}
  \acro{MD}{molecular dynamics}
  \acro{MFA}{mean-field approximation}
  \acro{MPTPS}{million particle time steps per second}
  \acro{MPI}{message passing interface}
  \acro{MSD}{mean-squared displacement}
  \acro{MT}{Mersenne-Twister}
  \acro{ODT}{order-disorder-transition}
  \acro{ORNL}{Oak Ridge National Laboratory}
  \acro{P2VP}{poly(2-vinylpyridine)}
  \acro{PAN}{poly(acrylonitrile)}
  \acro{PB}{poly(butadiene)}
  \acro{cPB}{\cis-polybutadiene}
  \acro{PCG}{permuted congruential generator}
  %\acro{PE}{polyethylene}
  \acro{PEE}{poly(ethylethylene)}
  \acro{PEMA}{poly(ethyl methacrylate)}
  \acro{PEO}{poly(ethylene oxide)}
  \acro{PEP}{poly(ethylene-propylene)}
  \acro{PI}{poly(isoprene)}
  \acro{PMMA}{poly(methyl methacrylate)}
  \acro{PRNG}{pseudo random number generation}
  \acro{PS}{polystyrene}
  \acro{RDF}{radial distribution function}
  \acro{RESPA}{reference system propagator algorithms}
  \acro{RNEMDS}{reverse nonequilibrium molecular dynamics simulation}
  \acro{SAW}{self-avoiding random walk}
  \acro{SAXS}{small angle X-ray scattering}
  \acro{SCFT}{self-consistent field-theory}
  \acro{SCMF}{single-chain-in-mean-field}
  \acro{SLSP}{slip-spring}
  \acro{SMC}{smart Monte-Carlo}
  \acro{SOMA}{SOft coarse grained MC Acceleration}
  \acro{SST}{strong-segregation theory}
  \acro{TEM}{transmission electron microscopy}
  \acro{TPS}{time steps per second}
  \acro{XDMF}{extensible data model and format}
  \acro{XML}{extensible markup language}
  \acro{XL}{cross-link}
  \acro{cPI}{cis-poly(isoprene)}
\end{acronym}

\end{document}